\documentclass[prb,aps,twocolumn,groupedaddress,showpacs,floatfix]{revtex4}
\usepackage{amsmath}
\usepackage{amssymb}
\usepackage{bm}
\usepackage{graphicx}
\usepackage{dcolumn}

\renewcommand{\Im}{{\, \mathrm{Im}}}

\newcommand{\tst}{\textstyle}
\newcommand{\be}{\begin{equation}}
\newcommand{\ee}{\end{equation}}
\newcommand{\OO}{{\mathcal{O}}}
\newcommand{\RR}{{\mathcal{R}}}
\newcommand{\TT}{{\mathcal{T}}}
\newcommand{\txthalf}{\textstyle{\frac{1}{2}}}
\newcommand{\ra}{\rightarrow}
\newcommand{\p}{^{\prime}}
\newcommand{\tr}{{\mathrm{tr}}}

\newcommand{\natom}{{N_{{\mathrm{atom}}}}}
\newcommand{\nnode}{{N_{{\mathrm{node}}}}}
\newcommand{\deltaN}{{\delta}N}

\begin{document}

\title{Coarse-grained molecular dynamics:\\
Nonlinear finite elements and finite temperature}

\author{Robert E. Rudd}\thanks{Corresponding author} 
\email[E-mail:~]{robert.rudd@llnl.gov}
\affiliation{Lawrence Livermore National Laboratory, Condensed Matter Physics Div., L-045, Livermore, CA 94551 USA }

\author{Jeremy Q. Broughton}
\affiliation{JPMorgan Chase Bank, New York, NY 10017 USA }

\begin{abstract}
Coarse-grained molecular dynamics (CGMD) is a technique developed
as a concurrent multiscale model that couples conventional 
molecular dynamics (MD) to a more coarse-grained description
of the periphery.  The coarse-grained regions are modeled on
a mesh in a formulation that generalizes conventional finite
element modeling (FEM) of continuum elasticity.  CGMD is 
derived solely from the MD model, however, and has no
continuum parameters.  As a result, it provides a coupling
that is smooth and provides control of errors that arise
at the coupling between the atomistic and coarse-grained
regions.  In this article, we elaborate on the formulation
of CGMD, describing in detail how CGMD is applied to anharmonic
solids and finite temperature simulations.  As tests of CGMD,
we present in detail the calculation of the phonon spectra
for solid argon and tantalum in 3D, demonstrating how CGMD provides a better
description of the elastic waves than that provided by FEM.  We
also present elastic wave scattering calculations that show
the elastic wave scattering is more benign in CGMD than FEM.
We also discuss the dependence of scattering on the properties
of the mesh.  We introduce a rigid approximation to CGMD that
eliminates internal relaxation, similar to the Quasicontinuum
technique, and compare it to the full CGMD.
\end{abstract}

\date{\today}

\maketitle


\section{Coupling of Length Scales}
\label{sect-intro}

The Science of Materials is foremost a study of structure.  
Once structure is determined other important 
issues such as dynamics and kinetics may be addressed.
Structure in materials is most effectively analyzed according to its length
scale.  Materials structure at different scales such as crystal structure,
crystal defect structure, microstructure and macrostructure has led to
the development of models at the atomic scale, nanoscale, microscale, and
macroscale, not to mention the mesoscale and a vast array of other
distinctions in scale.  These models work because the physics at one
scale decouples to a large extent from that at other scales, provided
there exists a sufficient separation of scales.  Then physical 
properties calculated at one scale may be passed to the next higher 
scale in a hierarchical approach that can be very effective.\cite{Moriarty, Hao}

There are systems of interest that are inherently multiscale
where the physics at one scale is strongly coupled to that
at other scales.\cite{PSS}  Turbulence is an excellent example, where
energy input through stirring at the macroscopic scale
cascades down through vorticity across a range of length
scales until it is ultimately dissipated at the shortest
length scales.  The size of the vortices varies continuously,
and while there are length scales with distinct physics,
the boundaries between them are blurred.  As a result 
hierarchical models have been largely unsuccessful, and
turbulence remains a hard problem.\cite{turb}  
This situation is in marked contrast to low Reynolds number flow,
in which the physics at small length scales
can be encoded in a few parameters, which may be computed and 
then fed into simulations at a larger scale.  In this way, it 
has been possible to start with {\em ab initio} electronic structure 
calculations of H$_2$O and through a sequence of scales end up 
with a description of tides in Buzzard's Bay.\cite{Clementi}

Many other examples of strongly coupled multiscale
systems exist.  Ironically, the advent of Nanoscience
and the current focus on structures of one particular
scale, the nanoscale, has led to the need to understand
a class of strongly coupled multiscale systems.  
Consider epitaxial quantum dots, for example.\cite{Bimberg,EQDprl,PLiu}
The quantum dot consists of a dome of semiconductor that
forms during heteroepitaxy.  The dome itself is
typically a few nanometers to tens of nanometers
across, but its size, shape and location are affected by
the presence of other structures during growth, even
those microns away.  Another
example is a Nano-Electro-Mechanical System (NEMS)
resonator.\cite{Roukes,JMSM,DTM99,IJMCE} It consists of a semiconductor
bar about 50~nm wide and a fraction of a micron long
attached to the substrate at both ends.  The bar
itself is clearly nanoscale, and yet as it resonates,
the oscillating strain fields extend far into
the substrate.  Of course, there are many other
examples in metallurgy and solid state physics,
as indicated above.  Remarkably, similar effects are
beginning to be appreciated in the study of soft 
materials, chemistry and even biology.

These systems are examples of what could be termed
embedded nanomechanics. \cite{MRS01}  The mechanical properties
of the nanoscale structure is clearly important,
and these properties may be quite different from
what would be predicted according to conventional
macroscopic mechanics, but also important is the
way the nanoscale structure is coupled to its
surroundings.
Embedded nanomechanical structures are too small 
to be modeled reliably with conventional continuum 
elastic theory and finite elements, and too large to 
be modeled by conventional atomistics.  
Even in single crystals, sub-micron dynamical regions 
bounded by surfaces or interfaces are affected by 
Angstrom- and nano-scale physics which causes deviations
from continuum
elastic theory;\cite{MBKV} dynamical regions larger than 
0.1 micron cubed exceed the current limit of about one billion 
atoms for atomistic simulation of solids on a supercomputer.\cite{FaridWhite}  
The atomistic effects are 
compounded in materials with local defects or cracks that couple
to long-range strain fields.\cite{CoLS}
The situation is not entirely
intractable, however, because the most important atomistic effects
are often localized to small regions of the system: surfaces, defects,
regions of large deflection or internal strain, and regions of
localized heating perhaps due to friction.  
The challenge is to develop a robust model for such an
inhomogeneous system which captures the important 
atomistic effects without the prohibitive computational cost of 
a brute force atomistic simulation for the entire system.  
In this article we focus on the link between 
the micron scale and the nanoscale and develop a model, 
coarse-grained molecular dynamics (CGMD),\cite{CGMD} which bridges 
the disparate scales seamlessly.

\begin{figure}[t]
\includegraphics[width=0.5\textwidth]{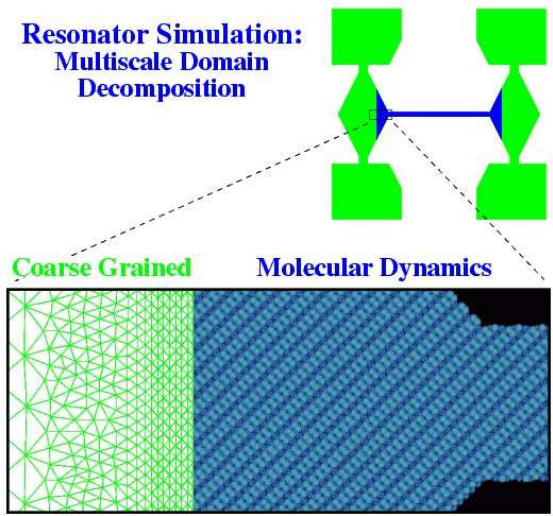}
\caption{(color online) Schematic diagram of a coarse-grained simulation of
a NEMS silicon microresonator.$^{4-6}$
The coarse-grained (CG) region comprises most of the
volume, but the molecular dynamics (MD) region contains 
most of the simulated degrees of
freedom.  Each sphere shown in the MD region represents an atom.
Note that the CG mesh is refined to the atomic scale
where it joins with the MD lattice.
\label{fig-cols}}
\end{figure}

The choice to use atomistic models at the finest resolution is
motivated in some cases by the fact that the inherent length scale
of the process of interest is the interatomic spacing and in other
cases by the ability to derive interatomic potentials from quantum
mechanics and hence built a model from first principles.  Yet 
another motivation is that the processes of interest may be
thermally activated, and molecular dynamics provides a means
to simulate the thermal effects directly.  Entropic and thermal
effects are often paramount in soft matter systems, and in 
hard matter thermal activation is important in defect diffusion,
the motion of dislocations in metals with high Peierls barriers
and many nucleation phenomena.  Temperature is important in 
other ways, too, such as in inducing phase transitions.  Also the
population of phonons increases with temperature, causing
thermal expansion, changes (typically softening) in the elastic
constants and dislocation drag at high strain rates.  These
are but a few well-known examples of the important role 
temperature plays, and thus in our development of multiscale
models we search for methodologies capable of handling
non-zero temperatures.

The variation of the strain field/atomic displacements in 
inhomogeneous solid systems suggests the use
of different computational methodologies for different regions,
as we mentioned above.  The challenge is to meld them
into a seamless, monolithic simulation.  
The first such proposal implemented a coupling between 
molecular dynamics\cite{Alder,AT} (MD) 
and a finite element model\cite{earlyFEM,FE} (FEM) implementation 
of continuum elastic theory using stress consistency as the
boundary condition at the interface.\cite{KGF}  
More recently, a dynamical instability in the original formulation
has been eliminated through the use of a mean force boundary
condition together with uniform symplectic time evolution.\cite{CLS}
In both of these formulations, the same constitutive relation
is used regardless of the size of the cells in the FEM mesh,
leading to a discontinuity at the atomic limit.

At its heart the FEM description of such a system relies on the
ability to improve the accuracy of the simulation by 
going to a finer mesh.\cite{FE}
A mesh of varying coarseness is chosen, adaptively or by fiat, such
that no single region contributes disproportionately to the error.  
These errors typically result from large strain, velocity, or
other gradients which violate 
the discrete expression for the integral of the elastic energy density 
of a continuous medium.  This approximation may be improved by mesh 
refinement.  There is a limit, however.
As the mesh size approaches the atomic scale, the constitutive 
equations have significant errors because the expression for
the elastic energy does not represent localized bonds and the
standard distributed mass expression for the kinetic energy 
does not account for the fact that essentially all of the
mass is localized in the nuclei, at least four order of
magnitude smaller than the interatomic spacing.
At this point, the physics of the governing equations
is wrong, and further mesh refinement does not help.  

The approach of Refs.\ \onlinecite{KGF} and \onlinecite{CLS} 
to improve this situation replaces the FEM equations
of motion on regions of the mesh that are atomic-sized with MD
equations of motion and implements a hand shaking between
the MD and FEM regions.  Although this technique is remarkably
successful, the union is not perfectly seamless.
In the FEM cells approaching the atomic limit, the energy
density varies smoothly within each cell, whereas on the other
side of the interface, the MD energy is effectively localized 
to interatomic bonds.
The short-wavelength modes of the system are able to probe
this discrepancy, leading to errors that grow with the
wavenumber.

The quasicontinuum technique\cite{Phillips,KnapOrtiz,MillerTadmor} 
offers another approach 
to this problem.  It is a zero temperature relaxation technique 
in which the elastic energy used in the FEM region is computed by
applying the FEM interpolated displacement field 
(through the Cauchy-Born rule) 
to a reference system of atoms 
interacting by MD forces.  This is a very nice
idea, but it has a number of difficulties in practice.  
The atoms in the reference system are taken to be at their zero
temperature locations--no fluctuations are allowed.
Thus, many degrees of freedom are summarily set to zero
(although finite temperature versions of the quasicontinuum technique
are currently under development\cite{finiteTQC}). 
Also, the implementation of the quasicontinuum technique suffers
from discontinuities (``ghost forces'') due to the mismatch of the 
displaced reference systems from cell to cell and non-locality
of atomic bonding.\cite{Curtin}

Other concurrent multiscale models have been proposed
recently. 
There are several nice review articles on this subject
to which we direct interested readers.\cite{Curtin,Liu,MillerTadmor}
The relationships of several multiscale models to CGMD are of
particular note.
A finite temperature dynamical model based on renormalization
group ideas has been proposed by Curtarolo and Ceder.\cite{Curtarolo}
The finite temperature coarse-graining approach based on
Monte Carlo calculations has been developed recently by
Wu et al.\cite{UNL}
The bridging scale decomposition is another approach for
coupling atomistic and continuum simulations due to 
Wagner and Liu that provides a coupling between FEM
and atomistics that does not require the FEM mesh to
be refined to the atomic level\cite{NW}, and it has been 
applied to finite temperature simulations by Park et al.\cite{Park}
The projection techniques of the bridging scale decomposition 
and the ensemble averages that they use are 
closely related to the techniques of CGMD introduced earlier.
Also, we note that the assumption that no defect (dislocation) 
propagation takes place from an atomistic region into a 
finite element or coarse-grained region has been 
relaxed through the development of the CADD method, albeit so
far just in two dimensions.\cite{CADD}

We have proposed a replacement for conventional finite
elements suitable for a mesh which is atomic sized in
some regions.\cite{CGMD}  This technique, CGMD, effectively 
provides the scale-dependent constitutive equations 
needed at the interface.  In the atomic limit it is guaranteed
to reproduce the atomistic equations of motion.
This enables MD regions to be coupled seamlessly
to regions of generalized FEM, bringing the full power of MD to bear
on important parts of the system without the computational overhead
of MD in other large, but physically less complex regions.  
The CGMD procedure is based 
on a statistical coarse graining prescription.
While various aspects of CGMD have been introduced
previously, this is the first article to present
the model in great detail.

This kind of multi-scale simulation poses a number of challenges.
First, the model must have a well-behaved, physical response to 
stationary strain fields, slowly varying in position, that
extend into the CG region--there should be no
ghost forces.  Second, the model must have sensible 
thermodynamics in equilibrium.  The effect of short-wavelength 
modes cannot be set to zero unless their energy is well above the thermal
energy.  Third, the system must have realistic dynamics, free of
pathologies due to influences in the central MD region propagating out
to unphysical interfaces, reflecting and propagating back into
the central region to cause unphysical effects.  Fourth, 
the model should exhibit well-behaved nonequilibrium 
thermodynamics, with a sensible
response when low-lying modes are driven
out of equilibrium.
Finally, the methodology needs to be amenable to a practical
implementation in terms of being able to utilize the broad
spectrum of MD models in use, including many-body interatomic 
potentials that extend beyond nearest neighbors, computationally
efficient domain decompositions for parallel distributed
memory computers, visualization, etc.
In this article we describe in detail how CGMD is implemented
in order to meet these challenges.

In particular, we provide the first detailed description 
of the way CGMD may be applied to anharmonic solids and 
finite temperature simulations.  Previously, we have 
described how CGMD is formulated for harmonic lattices \cite{CGMD},
including finite temperature contributions.  We have also
indicated how anharmonic effects are included, but the
details have not been given.  We give the details here,
and explore their implications.  We also introduce a 
rigid approximation to CGMD that eliminates internal 
relaxation, simplifying the formulation and reducing
the computation cost of CGMD.  We examine the physical
difference between CGMD and its rigid approximation.
As tests of CGMD,
the calculation of the phonon spectra for solid argon 
and tantalum
in 3D are presented.  These calculations provide a test
of the quality of the representation of elastic wave energetics.
We also present elastic wave scattering calculations, a test
of how CGMD behaves on an irregular mesh.  Whenever possible
we present analytic formulas that contain a wealth of 
information about the performance of CGMD in as much
generality as possible (explicitly showing dependence 
on the interatomic potential, atomic masses, crystal
lattice and mesh structure).  Of course, realistic
simulations involve numerical assembly and integration
of the CGMD equations of motion.

\section{Coarse Graining Prescription}
\label{sect-cg}

Consider a system of MD 
atoms in a solid, crystalline or amorphous, 
and a coarse-grained (CG) mesh partitioning
the solid into cells (cf.\ Fig.\ \ref{fig-cols}).  
The mesh size may vary, so
that in important regions a mesh node is assigned to each equilibrium atomic
position, whereas in other regions the cells contain many atoms and
the nodes need not coincide with atomic sites.
CGMD offers a way to reduce the atomistic coordinates
to a much smaller set of degrees of freedom associated with
the displacement field at the nodes of the CG mesh,  and the
equations of motion for this mean displacement field.
In particular, the energy functional 
for the CG system is defined as a constrained ensemble average
of the atomistic energy under fixed thermodynamic conditions.
The equations of motion are Hamilton's equations for this conserved 
energy functional and in principle additional random and 
dissipative forces due to fluctuations.

The classical ensemble must obey the constraint 
that the position and momenta 
of the atoms are consistent with the mean displacement and momentum 
fields.  Let the displacement of atom $\mu$ be
${\mathbf{u}}_\mu = {\mathbf{x}}_\mu - {\mathbf{x}}_{\mu 0}$
where ${\mathbf{x}}_{\mu 0}$ is its equilibrium position.
The displacement of mesh node $j$ is a weighted average of the atomic
displacements
\begin{equation}
{\mathbf{u}}_j = \sum _{\mu} f_{j\mu } \, {\mathbf{u}}_{\mu} ,
\label{weighting}
\end{equation}
where $f_{j\mu }$ is a weighting function, related to 
the microscopic analog of FEM interpolating functions below.
An analogous relation applies to the momenta ${\mathbf{p}}_\mu$.  
Since the nodal
displacements are fewer or equal to the atomic positions in number,
fixing the nodal displacements and momenta does not (necessarily) 
determine the atomic coordinates entirely.  Some subspace
of phase space remains, corresponding to 
degrees of freedom that are missing from the mesh.  We define
the CG energy as the average energy of the canonical
ensemble on this constrained phase space:
\begin{eqnarray}
E({\mathbf{u}}_k,\dot{{\mathbf{u}}}_k)
 & = & 
\langle \ H_{MD} \  \rangle _{{\mathbf{u}}_k,\dot{{\mathbf{u}}}_k}
 \\
 ~ & = & 
 \int \! d{\mathbf{x}}_{\mu}  
d{\mathbf{p}}_{\mu} \ 
H_{MD} \, e^{-\beta H_{MD}} \  \Delta / Z, 
\label{cgHam}
\\
Z({\mathbf{u}}_k,\dot{{\mathbf{u}}}_k) 
 & = & 
\int \! d{\mathbf{x}}_{\mu}  
d{\mathbf{p}}_{\mu} \  
e^{-\beta H_{MD}} \  \Delta , 
\label{partFunc}
\end{eqnarray}

\vspace{-6mm}
\begin{eqnarray}
\Delta  & = & {\tst
 \prod _j  \,
\delta \left({\mathbf{u}}_j - \sum _{\mu } 
{\mathbf{u}}_{\mu} f_{j\mu } \right) 
 ~~ \delta \left( \dot{{\mathbf{u}}}_j - 
\sum _{\mu } \frac{{\mathbf{p}}_{\mu} \, f_{j\mu }}{m _{\mu}} \right) ,
}
\label{cgDef}
\end{eqnarray}
where $\beta = 1/(kT)$ is the inverse temperature, $Z$ is the
partition function and 
$\delta ({\mathbf{u}})$ is a three-dimensional delta function.
The delta functions enforce the mean field constraint (\ref{weighting}).
Note that Latin indices, $j,k,\ldots$, denote mesh nodes and
Greek indices, $\mu ,\nu, \ldots$, denote atoms.
The energy (\ref{cgHam}) is computed below (Eq.\ (\ref{cgHamHarm})).

When the mesh nodes and the atomic sites are identical, 
$f_{j\mu }=\delta _{j\mu }$ and the
CGMD equations of motion agree with the atomistic
equations of 
motion. \cite{footnote-Drange}
As the mesh size increases some short-wavelength
degrees of freedom are not supported by the coarse mesh.  These
degrees of freedom are not neglected entirely, because their
thermodynamic average effect has been retained.  This approximation
is expected to be good provided the system is initially in 
thermal equilibrium, and changes to the system would only produce
adiabatic changes in the missing degrees of freedom.  In
particular, the relaxation time of those degrees of freedom
should be fast compared to the driving forces in the CG region.
As long as this condition is satisfied, the long wavelength modes 
may be driven out of equilibrium without problems. \cite{footnote-adiabat}

We have written the CG energy as an internal energy, 
a function of the entropy, $S$, rather than the temperature. 
This is designed for systems in which the
short wavelength modes change adiabatically.  This is a 
good approximation, for example, when long wavelength
elastic waves propagate through a solid in the linear 
regime at finite temperature.\cite{LandauLifschitz,Braginsky}  
In other systems, the short wavelength modes may be 
in contact with a heat bath, so that their
evolution is isothermal rather than isentropic.  
For example, the electron gas in metals can act as a heat
bath on time scales longer than the thermal relaxation time.  
Then the Helmholz free energy, 
\begin{equation}
F({\mathbf{u}}_k,\dot{{\mathbf{u}}}_k)=-kT\, \log Z,
\label{Helmholz}
\end{equation}
should be 
used rather than the internal energy.  In this case, the
ensemble average behavior of the CG collective modes is
exactly the same as that of the corresponding averaged
atomic modes in the underlying atomistic system:
\begin{eqnarray}
\langle {\mathbf u}_{j_1} \cdots {\mathbf u}_{j_n} \rangle & = &
\int \! d{\mathbf u}_j d\dot{{\mathbf u}}_j \,
({\mathbf u}_{j_1} \cdots {\mathbf u}_{j_n} ) \, e^{-\beta F} 
\label{corr1} \\
& = & f_{j_1\mu _1} \cdots f_{j_n\mu _n} \times \nonumber \\
& & ~~~
\int \! d{\mathbf u}_{\mu} d{\mathbf p}_{\mu} \,
{\mathbf u}_{\mu_1} \cdots {\mathbf u}_{\mu_n}  \,
e^{-\beta H_{MD}} , 
\label{corr2} 
\end{eqnarray}
which follows from plugging in the expression for $F$ (\ref{Helmholz})
and (\ref{partFunc}) into (\ref{corr1}) and integrating the delta
functions.\cite{Dupuycalc}
This equation shows the equivalence of all unnormalized correlation functions,
but since the partition functions (zero point functions) are identical,
the normalized correlation functions are the same, as well.  The
emergence of the canonical distribution in other cases requires
a treatment of thermal relaxation processes (cf.\ Section \ref{sec-thermal}).
It should be noted that even in the isothermal ensemble the
faithfulness of correlation functions applies only to equal-time
correlation functions at equilibrium, and consideration of dissipative
processes are needed to reproduce interesting correlation
functions such as the autocorrelation function 
$\langle \dot{u}_i(0) \dot{u}_i(t) \rangle$ associated
with the fluctation-dissipation theorem.

To end this section, we note that the definition of the
CGMD energy may appear to neglect the well-known quantum mechanical
contributions to lattice dynamics.  Phonons are bosons, after all,
and they should obey Bose-Einstein statistics.  The definition
of the CGMD energy (\ref{cgHam}) is clearly a classical
expression based on Boltzmann statistics.  To what extent
can it be expected to be valid?  The reason the classical
expression is valid for most of the conceivable applications
of CGMD is that the Bose-Einstein distribution most strongly affects 
the lowest states; i.e., exactly the states that are retained
explicitly in the CGMD Hamiltonian.  The higher energy states
have low occupation in equilibrium, and are not affected 
significantly by strong quantum effects such as Bose condensation.
The CGMD Hamiltonian is therefore expected to be a good description 
of the coarse-grained system.  It may be necessary to use 
a path integral, or other quantum mechanical version, of MD
to treat the retained degrees of freedom at sufficiently
low temperature,\cite{PIMD} but the internal degrees of
freedom are described well by Eq.\ (\ref{cgHam}).

\section{Shape Functions}
\label{sect-shape}

In addition to the general framework we have presented for CGMD, a specific
choice of the weighting functions is required for calculations.
They result from the introduction of a set of shape functions
$\left\{ N_j({\mathbf{x}})\right\} _{j=1}^{N_{{\mathrm node}}}$
on the mesh from which the interpolated fields are constructed.
The shape functions have the following properties:
\renewcommand{\theenumi}{\roman{enumi}}
\begin{enumerate}
\item $N_j({\mathbf{x}}_k) = \delta _{jk}$,
\label{shapecriteria1}
\item ${\displaystyle \sum _{j=1}^{N_{{\mathrm node}}}
   N_j({\mathbf{x}}) = 1}$,
\label{shapecriteria2}
\item C$^0$ continuity.
\label{shapecriteria3}
\end{enumerate}
The first property states that the functions are
normalized and local on the mesh nodes, ${\mathbf{x}}_k$. 
The second states that the functions form a partition of unity,
so the center of mass mode is represented.  The third states
that the functions are continuous but their derivatives need 
not be.  This continuity guarantees that the elastic energy, 
proportional to an
integral of the square of the strain $({\mathbf{\partial u}})_{sym}$, 
is well-behaved in the continuum limit.  
The interpolated displacement field is then defined by
${\mathbf{u}}({\mathbf{x}}) = \sum {\mathbf{u}}_j \, N_j({\mathbf{x}})$.
Often there are additional considerations, such as the need to 
refine the mesh onto a particular crystal lattice at the MD/CG
interface. \cite{ICCN}

Given any set of atomic displacements we can find the
displacement field represented on the CG mesh which best fits these 
data in the least squares sense:  
\begin{equation}
\chi ^2 = \sum _{\mu} \left| 
{\mathbf{u}}_{\mu} - \sum _j {\mathbf{u}}_j \, N_{j\mu} 
\right| ^2,
\end{equation}
where $N_{j\mu} = N_j({\mathbf{x}}_{\mu 0})$.
This $\chi ^2$ error is minimized by 
\begin{eqnarray}
{\mathbf{u}}_j & = &  \sum _\mu f_{j\mu } \, {\mathbf{u}}_\mu , 
\label{ujproject} \\
f_{j\mu } & = & \sum _k 
\left( \sum _{\nu } N_{j\nu } N_{k\nu } \right) ^{-1} N_{k\mu} 
\end{eqnarray}
[cf.\ (\ref{weighting})].
This equation defines the weighting function $f_{j\mu }$ of (\ref{weighting}) 
in terms of the interpolating function $N_j({\mathbf x})$.  We note
that recently
this relationship introduced in CGMD has been generalized for use
in the bridging scale, and other L$^2$ projection techniques.\cite{Liu}

The formulation we have described is appropriate to retain the
low-lying acoustic phonon modes in the coarse-grained system.
In some cases it is desirable to retain the long-wavelength
optical phonons, as well.  For example, in the study of III-V
epitaxial quantum dots, internal relaxation of the zinc blende
structure in the strained dots leads to important changes in
the optical spectra of the dots.\cite{WilliamsonZunger}
If it is important to model the optical phonons or to capture
the internal relaxation in a crystal lattice with a multiple-atom
basis, each interpolation
function should carry a band index, $a$, in addition to the nodal
index, $j$: $N_j^{(a)}({\mathbf{x}})$.  Then the
basis requirements are somewhat different.  The functions
should be local and normalized within each band.  They
should be C$^0$ continuous apart from variations with
the unit cell.  And they should form a generalization of 
the partition of unity.  In particular, the requirement of
forming a partition of unity is the requirement that 
uniform displacement of the system be represented in the
basis of shape functions.  That translation invariance is
responsible for the ${\mathbf{k}}=0$ acoustic-mode phonons 
having zero energy.  We generalize the partition of unity
requirement to the requirement that all of the ${\mathbf{k}}=0$
phonon, both acoustic and optical, be represented.  In
particular, denote the 
displacement associated with the ${\mathbf{k}}=0$ 
phonon for band $a$ as ${\mathbf{u}}^{(a)}_{\mu}$, normalized
such that
\begin{equation}
\sum _{\mu \in {\mathrm{unit~cell}}} 
\left| {\mathbf{u}}^{(a)}_{\mu} \right| ^2 = N_{{\mathrm{basis}}}
\end{equation}
where $N_{{\mathrm{basis}}}$ is the number of atoms in the
Wigner-Seitz unit cell.  Then the shape functions can be
defined as
\begin{equation}
N_j^{(a)}({\mathbf{x}}_{\mu 0}) = {\mathbf{u}}^{(a)}_{\mu} N_j({\mathbf{x}}_{\mu 0})
\end{equation}
where $N_j({\mathbf{x}})$ is a conventional shape function, such
as linear interpolation.  The generalized partition of unity
requirement is that 
\begin{equation}
\sum _j N_j^{(a)}({\mathbf{x}}_{\mu 0}) = {\mathbf{u}}^{(a)}_{\mu 0}.
\end{equation}
Note that in the case of a monatomic
unit cell this shape-function basis is a linear combination 
of the shape functions we have discussed above.  In that case
${\mathbf{u}}^{(a)}_{\mu}$ is the same for all lattice sites,
and the orthonormal vectors corresponding to the three acoustic-mode
phonons span three-dimensional space.
We do not discuss an example of a polyatomic CGMD including optical
phonons explicitly, but the
CGMD formalism continues to work. 
It should be emphasized, however,
that even in polyatomic materials this band-index extension 
may not be needed to capture the mechanical response of interest.

\section{The CGMD Hamiltonian}
\label{sect-cgmdHam}

We now turn to the calculation of the CGMD energy.

\subsection{Harmonic Lattices}
\label{subsec-harmLatt}

The CG energy (\ref{cgHam}) may be computed in closed
form using analytic techniques in the case of a harmonic lattice.  
The expression was originally presented in Ref.\ \onlinecite{CGMD}.
We take the form of the atomistic Hamiltonian 
to be
\begin{equation}
H_{MD} = \sum _{\mu} \frac{{\mathbf{p}}^2_{\mu}}{2m_{\mu}} +  \sum _{\mu} 
E^{\mathrm{coh}}_{\mu} + \sum _{\mu ,\nu} 
\frac{1}{2} {\mathbf{u}}_{\mu }  \cdot D_{\mu \nu}
{\mathbf{u}}_{\nu }  ,
\label{harmMD}
\end{equation}
where $E^{\mathrm{coh}}_{\mu}$ is the cohesive energy
of atom $\mu$ and $D_{\mu \nu}$ is the dynamical matrix.  It acts
as a tensor on the components of the displacement vector at each
site.  We re-express the CG energy (\ref{cgHam}) using a parametric
derivative of the log of the constrained partition function (\ref{partFunc}),
\begin{eqnarray}
E({\mathbf{u}}_k,\dot{{\mathbf{u}}}_k)
 & = & 
 - \partial _{\beta} \log Z({\mathbf{u}}_k,\dot{{\mathbf{u}}}_k;\beta) ,
\label{cgHam2} 
\end{eqnarray}
and we introduce the Fourier transform representation of
the delta function (a form of Lagrange multiplier) 
to simplify the constraint 
(\ref{cgDef})
\begin{eqnarray}
\Delta & = & \int \left( \frac{d\lambda }{2\pi} \right)^{3N_{node}} 
\, e^{i {\mathbf{\lambda}} _j \cdot ( {\mathbf{u}}_j - 
f_{j\mu} {\mathbf{u}}_{\mu} )} 
\times \nonumber \\
 & & ~~~~
\int \left( \frac{d\lambda ^{\prime} }{2\pi} \right)^{3N_{node}} \,
e^{i {\mathbf{\lambda}} _k ^{\prime} \cdot ( \dot{{\mathbf{u}}}_k - 
f_{k\nu} {\mathbf{p}}_{\nu} / m_{\nu})} ,
\end{eqnarray}
where here, and in what follows unless stated otherwise,
the repeated indices are summed.
Expressed in this way, the constrained partition function for the
harmonic lattice is a Gaussian integral. 
The complicated
domain of integration in (\ref{cgHam}) resulting from the constraints 
is replaced by a simple domain plus some extra integrals.
This standard technique gives an expression
which may be evaluated in closed form.  

For pedagogical purposes we present the calculation of the
CG potential energy here in some detail.  The calculations of
each of the other CG energy terms, the CG kinetic and 
anharmonic potential terms, follow the same basic approach.
It may be helpful therefore to present the calculation of 
the CG harmonic potential energy in detail, and readers who are not 
interested in this algebra may skip ahead to the next paragraph
at Eq.\ (\ref{cgHamHarm}).  

In order to get a closed-form expression for the CG potential
energy, we make use of the well-known exact formula for the integral
over all space of a Gaussian times an arbitrary polynomial
prefactor.  
In the interest of compact notation, we combine the atomic and
spatial indices and consider the displacements and the dynamical 
matrix to be objects in 3$\natom$-dimensional space.
Similarly, we take all CG variables to live in 3$\nnode$-dimensional 
space.  The form for the Gaussian integral is known for these
high dimensional spaces, and we evaluate the integral (\ref{partFunc})
through two successive Gaussian integrations: first an integral
over the MD phase space and then an integral over the 
Lagrange multiplier space.  
Using (\ref{cgHam2}), we need to calculate the 
CG partition function.  It factorizes into kinetic
and potential parts, $Z=Z_{{\mathrm{kin}}}Z_{{\mathrm{pot}}}$.
We focus on the potential energy part of the CG partition function: 
\begin{eqnarray}
Z_{{\mathrm{pot}}}(u_k; \beta) & = & 
 \int \!  du \! \int d\lambda \,
 e^{-\frac{1}{2} \beta u_{\mu} D_{\mu \nu} u_{\nu} + 
     i \lambda_j \left( u_j - f_{j\mu} u_{\mu} \right)} 
  \nonumber \\
  \label{CGpot1}
\end{eqnarray}
where $du = (du)^{3\natom}$ and 
$d\lambda = \left( \frac{d\lambda }{2\pi} \right)^{3N_{node}}$.
Here, and throughout this derivation, we suppress the
cohesive energy by choosing the zero of energy such that
the cohesive energy is zero; we then restore the
cohesive energy in the final formulas.
We first compute the integral over $du$ by completing
the square in the argument of the exponential.  Let
\begin{equation}
\tilde{u}_{\mu} = u_{\mu} - i D_{\mu \nu}^{-1} f_{j\nu} \lambda_j / \beta
\label{firstshift}
\end{equation}
where we assume that the matrix inverse $D_{\mu \nu}^{-1}$ exists after
a suitable regularization to deal with the zero eigenvalues
(we will return to this point).
The shift (\ref{firstshift}) leaves the measure $du$ invariant,
so 
\begin{eqnarray}
Z_{{\mathrm{pot}}}(u_k; \beta) & = & 
 \int \!  d\tilde{u} \, 
 e^{-\frac{1}{2} \beta \tilde{u}_{\mu} D_{\mu \nu} \tilde{u}_{\nu}} \times  
  \label{CGpot2} \\
  & & ~~~~~ \int d\lambda \, e^{ i \lambda_j  u_j + 
    \frac{1}{2} \lambda_j f_{j\mu} D_{\mu \nu}^{-1} f_{k\nu} \lambda_k / \beta}
  \nonumber 
\end{eqnarray}
where the integral has now split into two independent factors.
The Gaussian integral over $d\tilde{u}$ is elementary\cite{GaussianInt}:
\begin{eqnarray}
 \int \!  d\tilde{u} \, 
 e^{-\beta \tilde{u}_{\mu} D_{\mu \nu} \tilde{u}_{\nu} }
 & = &
 \left( 2 \pi \beta ^{-1} \right) ^{3\natom/2} 
 \left( {\det}^{\prime} D \right) ^{-1/2} \nonumber \\
 & = & C_1 \, \beta ^{-3\natom/2}
\end{eqnarray}
where we have used $\beta ^{-1} = kT$ 
and ${\det}^{\prime}$ denotes the determinant
without the zero eigenvalues.
On the second line we note the
simple power-law dependence on $\beta$;
the constant factor $C_1$ is ultimately irrelevant.
In order to simplify notation, we define
$K_{jk} = \left( f_{j\mu} D_{\mu \nu}^{-1} f_{k\nu} \right) ^{-1}$,
where again a suitable regularization is implied on the 
right-hand side.  Then we have
\begin{eqnarray}
Z_{{\mathrm{pot}}}(u_k; \beta) & = & C_1 \, \beta ^{-3\natom/2} \, \times
  \label{CGpot3} \\
  & & ~~~~~ \int d\lambda \, e^{ i \lambda_j  u_j + 
    \frac{1}{2} \lambda_j K^{-1}_{jk} \lambda_k / \beta } 
  \nonumber
\end{eqnarray}
Now we compute the next integral, again by
introducing a suitable shift in the variables:
\begin{eqnarray}
\tilde{\lambda}_j = \lambda_j - i \beta K_{jk} u_k
\end{eqnarray}
so that
\begin{eqnarray}
Z_{{\mathrm{pot}}}(u_k; \beta) & = & C_1 \, \beta ^{-3\natom/2} \, \times
  \label{CGpot4} \\
  & & ~~~~~ \int d\tilde{\lambda} \, 
    e^{ \frac{1}{2} \tilde{\lambda}_j K^{-1}_{jk} \tilde{\lambda}_k / \beta } 
    \, e^{ - \frac{1}{2} \beta u_j K_{jk} u_k } 
  \nonumber
\end{eqnarray}
Again the Gaussian integral is elementary
\begin{eqnarray}
 \int \!  d\tilde{\lambda} \, 
 e^{\frac{1}{2} \tilde{\lambda}_j K^{-1}_{jk} \tilde{\lambda}_k / \beta }
 & = &
   C_2 \, \beta ^{3\nnode/2}
\end{eqnarray}
where only the dependence on $\beta$ is 
relevant.  \cite{footnote-infinity}
Finally, we have the expression we need:
\begin{eqnarray}
Z_{{\mathrm{pot}}}(u_k; \beta) & = & C_1 \, C_2 \, 
  \beta ^{-3(\natom-\nnode)/2} \, \times
  \nonumber \\
  & & ~~~~~~~~ e^{ - \frac{1}{2} \beta u_j K_{jk} u_k } 
  \label{CGpot5}
\end{eqnarray}
The CG (harmonic) potential energy is then
\begin{eqnarray}
E_{pot}(u_k) & = & -\partial _{\beta} \log Z _{pot} \\
            & = & \frac{3}{2} (\natom-\nnode) kT 
                    + \frac{1}{2} u_j K_{jk} u_k
  \label{CGpotEharm}
\end{eqnarray}
This expression shows that the CG potential energy is
exactly given by the sum of two contributions.  First,
each degree of freedom that has been integrated out
contributes a thermal energy of $\frac{1}{2}kT$ to
the potential energy. And second, the remaining CG
degrees of freedom experience a harmonic potential
of the form $\frac{1}{2} u_j K_{jk} u_k$.  The 
calculation of the kinetic energy is essentially
identical in form, but a bit more simple due to the
diagonal mass matrix.  We note that while this
derivation is mathematically elegant, it 
finesses many subtleties in a way that may not
give the reader complete confidence in the
derivation; e.g. we have ignored the fact
that the dynamical matrix is singular and
we have not been very careful about the values
of the irrelevant constants $C_1$ and $C_2$.
We present a more careful derivation in 
Appendix \ref{app-CGderiv} that has the distinct 
benefit that the reason for, and extent of, the 
non-locality of the stiffness matrix is readily 
apparent.  We now return to the CG energy,
and once again separate the spatial
and nodal indices for the following.

The full CG energy (\ref{cgHam2}) for a monatomic harmonic solid of
$\natom$ atoms coarse grained to $\nnode$ nodes is then
given by
\begin{equation}
E({\mathbf{u}}_k,\dot{{\mathbf{u}}}_k) = U_{{\mathrm{int}}} 
  + \frac{1}{2} \sum _{j,k} \left( 
M_{jk} \, \dot{{\mathbf{u}}}_j \cdot \dot{{\mathbf{u}}}_k
+ {\mathbf{u}}_j \cdot K_{jk} {\mathbf{u}}_k \right) , 
\label{cgHamHarm} 
\end{equation}
where the contribution of the internal degrees of freedom is
\begin{eqnarray}
U_{{\mathrm{int}}} & = &  \natom E^{\mathrm{coh}} + 3(\natom-\nnode)kT. 
\label{Uint}
\end{eqnarray}
$M_{jk}$ and $K_{jk}$ are defined as follows.
The mass matrix is 
\begin{eqnarray}
M_{jk} & = & \left( \sum_{\mu} f_{j\mu } \, m _{\mu}^{-1} 
\, f_{k\mu } \right) ^{-1} \label{Mdef} \\
  & = & m \, \sum_{\mu} N_{j\mu } \, N_{k\mu } ~~~~({\mathrm{monatomic}}),
\label{Mdefmono}
\end{eqnarray}
where the second line applies to monatomic solids with atomic mass $m$.
This may be expressed in matrix notation as
\begin{eqnarray}
M_{jk} & = & \left[ (N N^T) 
\left( N \, M_{MD}^{-1} \, N^T \right) ^{-1} 
  (N N^T) \right] _{jk} \label{Mdefmat} \\
  & = & m \, (N N^T)_{jk}  ~~~~({\mathrm{monatomic}}),
\label{Mdefmonomat}
\end{eqnarray}
where the MD mass matrix is 
$M^{MD}_{\mu \nu} = m_{\mu} \delta _{\mu \nu}$.
Note that each of the quantities in parentheses ()
in Eq.\ (\ref{Mdefmat}) is an $N_{node}\times N_{node}$
square matrix.

At times, it may be desirable to use a diagonal approximation
of the mass matrix, often called the lumped mass matrix in the
FEM literature.  In CGMD it is given by\cite{PSS}
\begin{eqnarray}
M_{ij}^{{\mathrm{lump}}} & = & \delta _{ij} \, \sum _{\mu} N_{j\mu} \, m_{\mu} 
\label{CGMDlumpM}
\end{eqnarray}

The CGMD stiffness matrix is given formally by 
\begin{eqnarray}
K_{jk} & = & \left(
\sum_{\mu \nu} f_{j\mu } \, D^{-1} _{\mu \nu} \, f_{k\nu } \right)^{-1} 
\label{Kdef}
\\
 & = & \left[ (N N^T) \left( N \, D^{-1} \, N^T \right) ^{-1} 
              (N N^T) \right] _{jk} ,
\label{Kdef2}
\end{eqnarray}
where each of the entries on the second line is a matrix.
The inverses in (\ref{Mdef}), (\ref{Mdefmat}),
(\ref{Kdef}) and (\ref{Kdef2}) are matrix inverses.

Remarkably, there is another way to write the
CGMD mass and stiffness matrices that does not require
two inverses.  These forms are derived in 
Appendix \ref{app-CGderiv}, and the notation is explained there.
In particular, they are given by
\begin{eqnarray}
K_{jk} & = & N_{j\mu} D_{\mu\nu} N_{k\nu} 
   - D^{\times}_{j\mu} \tilde{D}_{\mu\nu}^{-1} D^{\times}_{k\nu}
\label{stiffA1} \\
M_{jk} & = & N_{j\mu} m_{\mu} N_{k\mu} 
   - M^{\times}_{j\mu} \tilde{M}_{\mu\nu}^{-1} M^{\times}_{k\nu}
\label{massA1} 
\end{eqnarray}
We do not currently know of an easy way to show the
equivalence of Eqs.\ (\ref{Kdef}) and (\ref{stiffA1}),
but we have checked that they are numerically equal.
Each form has its preferred uses.  In the calculation of
the CGMD spectra and any other application in which
the stiffness matrix in reciprocal space is needed,
 Eq.\ (\ref{Kdef}) is advantageous.  It is formally
simpler, but it suffers from requiring two inverses
and from formal singularities due to the zero modes
of the dynamical matrix.  Both of these drawbacks
disappear in reciprocal space, where the dynamical
matrix is diagonal (in the Fourier transform of
the nodal indices).  On the other hand, Eq.\ (\ref{stiffA1})
is well defined, and it only requires one inverse.
Typically, the inverse in the atomic indices is
taken in reciprocal space making use of the 
perfect crystal space group symmetry, whereas
the second inverse (\ref{Kdef}) is in the nodal
indices for which reciprocal space offers an
advantage only in special cases where the mesh
is uniform.  For irregular meshes, Eq.\ (\ref{stiffA1})
is preferred.

Consider the form of the second expression for
the stiffness matrix (\ref{stiffA1}).  The first
term represents a form of coarse graining in which
each atom is forced to be exactly at the position
defined by the interpolation function.  Within
the context of CGMD, we will refer to the
approximation where the other terms are
neglected as the rigid approximation.  To be
precise,
\begin{equation}
P^{{\mathrm{\perp}}}_{\mu\nu} \rightarrow 0 ~~~{\mathrm{Rigid~Approximation}}
\end{equation}
by which we mean that $P^{{\mathrm{\perp}}}_{\mu\nu}$,
defined in Eq.\ (\ref{pperp})
is set to zero in all of the subsequent CGMD formulas.
For instance, both $D^{\times}_{j\nu}$ and
$M^{\times}_{j\nu}$ vanish in the rigid approximation,
so only the first term survives in Eqs.\ (\ref{stiffA1}) and (\ref{massA1}) .  

Now let us consider the second term in the stiffness matrix 
(\ref{stiffA1}).  This kind of term has not been
discussed in the context of concurrent multiscale
modeling previously, and it is very interesting.
It is in the form of a lattice Green function 
contribution to the stiffness.  According to
our principle of microscopic-macroscopic
correspondence, the atomistic degrees of
freedom are assumed to be a best fit to
the coarse-grained degrees of freedom.  Even
at zero temperature, this requirement does not necessarily
mean that the atoms will be position exactly
where the interpolation function would put them.
Instead, they typically relax to a lower energy
configuration.  This relaxation in the short
wavelength degrees of freedom introduces a
non-local coupling between the coarse-grained
degrees of freedom through the Green function
$\tilde{D}_{\mu\nu}^{-1}$.  The appearance of
a Green function in a relaxation problem is
natural.  Consider a system governed by
the elastic energetics 
$E=\frac{1}{2} u_{\mu} D_{\mu\nu} u_{\nu} - f_{\mu} u_{\mu}$.
The minimal energy state is $u_{\mu}=D_{\mu\nu}^{-1} f_{\mu}$
with the energy 
$E_{{\mathrm{min}}}=-\frac{1}{2} u_{\mu} D_{\mu\nu}^{-1} u_{\nu}$.
The Green function thus arises naturally in the energy
of the relaxed state.  In CGMD, it is the internal modes
that can relax, and so it is the Green function of these
internal modes that enters the CGMD Hamiltonian 
and introduces the non-locality.
This kind of weak non-locality is not present in
finite element modeling, but it is entirely
physical.  In fact, continuum formulations of non-locality
in elasticity have been introduced to account for size
effects in dislocations, crack tips and other nanoscale
structures.\cite{nonLoc,nonLoc2}  It may be possible to neglect
the non-locality for a particular application, but it is real
and it arises naturally in CGMD.

The energy (\ref{cgHamHarm}) 
contains terms representing the average kinetic and 
potential energies, plus the thermal energy term expected from
the equipartition theorem for the modes that have been integrated
out.  
As mentioned above, this Hamiltonian continues to work for polyatomic 
solids, 
in which the optical modes may be coarse grained in various ways
to represent different physics.

\subsection{No Ghost Forces}

One goal of concurrent multiscale modeling is to have the atoms
in the atomistic region behave as closely as possible to the way
they would if the atomistics extended throughout the system.
Deviations from this ideal have been termed ghost forces.  Some
deviation is inevitable, but two kinds of deviation have received
particular attention as pathologies.  Both are at zero temperature.
First, if the displacement with respect to the equilibrium lattice
is zero throughout the system, then the atoms at the interface
should experience no force.  Second, if the displacement 
corresponds to uniform strain and the uniformly strained
atomistic system is in equilibrium, then the atoms at the
interface in the concurrent multiscale model should experience
no force
(See the recent review article by Curtin and Miller\cite{Curtin}).

CGMD does not suffer from the first type of ghost force, as
can be seen immediately from the absence of terms linear in
${\mathbf{u}}_j$ in Eq. (\ref{cgHamHarm}).  The second type
of ghost force is also absent from CGMD, provided the 
strain is admissible in the space of shape functions.  This
property should be clear from the construction of CGMD, 
where if the best fit interpolation function reproduces
the uniform strain at zero temperature, the delta functions
impose that the CGMD energy agree with the MD energy exactly.

To be more precise, the strained state described by the
underlying atomistic displacement ${\mathbf{u}}_{\mu}$ 
is admissible if there exists a set of nodal displacements
${\mathbf{u}}_{j}$ such that
\begin{eqnarray}
{\mathbf{u}}_{\mu} & = & \sum _j {\mathbf{u}}_{j} N_j(x_{\mu 0}) \\
 & = & {\mathbf{u}}_{\nu} f_{j\nu} N_{j\mu} \\
 & = & P^{CG}_{\mu \nu} {\mathbf{u}}_{\nu}
\end{eqnarray}
where we have used Eq.\ (\ref{ujproject}) in the second line 
to express the best-fit
${\mathbf{u}}_{j}$ in terms of ${\mathbf{u}}_{\mu}$.
The matrix $P^{CG}_{\mu \nu}$, defined in Appendix \ref{app-CGderiv}
in Eq.\ (\ref{PCG3}), must act as the identity on ${\mathbf{u}}_{\nu}$.
We assume that the uniformly strained atomistic system is in 
equilibrium,and hence $D_{\mu \nu} {\mathbf{u}}_{\nu} = 0$.  We
calculate the force on node $i$ (which may be an atom at the
interface) using Eqs.\ (\ref{stiffA1}) and (\ref{ujproject})
as
\begin{eqnarray}
-K_{ij} {\mathbf{u}}_{j} & = & - 
 N_{i\mu} D_{\mu\nu} N_{j\nu} \left( f_{j\rho}{\mathbf{u}}_{\rho} \right)
   - D^{\times}_{i\mu} \tilde{D}_{\mu\nu}^{-1}  
     D^{\times}_{j\nu} \left( f_{j\rho} {\mathbf{u}}_{\rho} \right) \\
  & = & N_{i\mu} D_{\mu\nu} P^{CG}_{\nu \rho} {\mathbf{u}}_{\rho} 
   - D^{\times}_{i\mu} \tilde{D}_{\mu\nu}^{-1}
     P^{\perp}_{\nu\tau} D_{\tau \nu}  P^{CG}_{\nu \rho} {\mathbf{u}}_{\rho} \\
  & = & N_{i\mu} D_{\mu\nu} {\mathbf{u}}_{\nu}
   - D^{\times}_{i\mu} \tilde{D}_{\mu\nu}^{-1}
     P^{\perp}_{\nu\tau} D_{\tau \nu} {\mathbf{u}}_{\nu} \\
  & = & 0
\end{eqnarray}
Thus the atoms (and nodes) of the coarse-grained system experience no force.
In going from the second to the third line we have used the admissibility
condition that $P^{CG}_{\nu \rho}$ act as the identity matrix 
on ${\mathbf{u}}_{\rho}$, and in the following line we used that 
${\mathbf{u}}_{\nu}$ is in equilibrium and so 
$D_{\tau \nu} {\mathbf{u}}_{\nu}=0$.  This derivation proves that
any admissible equilibrium atomistic configuration is also an
equilibrium CGMD configuration.  The derivation continues to
work once anharmonic forces are included, as discussed below,
since the terms in the CGMD energy up to second order in displacements 
are the same.  Thus, CGMD is free of ghost forces
in both senses of the term.

\subsection{Anharmonic Lattices}

We have formulated CGMD for an underlying anharmonic
Hamiltonian in perturbation theory, assuming again 
negligible diffusion in the CG region.  With an anharmonic potential 
the higher frequency modes comprising the heat bath do not decouple, 
and they introduce temperature-dependent effects such as
thermal expansion and thermal softening of the elastic constants. 
The basic idea of how anharmonicity
is treated was presented in Ref.\ \onlinecite{PSS}.  The details
are presented here for the first time.

In this subsection we develop a formal analysis of the contributions
of the anharmonic interatomic forces to the CGMD energy and equations
of motion.  This analysis provides insight into how effects in the
coarse-grained system are linked to their atomistic origins through
analytic formulas, and hence are very powerful.  On the other hand,
the formulas are sufficiently complicated that a different approach
is employed in practice, and we stress this point.  The formal
developments that follow take the perfect $T=0$K lattice as the
reference state; in practice, it is much more useful to take
the lattice at the temperature of interest to be the reference
state.  In that case, the harmonic theory presented above may
be used, with anharmonic effects from the lattice entering into
the reference state in a quasi-harmonic approach.  The thermal
expansion of the lattice and the thermal softening of the dynamical
matrix capture the anharmonic forces.\cite{CGMD}  The finite temperature
lattice constant and dynamical matrix are inputs to the CGMD
formulation, precomputed in conventional MD calculations.  While
that approach is very effective in practice, it is more satisfying
from a theoretical point of view to have a direct, analytic theory
of the thermal effects in CGMD, and it is to this development we
turn now.

The CGMD energy (\ref{cgHam2}) is computed by perturbation theory
about the harmonic Hamiltonian (\ref{harmMD}).  Specifically,
\begin{equation}
H_{MD} = H_h + H^{\prime}
\end{equation}
where $H_h$ is the harmonic Hamiltonian (\ref{harmMD}) and 
$H^{\prime}$ consists of the anharmonic corrections, which are assumed
to be small.  This is a good approximation in silicon, and
many other materials of interest below their melting point
(i.e.\ at low homologous temperatures).  
The perturbation may be written explicitly as
\begin{equation}
H^{\prime} = \sum _{n=3} ^{\infty} \frac{1}{n!} 
\sum D^{a_1 \ldots a_n}_{\mu _1 \ldots \mu _n}
u^{a_1}_{\mu _1} \cdots u^{a_n}_{\mu _n} 
\label{pert}
\end{equation}
where the $\mu$'s label the atoms, as before, and the $a$'s label
components of the vectors.
The higher order dynamical matrices 
$D^{a_1 \ldots a_n}_{\mu _1 \ldots \mu _n}$
may be taken to be completely symmetric in
their indices from the form of Eq.\ (\ref{pert}).
Below we occasionally use the notation $D^{(n)}$ to
represent $D^{a_1 \ldots a_n}_{\mu _1 \ldots \mu _n}$
schematically in places where the additional concision
should not lead to confusion.
This perturbation theory is a low temperature 
expansion, as seen by switching to the scaled coordinates
\begin{eqnarray}
\tilde{{\mathbf u}}_{\mu} & = & \frac{{\mathbf u}_{\mu}}{\sqrt{kT}} , 
   ~~~~~~
\tilde{{\mathbf u}}_{j}  =  \frac{{\mathbf u}_{j}}{\sqrt{kT}} , 
  \\
\tilde{{\mathbf p}}_{\mu} & = & \frac{{\mathbf p}_{\mu}}{\sqrt{kT}} , 
   ~~~~~~
\dot{\tilde{{\mathbf{u}}}}_{j}  =  \frac{\dot{{\mathbf{u}}}_{j}}{\sqrt{kT}} , 
  \\
\tilde{{\mathbf \lambda}}_{j} & = & \sqrt{kT} \, {\mathbf \lambda}_{j} , 
   ~~~~~~
\tilde{{\mathbf \lambda}}_{j}\p  =  \sqrt{kT} \, {\mathbf \lambda}_{j} . 
\end{eqnarray}
Then
\begin{eqnarray}
E & = & U_{{\mathrm{int}}} ^h
 - \partial _{\beta} \log \int \! d\tilde{{\mathbf{u}}}_{\mu}  
d\tilde{{\mathbf{p}}}_{\mu} \ 
 e^{-\tilde{H}_{h}}\, e^{-H\p } \  \Delta , 
\label{anhar} \\
\tilde{H}_{h} & = & 
\sum _{\mu} \frac{\tilde{{\mathbf{p}}}^2_{\mu}}{2m} + \sum _{\mu ,\nu} 
\frac{1}{2} \tilde{{\mathbf{u}}}_{\mu }  \cdot D_{\mu \nu}
 \tilde{{\mathbf{u}}}_{\nu }  , \\
\tilde{H}^{\prime} & = & \sum _{n=3} ^{\infty} \frac{1}{n!} 
\sum \beta ^{(2-n)/2} D^{a_1 \ldots a_n}_{\mu _1 \ldots \mu _n}
\tilde{u}^{a_1}_{\mu _1} \cdots \tilde{u}^{a_n}_{\mu _n} 
\label{anharexp}
\end{eqnarray}
where $U_{{\mathrm{int}}} ^h$ is the internal
energy for the CG harmonic lattice (\ref{cgHamHarm}).
The temperature dependence is now in $U_{{\mathrm{int}}}^h$,
$\Delta$,
and in the perturbation $\tilde{H}^{\prime}$, but not in $H_{h}$. 

To evaluate this integral, the exponential involving $\tilde{H}^{\prime}$
is expanded in a Taylor series in $\tilde{u}$ and/or $kT$.  
The resulting integrals are
elementary.  It is common practice to use a diagrammatic representation
of the integrals to facilitate bookkeeping in this kind of 
expansion.\cite{Fdiagrams}  The log 
in Eq.\ (\ref{anhar}) 
is then produced by restricting to connected graphs.\cite{Ramond} 
Similar perturbative approaches have been used in many contexts; the application
most relevant to the current study is the Self-Consistent Phonon
Approximation used in lattice dynamics.\cite{SCPA,Wallace}

The resulting form of the CG energy for the anharmonic lattice
is
\begin{eqnarray}
E({\mathbf{u}}_k,\dot{{\mathbf{u}}}_k) & = & U_{{\mathrm{int}}} 
  + \sum _{j,k} \frac{1}{2} \left( 
M_{jk} \, \dot{{\mathbf{u}}}_j \cdot \dot{{\mathbf{u}}}_k
+ {\mathbf{u}}_j \cdot K_{jk} {\mathbf{u}}_k \right) + \nonumber \\
& & ~~~~~~
\sum _{n=1} ^{\infty} \frac{1}{n!} 
\sum K^{a_1 \ldots a_n}_{j _1 \ldots j _n} (T) \, 
u^{a_1}_{j _1} \cdots u^{a_n}_{j _n} 
\label{cgHamAnharm}
\end{eqnarray}
where now $U_{{\mathrm{int}}}$ is a complicated function of $\beta$,
as are the stiffness coefficients.  Our goal is to calculate this
expansion analytically at each order of perturbation theory.
Since the diagrammatic approach may not be familiar to all
readers, we will not use it for the derivations, but we 
return to it below.  For now, consider the integral
needed to calculate the CGMD energy up to second
order in $D^{(3)}$ and first order in $D^{(4)}$.  These
first few terms in the perturbation series will capture
the leading thermal and non-linear effects.  Higher
order terms could be calculated similarly, if needed.
To this order, the CGMD potential energy expansion is
\begin{eqnarray}
U & = & U_{{\mathrm{int}}} ^h
 - \partial _{\beta} \log \int \! d\tilde{{\mathbf{u}}}_{\mu}  
d\tilde{{\mathbf{p}}}_{\mu} \ 
 \left[ 1  
\right. 
\label{anhar6} 
 \\
& & 
-\frac{\beta ^{-1/2}}{3!} D^{a_1 a_2 a_3}_{\mu _1 \mu _2 \mu _3} \tilde{u}^{a_1}_{\mu _1} \tilde{u}^{a_2}_{\mu _2} \tilde{u}^{a_3}_{\mu _3}
\nonumber \\
& &
+
\frac{1}{2} \beta ^{-1} \left( \frac{1}{3!} D^{a_1 a_2 a_3}_{\mu _1 \mu _2 \mu _3} \tilde{u}^{a_1}_{\mu _1} \tilde{u}^{a_2}_{\mu _2} \tilde{u}^{a_3}_{\mu _3} \right)^2
\nonumber \\
& &
\left.
-
\frac{1}{4!} \beta ^{-1} D^{a_1 a_2 a_3 a_4}_{\mu _1 \mu _2 \mu _3 \mu_4} \tilde{u}^{a_1}_{\mu _1} \tilde{u}^{a_2}_{\mu _2} \tilde{u}^{a_3}_{\mu _3}  \tilde{u}^{a_4}_{\mu _4}
+ \cdots
\right] \, e^{-\tilde{H}_{0}} \  \Delta , 
\nonumber 
\end{eqnarray}
where the polynomial terms have resulted from expanding the
anharmonic exponential (\ref{anharexp}) in a Taylor series.
In classic lattice dynamics analysis, a similar expansion is
used.  There the cubic term is odd in $\tilde{u}$ and integrates to zero, while
the other two terms give non-zero contributions 
to the heat capacity of the order $kT$.\cite{Wallace}
In CGMD, each of the three terms gives a non-zero contribution, as we will now show.

Integrals of the form (\ref{anhar6}), a polynomial times 
a Gaussian, may be evaluated in closed form.  The algebra
can be tedious, so often generating functions are used
to simplify the bookkeeping.  The CGMD generating function
is defined to be
\begin{eqnarray}
Z_h[J] & = & 
 Z_{{\mathrm{kin}}} \, \int \! d\tilde{{\mathbf{u}}}_{\mu}  
e^{-
\frac{1}{2} \tilde{{\mathbf{u}}}_{\mu }  \cdot D_{\mu \nu} \tilde{{\mathbf{u}}}_{\nu } 
} \  
e^{\tilde{{\mathbf{J}}}_{\mu } \cdot \tilde{{\mathbf{u}}}_{\mu }
}
\, \Delta 
\end{eqnarray}
where $Z_{{\mathrm{kin}}}$ is the part of the partition
function coming from the kinetic energy and including
the nodal velocity constraints.  The subscript on $Z_h$
indicates that it is the generating function corresponding
to the harmonic theory.  Generating functions
are useful because derivatives with respect to $\tilde{J}_{\mu}$
bring down factors of $\tilde{{\mathbf{u}}}_{\mu }$,
\begin{equation}
\left.  
\partial_{\tilde{{\mathbf{J}}}_{\lambda }}  Z_h[J] \right| _{J=0}
 = Z_{{\mathrm{kin}}} \, \int \! d\tilde{{\mathbf{u}}}_{\mu}
e^{-
\frac{1}{2} \tilde{{\mathbf{u}}}_{\mu }  \cdot D_{\mu \nu} \tilde{{\mathbf{u}}}_{\nu }
} \
\tilde{{\mathbf{u}}}_{\lambda }
\, \Delta
\end{equation}
so they provide a simple way of calculating the integrals
of Gaussians multiplied by a polynomial prefactor.
The CGMD internal energy (\ref{anhar6}) may therefore be
computed from the generating function as
\begin{eqnarray}
U & = & U_{{\mathrm{int}}} ^h
 - \partial _{\beta} \log
 \left\{   
\left.  
\exp \left[ -\beta H^{\prime} ( \partial_{\tilde{{\mathbf{J}}}} ) \right]
\, 
Z_h[J] \right| _{J=0}
\right\} 
\label{anhar6nabla} 
\end{eqnarray}
where the factor 
$\exp \left[ -\beta H^{\prime} ( \partial _{\tilde{{\mathbf{J}}}} ) \right]$
is intended to be expanded in its Taylor series for practical calculations.

We return to the anharmonic terms below [cf.\ Eq.\ (\ref{anharG})],
but first we compute the partition function.
In order to compute $Z_h[J]$ we can 
complete the square and relate the result to $Z_h[0]$:
\begin{eqnarray}
Z_h[J] & = & 
 Z_{{\mathrm{kin}}} \, 
e^{\frac{1}{2} \tilde{{\mathbf{J}}}_{\mu } \cdot D_{\mu \nu}^{-1} \tilde{{\mathbf{J}}}_{\nu }} \,
\int \! d\tilde{{\mathbf{w}}}_{\mu}  
e^{-
\frac{1}{2} \tilde{{\mathbf{w}}}_{\mu }  \cdot D_{\mu \nu} \tilde{{\mathbf{w}}}_{\nu } 
} \times \\  
& & ~~~~
\prod \delta \left( {\mathbf{u_j}} 
    - f_{j\mu} \tilde{{\mathbf{w}}}_{\mu} - f_{j\mu} D_{\mu \nu}^{-1} \cdot \tilde{{\mathbf{J}}}_{\nu } \right)
\nonumber \\
 & = & e^{\frac{1}{2} \tilde{{\mathbf{J}}}_{\mu } \cdot D_{\mu \nu}^{-1} \tilde{{\mathbf{J}}}_{\nu }} \, Z_h[J=0; \tilde{{\mathbf{u}}}_j - f_{j\mu} D_{\mu \nu}^{-1} \cdot \tilde{{\mathbf{J}}}_{\nu }]
\\
 & = & Z_h[J=0] \, e^{-\frac{1}{2} \tilde{{\mathbf{J}}}_{\mu } \cdot {\mathbf{G}}_{\mu \nu} 
\cdot \tilde{{\mathbf{J}}}_{\nu } - 
\tilde{{\mathbf{H}}}_{\kappa } \cdot \tilde{{\mathbf{J}}}_{\kappa }  }
\label{generatingfunction}
\end{eqnarray}
where $\tilde{{\mathbf{w}}}_{\mu}=\tilde{{\mathbf{u}}}_{\mu} - 
D_{\mu \nu}^{-1} \cdot \tilde{{\mathbf{J}}}_{\nu }$.  
We have used Eq.\ (\ref{cgHamHarm}) to derive the third line.
The Green function ${\mathbf{G}}_{\mu \nu}$ and the 
external field ${\mathbf{H}}_{\mu}$ are given by
\begin{eqnarray}
{\mathbf{G}}_{\mu \nu} & = &   D_{\mu \kappa}^{-1} f_{j\kappa} K_{jk}
f_{k\lambda} D_{\lambda \nu}^{-1} - D_{\mu \nu}^{-1}  
\label{Gfunc} \\
{\mathbf{H}}_{\mu} & = & {\mathbf{u}}_j K_{jk}
f_{k\nu} D_{\nu \mu}^{-1} 
\label{exFld}
\end{eqnarray}
with $\tilde{{\mathbf{H}}}_{\mu} = \beta ^{1/2} {\mathbf{H}}_{\mu}$.

Using the projection matrices defined in Appendix \ref{app-CGderiv},
we can rewrite this expression in a more transparent form.
The projection matrices are given by
$P^{{\mathrm{CG}}}_{\mu\nu} = N_{j\mu} f_{j\nu}$
and $P^{{\mathrm{\perp}}}_{\mu\nu} = \delta _{\mu\nu} - 
P^{{\mathrm{CG}}}_{\mu\nu}$.  Using them 
together with the formula for $K_{ij}$ (\ref{Kdef}) we have
the alternate formulas
\begin{eqnarray}
{\mathbf{G}}_{\mu \nu} & = & P^{{\mathrm{\perp}}}_{\mu\rho}
\left(  \tilde{D}_{\rho \kappa}^{-1} f_{j\kappa} K_{jk}
f_{k\lambda} \tilde{D}_{\lambda \sigma}^{-1} - 
\tilde{D}_{\rho \sigma}^{-1} \right)  
P^{{\mathrm{\perp}}}_{\sigma\nu}
\label{Galt} \\
{\mathbf{H}}_{\mu} & = & {\mathbf{u}}_j N_{j\mu} + 
{\mathbf{u}}_j K_{jk} f_{k\nu} \tilde{D}_{\nu \lambda}^{-1} 
P^{{\mathrm{\perp}}}_{\lambda\mu}
\label{Halt} 
\end{eqnarray}
where $\tilde{D}_{\mu \nu}$ is defined by 
Eq.\ (\ref{Dtilde}). \cite{footnote-regular}
These formulas show that in the rigid approximation
${\mathbf{G}}_{\mu \nu}=0$ and 
${\mathbf{H}}_{\mu} = {\mathbf{u}}_j N_{j\mu}$.
The result is that
each occurrence of ${\mathbf{u}}_{\mu}$
in the MD potential energy is just replaced by 
$N_{j\mu}{\mathbf{u}}_j$ in R-CGMD.  Relaxation of
the unconstrained degrees of freedom makes the further
contributions involving $P^{{\mathrm{\perp}}}_{\mu\nu}$.

To be more explicit, the rigid approximation to CGMD
freezes the unconstrained (unresolved) modes.  Formally,
this is accomplished by setting the orthogonal projectors
to zero: $P^{{\mathrm{\perp}}}_{\mu\nu} \rightarrow 0$.
In this approximation the generating function
simplifies 
\begin{equation}
\left. Z_h[J] \right| _{\mathrm{rigid}} 
 = Z_h[0] \, e^{- \beta ^{1/2} \mu _j N_{j\mu} \tilde{J}_{\mu} }
\end{equation}
where all of the Green function contributions
have been eliminated.  Similarly, the mass and
stiffness matrices in $Z_h[0]$ involve only the
first terms in Eqs.\ (\ref{stiffA1}) and (\ref{massA1}).
The Rigid CGMD (R-CGMD) Hamiltonian is then given by
\begin{equation}
E({\mathbf{u}}_k,\dot{{\mathbf{u}}}_k) = U_{{\mathrm{int}}} 
  + \frac{1}{2} M_{ij} 
\, \dot{{\mathbf{u}}}_i \cdot \dot{{\mathbf{u}}}_j
+ U_{MD}( N_{j\mu}{\mathbf{u}}_j ) , 
\label{cgHamFull} 
\end{equation}
where $U_{MD}$ is the full, non-linear MD potential
energy evaluated with ${{\mathbf{u}}}_{\mu} = N_{j\mu}{\mathbf{u}}_j$,
and $M_{ij}=N_{j\mu} m_{\mu} N_{k\mu}$.  
The rigid approximation neglects 
thermal effects and relaxation of the internal
degrees of freedom (i.e., those that have been
integrated out).
This expression is a conserved
energy, and it is free from spurious forces commonly
referred to as `ghost forces' in discussions of
concurrent multiscale modeling.\cite{Curtin} 
It is also as computationally expensive as MD, because
unlike the stiffness matrix, the non-linear potential
cannot be pre-computed except perhaps for some
toy potentials.  Additional approximations
are needed.  One set of approximations based on
the use of representative atoms and the Cauchy-Born
rule leads to the
Quasicontinuum Method.\cite{Phillips}  Thus, the 
derivation above
shows the relationship between CGMD and the
Quasicontinuum Method, in particular that 
the Quasicontinuum Method is closely related
to the zero temperature rigid approximation of
CGMD.  Another successful concurrent multiscale
technique, Coupling of Length Scale (CLS) \cite{CLS}, 
can also be related to CGMD.  It involves taking the
large-$N$ asymptotic limit of the stiffness
matrix in the rigid approximation, where $N$
in this case is the number of atoms per element.
The large-$N$ limit yields continuum mechanics
within the element; i.e., a finite element
representation of continuum mechanics.  Both
CLS and the Quasicontinuum Method make additional
approximations at the MD/CG interface.

\begin{figure}[t]
\includegraphics[width=0.45\textwidth]{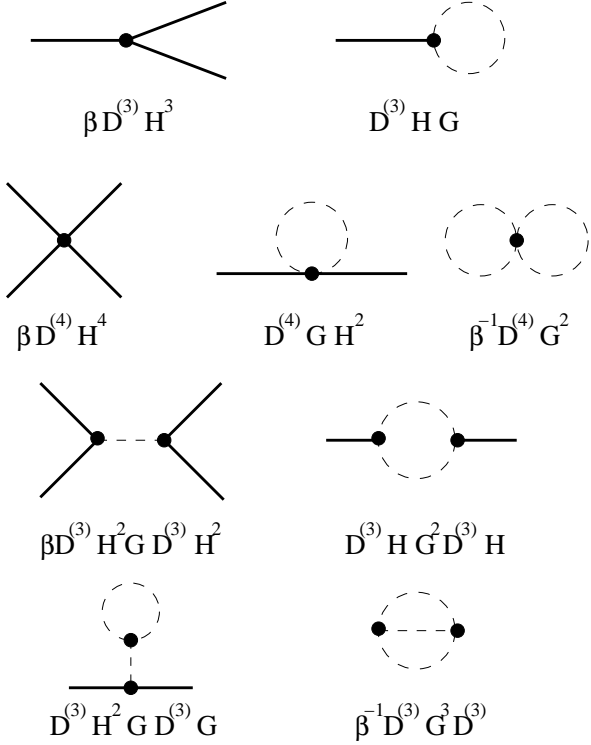}
\caption{Feynman diagrams of connected graphs
contributing to the anharmonic CGMD internal energy.  The
thick external legs represent factors of $H$, which in the
usual terminology correspond to dressed external fields.  
Physically, they are contributions from the CG fields with
some zero temperature relaxation of the internal degrees
of freedom.  
The dashed internal lines represent factors of $G$ that
arise from the fluctuations of the internal degrees of freedom.
These fluctuations are temperature dependent, and each 
factor of $G$ comes with a factor of $\beta ^{-1} (=kT)$.
\label{fig-feynman}}
\end{figure}

We now consider the physics of the terms in
the CGMD energy beyond the rigid approximation.
These are effects that arise due to the anharmonicity
of the interatomic potential, and 
include both thermal effects and non-linear relaxation.
The diagrammatic approach is a convenient and powerful
way to analyze perturbation theory at higher order,
such as the anharmonic contributions to CGMD.  
The quantities represented by the diagrams are typically 
generated from a relatively simple set of rules, known
as Feynman rules. Due to space limitations we cannot 
provide a thorough review of the rich mathematical structure 
encoded in Feynman diagrams, but direct the interested reader 
to one of the numerous texts on the subject.\cite{Ramond}
We have derived the Feynman rules
for CGMD, but they will not be presented here since
they are not actually needed.
Instead, we will employ the set of Feynman diagrams as 
a pedagogical device to consider the form of various
contributions to the CGMD energy.  The Feynman
diagrams up to second order in $D^{(3)}$ and first
order in $D^{(4)}$ are shown in Fig.\ \ref{fig-feynman}.
In the diagrams, the thick external legs represent
factors of ${\mathbf{H}}_{\mu}$; i.e., factors of 
the CG fields ${\mathbf{u}}_j$, suitably
dressed to account for elastic relaxation of the
internal degrees of freedom.  These factors are
playing the role of external fields in the
statistical field theory of the equilibrium
state of the internal degrees of freedom.
The thinner, dashed lines represent factors
of the Green function ${\mathbf{G}}_{\mu \nu}$.
The Green function accounts for thermal fluctuations of the
internal degrees of freedom, and indeed
each factor of ${\mathbf{G}}_{\mu \nu}$
brings with it a factor of $kT$.
Vertices in the graphs denote contracted tensor indices.

At a given level of perturbation theory, the
graphs pictorially represent the hierarchy
of contributions to the CGMD energy.  Although
graphs that split into more than one disconnected
sub-graph contribute to the partition function,
only connected graphs contribute to
the free energy.\cite{Ramond}  These are shown in 
Fig.\ \ref{fig-feynman}.  Consider first
the graphs with no external legs.  These
terms in the perturbation theory are
independent of ${\mathbf{u}}_j$; they
only contribute to the internal energy
resulting from degrees of freedom that 
have been integrated out.  Thus
they build up the non-trivial temperature
dependence of $U_{{\mathrm{int}}}$.  The
graphs at the other extreme--in particular
those that have no dashed lines and are
therefore independent of ${\mathbf{G}}_{\mu \nu}$--make
up the rigid approximation that was discussed above,
but now including some internal relaxation.

Using the generating function (\ref{generatingfunction}),
we have calculated
the CGMD internal energy (\ref{anhar6}) up to first order
in $D^{(4)}$ and second order in $D^{(3)}$:
\begin{eqnarray}
U & = & U_{{\mathrm{int}}} ^h
 - \partial _{\beta} \log Z \\
Z & = & \left(
1
-\frac{\beta ^{-1/2}}{6} D^{a_1 a_2 a_3}_{\mu _1 \mu _2 \mu _3} 
\partial _{\tilde{J}_{\mu _1}^{a_1}} \partial _{\tilde{J}_{\mu _2}^{a_3}}
\partial _{\tilde{J}_{\mu _3}^{a_3}}
\right.
\label{anharG} \\
& & ~
+
\frac{\beta ^{-1}}{72}
D^{a_1 a_2 a_3}_{\mu _1 \mu _2 \mu _3}
D^{a_4 a_5 a_6}_{\mu _4 \mu _5 \mu _6}
\partial _{\tilde{J}_{\mu _1}^{a_1}} \partial _{\tilde{J}_{\mu _2}^{a_3}}
\partial _{\tilde{J}_{\mu _3}^{a_3}} \partial _{\tilde{J}_{\mu _4}^{a_4}} 
\partial _{\tilde{J}_{\mu _5}^{a_5}} \partial _{\tilde{J}_{\mu _6}^{a_6}}
\nonumber \\
& & ~
\left.  \left.
-
\frac{\beta ^{-1}}{24} D^{a_1 a_2 a_3 a_4}_{\mu _1 \mu _2 \mu _3 \mu_4} 
\partial _{\tilde{J}_{\mu _1}^{a_1}} \partial _{\tilde{J}_{\mu _2}^{a_3}}
\partial _{\tilde{J}_{\mu _3}^{a_3}} \partial _{\tilde{J}_{\mu _4}^{a_4}} 
+ \ldots
\right) \, Z_h[\tilde{J}] \right| _{\tilde{J}=0}
\nonumber
\end{eqnarray}
where $Z_h[\tilde{J}]$ is given by Eq.\ (\ref{generatingfunction}).
The derivatives act on the Gaussian generating function 
(\ref{generatingfunction}) to give
\begin{eqnarray}
Z & = & \left[
1
+\frac{1}{2} D^{a_1 a_2 a_3}_{\mu _1 \mu _2 \mu _3}
G_{\mu _1 \mu _2}^{a_1 a_2} H_{\mu _3}^{a_3} 
\right.
\nonumber \\
& & ~
+\frac{1}{2} \left( 
\frac{1}{2} D^{a_1 a_2 a_3}_{\mu _1 \mu _2 \mu _3} 
G_{\mu _1 \mu _2}^{a_1 a_2} H_{\mu _3}^{a_3} 
\right) ^2
\nonumber \\
& & ~
+
\frac{\beta }{8}
D^{a_1 a_2 a_3}_{\mu _1 \mu _2 \mu _3}
D^{a_4 a_5 a_6}_{\mu _4 \mu _5 \mu _6}
G_{\mu _1 \mu _4}^{a_1 a_4}
H_{\mu _2}^{a_2} H_{\mu _3}^{a_3}
H_{\mu _5}^{a_5} H_{\mu _6}^{a_6}
\nonumber \\
& & ~
+
\frac{1}{4}
D^{a_1 a_2 a_3}_{\mu _1 \mu _2 \mu _3}
D^{a_4 a_5 a_6}_{\mu _4 \mu _5 \mu _6}
G_{\mu _1 \mu _4}^{a_1 a_4}
G_{\mu _2 \mu _5}^{a_2 a_5}
H_{\mu _3}^{a_3} H_{\mu _6}^{a_6}
\nonumber \\
& & ~
-
\frac{\beta ^{-1}}{12}
D^{a_1 a_2 a_3}_{\mu _1 \mu _2 \mu _3}
D^{a_4 a_5 a_6}_{\mu _4 \mu _5 \mu _6}
G_{\mu _1 \mu _4}^{a_1 a_4}
G_{\mu _2 \mu _5}^{a_2 a_5}
G_{\mu _3 \mu _6}^{a_3 a_6}
\nonumber \\
& & ~
+
\frac{1}{4}
D^{a_1 a_2 a_3}_{\mu _1 \mu _2 \mu _3}
D^{a_4 a_5 a_6}_{\mu _4 \mu _5 \mu _6}
G_{\mu _1 \mu _4}^{a_1 a_4}
G_{\mu _2 \mu _3}^{a_2 a_3}
H_{\mu _5}^{a_5} H_{\mu _6}^{a_6}
\nonumber \\
& & ~
-
\frac{\beta ^{-1}}{8}
D^{a_1 a_2 a_3}_{\mu _1 \mu _2 \mu _3}
D^{a_4 a_5 a_6}_{\mu _4 \mu _5 \mu _6}
G_{\mu _1 \mu _4}^{a_1 a_4}
G_{\mu _2 \mu _3}^{a_2 a_3}
G_{\mu _5 \mu _6}^{a_5 a_6}
\nonumber \\
& & ~
+
D^{a_1 a_2 a_3 a_4}_{\mu _1 \mu _2 \mu _3 \mu_4} \times
\nonumber \\
& & ~~~~~
\left(
-\frac{1}{4} G_{\mu _1 \mu _2}^{a_1 a_2} H_{\mu _3}^{a_3} H_{\mu _4}^{a_4} 
+
\frac{\beta ^{-1}}{8} G_{\mu _1 \mu _2}^{a_1 a_2}
G_{\mu _3 \mu _4}^{a_3 a_4}
\right)
\nonumber \\
& & ~
+ \ldots \left] 
e^{-\beta H^{\prime} ({\mathbf{u}}_{\mu}={\mathbf{H}}_{\mu})} \, Z_h[0]
\right.
\label{anharZ} 
\end{eqnarray}
%
Next we find the Helmholtz free energy $F=-kT \log Z$:
\begin{eqnarray}
F & = & F_{\mathrm{harmonic}} 
+ H^{\prime} ({\mathbf{u}}_{\mu}={\mathbf{H}}_{\mu})
\nonumber \\
& & ~
-
\frac{1}{8}
D^{a_1 a_2 a_3}_{\mu _1 \mu _2 \mu _3}
D^{a_4 a_5 a_6}_{\mu _4 \mu _5 \mu _6}
G_{\mu _1 \mu _4}^{a_1 a_4}
H_{\mu _2}^{a_2} H_{\mu _3}^{a_3}
H_{\mu _5}^{a_5} H_{\mu _6}^{a_6}
\nonumber \\
& & ~
-
\frac{1}{2}kT\, D^{a_1 a_2 a_3}_{\mu _1 \mu _2 \mu _3}
G_{\mu _1 \mu _2}^{a_1 a_2} H_{\mu _3}^{a_3} 
\nonumber \\
& & ~
-
\frac{1}{4} kT\,
D^{a_1 a_2 a_3}_{\mu _1 \mu _2 \mu _3}
D^{a_4 a_5 a_6}_{\mu _4 \mu _5 \mu _6}
G_{\mu _1 \mu _4}^{a_1 a_4}
G_{\mu _2 \mu _5}^{a_2 a_5}
H_{\mu _3}^{a_3} H_{\mu _6}^{a_6}
\nonumber \\
& & ~
-
\frac{1}{4} kT \,
D^{a_1 a_2 a_3}_{\mu _1 \mu _2 \mu _3}
D^{a_4 a_5 a_6}_{\mu _4 \mu _5 \mu _6}
G_{\mu _1 \mu _4}^{a_1 a_4}
G_{\mu _2 \mu _3}^{a_2 a_3}
H_{\mu _5}^{a_5} H_{\mu _6}^{a_6}
\nonumber \\
& & ~
+\frac{1}{4} kT \,
D^{a_1 a_2 a_3 a_4}_{\mu _1 \mu _2 \mu _3 \mu_4} 
G_{\mu _1 \mu _2}^{a_1 a_2} H_{\mu _3}^{a_3} H_{\mu _4}^{a_4} 
\nonumber \\
& & ~
-
\frac{1}{8} (kT)^2 \,
D^{a_1 a_2 a_3 a_4}_{\mu _1 \mu _2 \mu _3 \mu_4} 
G_{\mu _1 \mu _2}^{a_1 a_2} G_{\mu _3 \mu _4}^{a_3 a_4}
\nonumber \\
& & ~
+
\frac{1}{12} (kT)^2 \,
D^{a_1 a_2 a_3}_{\mu _1 \mu _2 \mu _3}
D^{a_4 a_5 a_6}_{\mu _4 \mu _5 \mu _6}
G_{\mu _1 \mu _4}^{a_1 a_4}
G_{\mu _2 \mu _5}^{a_2 a_5}
G_{\mu _3 \mu _6}^{a_3 a_6}
\nonumber \\
& & ~
+
\frac{1}{8} (kT)^2 \,
D^{a_1 a_2 a_3}_{\mu _1 \mu _2 \mu _3}
D^{a_4 a_5 a_6}_{\mu _4 \mu _5 \mu _6}
G_{\mu _1 \mu _4}^{a_1 a_4}
G_{\mu _2 \mu _3}^{a_2 a_3}
G_{\mu _5 \mu _6}^{a_5 a_6}
\nonumber \\
& & ~
+ \ldots 
\label{anharF} 
\end{eqnarray}
where $F_{\mathrm{harmonic}}$ is the CGMD free energy
for a harmonic crystal [cf.\ Eq.\ (\ref{cgHamHarm})]:
\begin{eqnarray}
F_{\mathrm{harmonic}} & = & F_{{\mathrm{int}}}
  + \frac{1}{2} \sum _{j,k} \left(
M_{jk} \, \dot{{\mathbf{u}}}_j \cdot \dot{{\mathbf{u}}}_k
+ {\mathbf{u}}_j \cdot K_{jk} {\mathbf{u}}_k \right)
\label{cgFreeEHarm} \\
F_{{\mathrm{int}}} & = & \natom E^{\mathrm{coh}} 
- 3(\natom-\nnode)kT \, \log (kT)
\label{Fint}
\end{eqnarray}
The free energy is suitable for isothermal CGMD
simulations.  On the other hand, the internal energy 
is appropriate
for adiabatic simulations, and hence more 
closely related to MD simulations without
a thermostat.  
It is given by $U=\partial _{\beta} \left( \beta F \right)$:
\begin{eqnarray}
U & = & E_{\mathrm{harmonic}} 
+ H^{\prime} ({\mathbf{u}}_{\mu}={\mathbf{H}}_{\mu})
\nonumber \\
& & ~
-
\frac{1}{8}
D^{a_1 a_2 a_3}_{\mu _1 \mu _2 \mu _3}
D^{a_4 a_5 a_6}_{\mu _4 \mu _5 \mu _6}
G_{\mu _1 \mu _4}^{a_1 a_4}
H_{\mu _2}^{a_2} H_{\mu _3}^{a_3}
H_{\mu _5}^{a_5} H_{\mu _6}^{a_6}
\nonumber \\
& & ~
+
\frac{1}{8} (kT)^2 \,
D^{a_1 a_2 a_3 a_4}_{\mu _1 \mu _2 \mu _3 \mu_4} 
G_{\mu _1 \mu _2}^{a_1 a_2} G_{\mu _3 \mu _4}^{a_3 a_4}
\nonumber \\
& & ~
-
\frac{1}{12} (kT)^2 \,
D^{a_1 a_2 a_3}_{\mu _1 \mu _2 \mu _3}
D^{a_4 a_5 a_6}_{\mu _4 \mu _5 \mu _6}
G_{\mu _1 \mu _4}^{a_1 a_4}
G_{\mu _2 \mu _5}^{a_2 a_5}
G_{\mu _3 \mu _6}^{a_3 a_6}
\nonumber \\
& & ~
-
\frac{1}{8} (kT)^2 \,
D^{a_1 a_2 a_3}_{\mu _1 \mu _2 \mu _3}
D^{a_4 a_5 a_6}_{\mu _4 \mu _5 \mu _6}
G_{\mu _1 \mu _4}^{a_1 a_4}
G_{\mu _2 \mu _3}^{a_2 a_3}
G_{\mu _5 \mu _6}^{a_5 a_6}
\nonumber \\
& & ~
+ \ldots 
\label{anharU} 
\end{eqnarray}
where $E_{\mathrm{harmonic}}$ is given by
Eq.\ (\ref{cgHamHarm}).
Now $T$ is understood to be a function of
the (constant) entropy and the state of 
deformation, although for many applications
since the deformation in the CG region is
small, it can be treated as constant.
For reference, the expressions for 
the Green function
$G_{\mu _1 \mu _2}^{a_1 a_2}$ and the
field $H_{\mu}^{a}$ are given
in Eqs.\ (\ref{Galt}) and (\ref{Halt}),
respectively.

We could continue to calculate the terms in the CGMD 
Hamiltonian to higher order in the MD anharmonic corrections
and/or the CGMD thermal perturbation expansion.  The number 
of terms grows rapidly, and we quickly reach the point
of marginal returns; i.e., the point at which the added
complexity is no longer rewarded with a commensurate
improvement in accuracy.  However, there are certain
kinds of contributions where improvements are possible.
In particular, it is important to capture the first 
non-trivial effects in the expansion.  The 
terms calculated thus far have contributed to the
harmonic Hamiltonian, the zero-temperature 
anharmonic terms and the energy of the internal
modes.  We have also calculated the leading contributions
in the free energy to the thermal expansion and 
temperature-dependence of the stiffness, i.e., the
terms proportional to ${\mathbf{u}}_j$ and 
${\mathbf{u}}_{j_1} \otimes {\mathbf{u}}_{j_2}$.
In the internal energy, these contributions are
higher order in the anharmonic lattice expansion,
so we have not calculated them yet.  Of course,
there are thermal corrections to the higher order
stiffnesses, as well, but they are not as important.

\begin{figure}[t]
\includegraphics[width=0.45\textwidth]{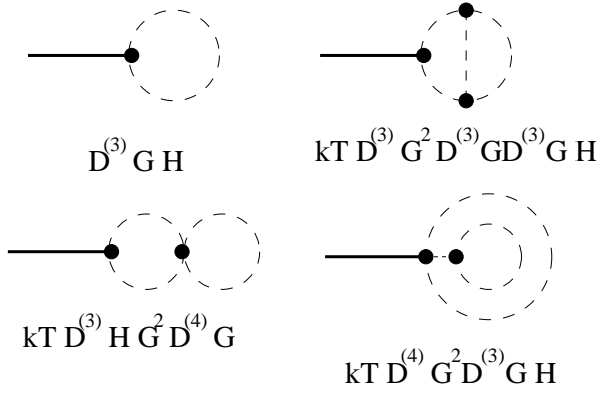}
\caption{The leading anharmonic contributions to the
partition function affecting the
thermal expansion in CGMD.
\label{fig-qh1}}
\end{figure}

We conclude this section with the calculation of the 
leading temperature-dependent, quasi-harmonic 
contributions to the CGMD Hamiltonian (internal
energy).  The graphs contributing the leading
temperature dependence to the one-point function
(governing thermal expansion) are shown in 
Fig.\ \ref{fig-qh1}.  The term ``one-point''
means a single leg, and hence a single factor
of ${\mathbf{u}}_j$.  Consider the lowest
order term in $kT$ in Fig.\ \ref{fig-qh1} as
it enters the free energy:
\begin{eqnarray}
- \frac{1}{2}kT\, D^{a_1 a_2 a_3}_{\mu _1 \mu _2 \mu _3}
G_{\mu _1 \mu _2}^{a_1 a_2} H_{\mu _3}^{a_3} & = &
- {\mathbf{f}}_j^0 {\mathbf{u}}_j 
\label{thermExp}
\end{eqnarray}
where we define the temperature-dependent ${\mathbf{f}}_j^0$ (not
to be confused with $f_{j\nu}$) to be:
\begin{eqnarray}
\left( {\mathbf{f}}_j^0 \right) _{a} & = & 
- \frac{1}{2}kT\, D^{a_1 a_2 a}_{\mu _1 \mu _2 \mu _3}
G_{\mu _1 \mu _2}^{a_1 a_2} 
f_{k\nu} D_{\nu \mu _3}^{-1} K_{jk}
\end{eqnarray}
where we have used the expression (\ref{exFld}) for $H_{\mu _3}^{a_3}$.
We have written the leading one-point term from
the free energy in Eq.\ (\ref{thermExp})
in a form familiar from finite element models
as an eigenstrain 
(cf.\ Section 2.12 of Hughes\cite{FE}).
As in conventional finite element formulations in
uniform temperature, this
term is a total difference (the discrete analog of
a total derivative), and it only enters through
the boundaries of the simulation.  
In a homogeneous system with free surfaces, the
result is expansion of the system as the 
temperature increases.  The expansion is 
proportional to $kT$ and $D^{(3)}$ to leading
order, as expected based on conventional
lattice dynamics.\cite{Wallace}
Note that since the linear term is a total difference,
the discussion of ghost forces given above continues 
to be valid in the case of thermal expansion.

\begin{figure}[t]
\includegraphics[width=0.45\textwidth]{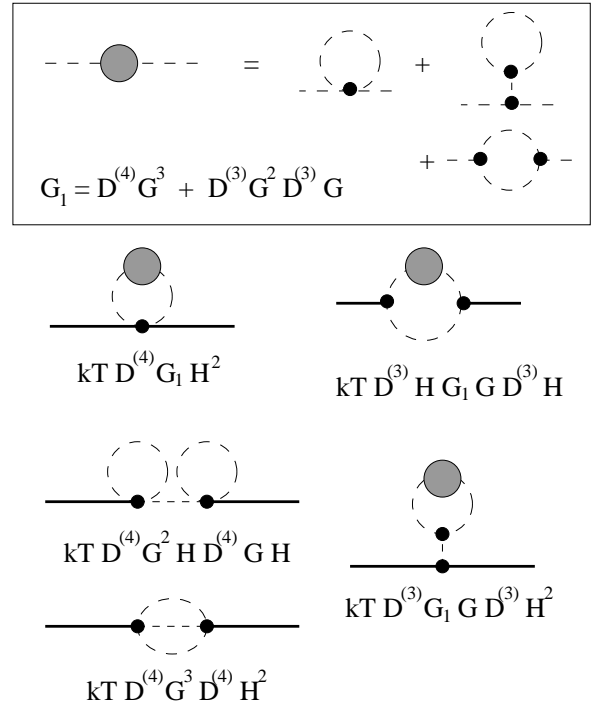}
\caption{The anharmonic contributions to the
partition function affecting the
temperature dependence of the adiabatic CGMD stiffness.
The Green function $G_1$ is represented by a dashed line
with a large filled circle.  It is defined in the upper box.
\label{fig-qh2}}
\end{figure}

The graphs contributing the leading
temperature dependence to the two-point function
governing the temperature dependence of the 
adiabatic elastic stiffness are shown in Fig.\ \ref{fig-qh2}.
The leading contributions to the isothermal
stiffness are the three graphs containing $G_1$
but with $G_1$ replaced by $G$.  As expected, the
first thermal contributions to the adiabatic stiffness
are of order $(kT)^2$ due to the third law of thermodynamics
(the Nernst theorem), whereas the first thermal contributions 
to the isothermal stiffness are of order $kT$.

Finally, we reemphasize that we have calculated properties
within the CGMD perturbation theory to show that the
theory is consistent and to gain some theoretical understanding
of the interplay of anharmonicity and temperature in
coarse-grained systems.  In practice, we take the reference
configuration to be the crystal lattice at a particular
temperature with the corresponding finite temperature
dynamical matrix.  It would be a tautology to compute
thermal expansion or thermal softening in CGMD: they agree 
with the MD result identically by construction.

\section{Thermal Fluctuations}
\label{sec-thermal}

In the previous Section we developed a description of the average
motion of the collective degrees of freedom.  We have shown that
the short wavelength modes contribute to the mean motion, even
though they are not explicitly present.  They have the additional
effect of inducing small fluctuations about this mean.  The CG
modes behave as if they were in a Brownian heat bath;  they
are jostled by small, random interactions with the (invisible, missing)
short wavelength modes.  In this section, we analyze these interactions
in detail.

Consider the fluctuations of the CG fields in harmonic MD 
in thermal and mechanical equilibrium,
\begin{eqnarray}
\langle {\mathbf u}_j {\mathbf u}_k \rangle  & = & 
f_{j \mu } f_{k \nu } Z^{-1} \int \! 
d{\mathbf{x}}_{\mu} d{\mathbf{p}}_{\mu} \
e^{-\beta H_{MD}} \  {\mathbf u}_{\mu} {\mathbf u}_{\nu} \\
& = & f_{j \mu } f_{k \nu } \, kT \, \partial _{D_{\mu \nu}}
\tr \log ( D ) \\
& = & kT \, f_{j \mu } D_{\mu \nu}^{-1} f_{k \nu } \\
& = & kT \, K_{j k }^{-1}
\end{eqnarray}
This calculation shows that the two point isothermal correlation function
of the CG displacements grows linearly with the temperature.  The result
is reasonable and could equally well be understood in terms of the
equipartition theorem.  As the temperature increases, the average
potential energy increases and so the mean amplitude of the harmonic
oscillations increases.  Note that the amplitude decreases as the
stiffness $K_{jk}$ increases.

We can repeat the calculation in harmonic CGMD again in 
thermal equilibrium for comparison:
\begin{eqnarray}
\langle {\mathbf u}_j {\mathbf u}_k \rangle _T & = & 
Z^{-1} \int \! 
d{\mathbf{u}}_{j} d{\dot{\mathbf{u}}}_{j} \
e^{-\beta H_{CGMD}} \  {\mathbf u}_{j} {\mathbf u}_{k} \\
& = & kT \, K_{j k }^{-1}
\end{eqnarray}
where $H_{CGMD}$ is given by Eq.\ (\ref{cgHamHarm}).
The result is exactly the same as the correlation function
of the CG fields in MD.  This equivalence of the correlation
functions holds for higher correlation functions as well,
due to Wick's theorem.\cite{Peskin}

While the equilibrium properties agree, there are differences
in the time autocorrelation functions of the displacement.
These differences result from the influence of the short
wavelength modes that are not represented on the mesh.  These
additional modes acts as a heat bath, exerting random and
dissipative forces on the CG fields.  The effect is entirely
analogous to Brownian motion, in which a large particle is
jostled by the thermal motion of the unseen atoms in a liquid
surrounding it. \cite{MRS}

There are several practical issues that arise in
the analysis of CGMD simulations regarding fluctuations.
In conventional MD, the mean atomic kinetic energy is
used to compute the temperature.  We now investigate
whether a similar connection holds in the CG region of
CGMD (naturally it continues to hold in the MD region
if it is in thermal equilibrium).  We consider the 
simplified case in which there is no large-scale 
motion such as center-of-mass motion; our derivation
continues to hold provided any such non-equilibrated
modes are subtracted from the nodal velocities prior
to analysis.  Consider the mean-squared velocity of
node $i$ in thermal equilibrium.
We use the partition function for CGMD to
calculate the expectation value of $\left| \dot{u}_i \right|^2$
in the canonical ensemble at thermal equilibrium:
\begin{eqnarray}
\langle \left| \dot{u}_i \right|^2 \rangle & = & 
  Z^{-1}_{{\mathrm{kin}}} \int \! d\dot{u} \, \left| \dot{u}_i^2 \right| 
e^{-\frac{1}{2} \beta M_{jk} 
\dot{{\mathbf{u}}}_j \cdot \dot{{\mathbf{u}}}_k} \\
 & = & 
  Z^{-1}_{{\mathrm{kin}}} \partial ^2 _{J_i} \left. \int \! d\dot{u} \, 
e^{-\frac{1}{2} \beta M_{jk}
\dot{{\mathbf{u}}}_j \cdot \dot{{\mathbf{u}}}_k -
\dot{{\mathbf{J}}}_j \cdot \dot{{\mathbf{u}}}_j} \right| _{J=0}\\
 & = & 
  \partial ^2 _{J_i} \left. 
e^{\frac{1}{2} \beta^{-1} M_{jk}^{-1}
\dot{{\mathbf{J}}}_j \cdot \dot{{\mathbf{J}}}_k } \right| _{J=0}\\
 & = & 
 3 kT \, M_{ii}^{-1} 
\label{tempVelo}
\end{eqnarray}
where the factor of 3 in the final line is the number of dimensions.
The potential energy contribution to the partition function
is irrelevant to this calculation and has been suppressed.
The important result expressed in Eq.\ (\ref{tempVelo}) is 
that the mean-squared velocity is directly
proportional to the temperature, just as it is in conventional
MD.  

In fact, the MD result is recovered by replacing $M_{ii}$
with the mass of atom $i$.  This calculation also applies to
concurrent multiscale models that use a lumped mass matrix.
For CGMD with a non-diagonal mass matrix, it is only the
diagonal of the mass matrix that affects the amplitude of
oscillation of a given node.  The off-diagonal terms do
introduce correlations between the velocities of neighboring
nodes that would not be present for a diagonal mass matrix.
Because of these correlations, the simple atomistic
relationship between the mean-squared velocity and the 
temperature that follows directly from the equipartition
theorem must be modified according to Eq.\ (\ref{tempVelo})
for use in CGMD.

Thus far in this article we have treated CGMD
as a system that conserves energy, and these random,
dissipative forces are absent.  In particular,
the evolution of the system generated by the
Hamiltonian (\ref{cgHam}) conserves the energy
given by that Hamiltonian.  Thermostats might
be added to simulate the electron-phonon coupling;
i.e., the interaction of the lattice vibrations
with the electronic degrees of freedom.  Such
additions violate energy conservation, \cite{footnote-Nose}
since energy can flow to and from the heat
bath and the system becomes an open system.

Even neglecting the electron-phonon coupling,
the coarse-grained system of solid mechanics
described by CGMD is an open system.  In the
full MD system energy can flow between modes
that would be retained in CGMD and those
that would be integrated out.\cite{PSS}  This 
interaction implies that the CGMD energy is conserved 
only on average, and that additional 
interactions are present in reality.  These
additional interactions take the form of
random and dissipative forces.  The form of 
these generalized Langevin forces may be 
computed using statistical mechanical
(Zwanzig-Mori) projection operator techniques,\cite{Zwanzig}
although it is beyond the scope of this article.
The resulting spatio-temporal memory
kernel has been described elsewhere.\cite{MRS}
The random, dissipative forces not only
act to bring the CG degrees of freedom
into equilibrium with the internal
degrees of freedom, but they act to
absorb short wavelength modes incident
on an interface where the mesh is
refined.  In principle, they also include
the propagators that reconstruct waves on
the far side of the CG region if the
mesh is refined again; in practice,
these propagators appear to be very
expensive to implement computationally.
In fact, one of the challenges of memory
kernels is their computational expense
both in terms of the memory required
to store the recent history and in 
terms of the demands they place
on parallelization to make the code suitable
for supercomputers.

\section{Implementation Details}
\label{sec-details}

In practice, CGMD is run much the way conventional MD would be.
The forces in the CG region are determined by the CGMD stiffness
matrix and the nodal displacements; the forces in the MD region
could be determined this way, as well, but since we have shown
that the forces in the MD region are just the usual MD forces,
the full MD potential is used to calculate the MD forces.
Using the accelerations, the velocity Verlet time integrator is
used to evolve the system in time.\cite{vVerlet}  The same
time step is used throughout the simulation.  In principle,
the natural frequency in the CG region is lower as the mesh
size increases, and a longer
time step could be used there;  in practice, the CG region
entails relatively little computational expense, and there
is little motivation to introduce a spatially varying
time step that could cause subtle problems.

One difference from MD and conventional FEM is that the 
topology of the CG mesh
is not allowed to change.  Neighboring nodes remain neighboring
nodes throughout the simulation.  The topology of the mesh is
determined by a cell list, which contains the nodes associated 
with each cell in the mesh, and a face list, which lists pairs of
neighboring cells.  An edge list, which lists pairs of
neighboring nodes, is also generated.

The stiffness matrix $K_{ij}$ is to be computed once at the start of
a simulation, and it remains unaltered during the subsequent
dynamics.  It does not matter whether atoms vibrate across cell
boundaries, as long as the crystal lattice topology does not
change and diffusion is negligible.  
The elements of $K_{ij}$ decrease exponentially with distance
from the diagonal, and in practice it is necessary to truncate
the stiffness matrix in order to control the memory and
CPU requirements for simulating large systems with irregular
meshes.  Both requirements scales as at least $O(N^2)$ if
the full stiffness matrix is retained, and this scaling can
be reduced to $O(N)$ if matrix elements are discarded if
$K_{ij} < \epsilon K_{ii}$ for some small number $\epsilon$.
This approach allows the simulation of billion atom systems (greatly
coarse-grained) on desktop 
workstations without approximation beyond those presented here. 

For $T\ne 0$ the finite temperature dynamical matrix should be
used for $D_{\mu \nu}$.  This quasi-harmonic approximation 
ensures a consistent thermodynamics, and it effectively sums
the two-legged diagrams of the finite-temperature anharmonic
perturbation theory (i.e., those terms second order in the
CG displacement).
For example, in ergodic systems the time average of the kinetic 
energy term in the CG energy (\ref{cgHamHarm}) is related to the 
temperature through a equipartition expression.  In general, the dynamical
matrix may depend on other macroscopic parameters, as well, such
as slowly varying external magnetic and electric fields. $D _{\mu \nu}$
should be evaluated under these conditions.  Also note that while
the harmonic approximation may be good in peripheral regions, it
may not be appropriate for the important regions.  We have shown
that the CGMD and MD equations of motion agree in regions where
the mesh coincides with the atomic sites.  In these regions, the
full MD potential may, and should, be employed, so that effects
such as diffusion and dislocation dynamics are allowed.

\subsection{Normal Modes and the Inverse of $D_{\mu \nu}$}

In order to simulate large CG regions, it is necessary
to take some measures to increase computational efficiency.
One such trick is to make use of the long-range order
in a single crystal to facilitate the computation of
the stiffness matrix.  The eigenstates of the dynamical
matrix $D_{\mu \nu}$ are plane waves.  For monatomic
lattices they correspond to the normal modes of the
system, the longitudinal and transverse phonons
in the acoustic and optical branches that are familiar
from lattice dynamics.\cite{BH,Wallace}  In reciprocal
space, where the basis elements are exactly these
plane waves, the dynamical matrix is diagonal.  The
inverse of the dynamical matrix is then trivial
to compute, and the subtlety of inverting a singular
matrix is eliminated because reciprocal space 
naturally factorizes into a direct product of
the three zero modes with ${\mathbf{k}}=0$ and all
of the other modes with non-zero eigenvalues.

To be specific, the equation for the stiffness
matrix (\ref{Kdef2}) becomes
\begin{eqnarray}
K_{ij}^{ab} & = & 
  (N N^T)_{im} \left[ N_m(-{\mathbf{k}}) \, 
   D^{-1}_{ab}({\mathbf{k}}) \, N_n({\mathbf{k}}) \right] ^{-1} 
              (N N^T)_{nj} ,
\label{stiffFourier}
\end{eqnarray}
where ${\mathbf{k}}=0$ is explicitly omitted from the sum.
Now the inner matrix inverse is the inverse of a 3$\times$3 matrix.

\subsection{Shape Functions}

This section discusses particular choices for
interpolation functions.  Compatible combinations of
these are also allowed, as in FEM with wedge and brick
elements, for example.
We emphasize how different choices for interpolation
functions meet the requirement of meshing the crystal
lattice at the MD/CG interface.
The usual linear interpolation functions 
for tetrahedral elements \cite{Huebner} 
are the simplest functions meeting the
three criteria (\ref{shapecriteria1}) - (\ref{shapecriteria3})
in Section \ref{sect-shape}.
They are defined such that $N_j({\mathbf{x}})$ is
1 at node ${\mathbf{x}}_j$, it goes linearly to zero at the
nearest-neighbor nodes, and it vanishes outside of the nearest cells.  
Suppose ${\mathbf x}$ is in the $k$th element with nodes ${\mathbf x}_{k_j}$
where $j=1,\ldots,4$. Then the interpolation functions are given
by the volumnal or natural tetrahedral coordinates:
\begin{eqnarray}
N_{k_j}({\mathbf x}) & = & {\mathbf x}\cdot \partial _{k_j} 
\log V_k({\mathbf x}_{k_1},\ldots,{\mathbf x}_{k_4})
 \\
V_k & = & \frac{1}{6} \left| 
\begin{array}{cccc}
1 & x_{k_1} & y_{k_1} & z_{k_1} \\
1 & x_{k_2} & y_{k_2} & z_{k_2} \\
1 & x_{k_3} & y_{k_3} & z_{k_3} \\
1 & x_{k_4} & y_{k_4} & z_{k_4} 
\end{array}
\right| 
\end{eqnarray}
where ${\mathbf x}_{k} = (x_{k} , y_{k} , z_{k})$ and
we have written the tetrahedral volume as a determinant.
The interpolation function for node $k_j$ is simply the 
volume of the tetrahedron formed by ${\mathbf x}$ and
the other three nodes divided by the volume of the entire
tetrahedral cell.  
These functions are clearly C$^0$ continuous and independent.
It is easily checked that they are linear and form a partition of unity.
They also have the desirable properties of locality and ease of use.  
The locality property is particularly
important for our applications, since the domains requiring an
atomistic treatment are localized to small regions of the system.   

Another basis we have found useful is the set of the longest 
wavelength normal modes.  These functions satisfy the less
stringent basis properties:
(i$^\prime$) linear independence, 
$\det N_j({\mathbf{x}}_k) \ne 0$, and (ii$^\prime$) representation
of unity, $1=\sum _{j=1}^{N_{{\mathrm node}}} c_j \, N_j({\mathbf{x}})$
for some constants $c_j$.
This basis provides a check of the CG Hamiltonian (\ref{cgHam}),
since these functions are the optimal basis for a regular
CG mesh---the phonon spectrum comes out exactly correct, 
apart from the missing short-wavelength modes.  
The disadvantage of this basis for irregular meshes
is that it is nonlocal and the short wavelength modes that should
be supported on the finer parts of the mesh are absent.  
In particular, the stiffness matrix elements decrease as
\begin{equation}
K_{ij} \sim \frac{1}{\Delta x_{ij}},
\end{equation}
instead of decreasing exponentially with distance in
the local basis case.
The short wavelength modes could be restored locally
through the use of a wavelet basis, in principle, 
but we have not implemented a wavelet-based version of CGMD.

In many cases, higher order polynomial interpolating functions are
the basis of choice.  Generalizations of the 8-node brick used
for hexahedral lattices \cite{Huebner} are particularly easy
to implement.  For example, a generalized 8-node brick is the 
element we used to calculate the CGMD spectra for solid argon
and tantalum
presented in Section \ref{subsec-argon} below.  First, consider
a simple cubic lattice.  The basic 8-node brick involves
interpolations functions of the form\cite{Huebner} 
\begin{equation}
N_1^{{\mathrm{cubic}}}( \xi _a )  =  
{\tst \frac{1}{8} (1+\xi _1) \, (1+\xi _2) \, (1+\xi _3) }
\label{scCorner} 
\end{equation}
where throughout this section we use the scaled coordinates
$\xi _a\in [-1,1)$.  $N_1$ is associated with the 
corner node at ${\vec{\xi}}=$ (1,1,1).
The other 7 interpolation functions
are generated by the action of the point group $O_h$ on
$N_1$.  

The cubic elements may be applied to a variety of
hexahedral elements by mapping the real space
coordinates onto the $[-1,1)^3$ cube in the 
coordinate space  
using the standard multi-linear coordinate transformation
often called ``natural coordinates'' for the hexahedron 
in the finite element literature.\cite{Huebner}
For our purposes, some especially
important cases are the monatomic Bravais lattices,
such as face-centered cubic (fcc) and body-centered cubic (bcc)
lattices.  Suppose ${\mathbf{a}}_a$ are the basis vectors
in real space, and ${\mathbf{b}}_a$ are the reciprocal
basis vectors such that 
${\mathbf{a}}_a \cdot {\mathbf{b}}_b = \delta _{ab}$.
For example, in the fcc lattice the basis vectors could
be chosen to be 
${\mathbf{a}}_1=(0,\frac{1}{2},\frac{1}{2})a$,
${\mathbf{a}}_2=(\frac{1}{2},0,\frac{1}{2})a$ and
${\mathbf{a}}_3=(\frac{1}{2},\frac{1}{2},0)a$,
and 
${\mathbf{b}}_1=(-1,1,1)/a$,
${\mathbf{b}}_2=(1,-1,1)/a$ and
${\mathbf{b}}_3=(1,1,-1)/a$,
where $a$ is the lattice constant.  Then interpolation
functions on the Bravais lattice are given by
\begin{equation}
N_j^{\mathrm{Bravais}}({\mathbf{x}})  =  
N_j^{\mathrm{cubic}}(\xi_a=2{\mathbf{b}}_a\cdot{\mathbf{x}}-1)
\label{BravaisShape}
\end{equation}
using the shape functions defined in Eq.\ (\ref{scCorner}).
Shape functions for the bcc lattice can be constructed 
in the same way, with the basis vectors
be chosen to be 
${\mathbf{a}}_1=(\frac{1}{2},\frac{1}{2},-\frac{1}{2})a$,
${\mathbf{a}}_2=(\frac{1}{2},-\frac{1}{2},\frac{1}{2})a$ and
${\mathbf{a}}_3=(-\frac{1}{2},\frac{1}{2},\frac{1}{2})a$,
and 
${\mathbf{b}}_1=(1,1,0)/a$,
${\mathbf{b}}_2=(1,0,1)/a$ and
${\mathbf{b}}_3=(0,1,1)/a$,
where $a$ is the lattice constant.

One drawback of the fcc and bcc shape functions (\ref{BravaisShape})
is that they break the point group symmetry of the lattice.
Acting on the mesh with an element of the point group
returns a new mesh with the same nodes but often a different
set of cell boundaries.  One example is $C_4^z$, the 90$^\circ$
rotation about $z$.  It changes the cell boundaries, as
can be seen by its action on 
$\frac{1}{2}({\mathbf{a}}_1+{\mathbf{a}}_2)$ [i.e.,
${\vec{\xi}}= (\frac{1}{2},\frac{1}{2},0)$].  $C_4^z$
maps this face point to ${\vec{\xi}}'= (1,0,-\frac{1}{2})$,
a point on an edge of the original mesh.  Another way
to understand this symmetry breaking is that we made a
choice when we selected the basis 
${\mathbf{a}}_1=(0,\frac{1}{2},\frac{1}{2})a$,
${\mathbf{a}}_2=(\frac{1}{2},0,\frac{1}{2})a$ and
${\mathbf{a}}_3=(\frac{1}{2},\frac{1}{2},0)a$.
Had we selected another basis, say
${\mathbf{a}}'_1=(-\frac{1}{2},0,\frac{1}{2})a$,
${\mathbf{a}}'_2=(0,\frac{1}{2},\frac{1}{2})a$ and
${\mathbf{a}}'_3=(-\frac{1}{2},\frac{1}{2},0)a$,
then the mesh would have been different. It would 
have different cell edges and faces, 
even though the cell nodes would be the same.
More importantly for our purposes, a displacement field 
interpolated using one set of shape functions
cannot, apart from a few special cases, be represented
exactly with the rotated shape functions.  The
results of CGMD modeling then depend to some
extent on the choice of basis and associated
shape functions.

The symmetry breaking is small and
of no consequence in most applications; however, 
we are interested in using wave spectra
as a test of CGMD, plotting the spectra
along high symmetry directions in the
Brillouin zone.  The symmetry breaking
is somewhat troublesome in this case
because high symmetry directions no longer
possess the high symmetry and directions
that are supposed to be equivalent by symmetry 
are not.  We have developed a symmetrization procedure 
to eliminate the effects completely.  Its use
is limited to applications where high symmetry
is important, such as wave spectra, so we
present it below where the spectra are calculated.

Another approach to this problem is to introduce
polynomial bases that respect the point group symmetry.
The well-known 
serendipity functions\cite{serendip,FE} are an example of a 
minimal polynomial basis that respects cubic symmetry.  
The serendipity functions may be generalized in a way
that makes them suitable for a cubic fcc or bcc cell
such that there is a node for each atom in the cubic
unit cell, so that the MD degrees of freedom are recovered 
in the atomic limit.  These are different than the usual
serendipity functions and, indeed, are not as well suited
for conventional FEM applications because of the location
of the nodes (e.g.\ the bcc element has an internal
node).\cite{FE}  They do, however, meet our need to match the atomic 
lattice and preserve symmetries in the atomic limit.
For example, the face-centered cubic (fcc) serendipity functions 
have nodes at
the corners and the at the middle of the faces of a cube.
To the best of our knowledge, this kind of FEM 
interpolation function has not been used previously. 
They are well-suited to Bravais lattices, because of their
simple action under the point group symmetry.  Consider a
cubic unit cell of an fcc lattice with local coordinates 
$\xi _a$, where $-1\le \xi _a\le 1$ for $a=1,2,3$.  Define
the function
\begin{eqnarray}
N_1( \xi _a ) & = & \frac{1}{8} (1+\xi _1) \, (1+\xi _2)
\, (1+\xi _3) \times 
\label{fccCorner} \\
& & ~~~ \left[ 2 ( \xi _1 +\xi _2 +\xi _3 -1 ) - 
( \xi _1 \xi _2 + \xi _2 \xi _3 + \xi _3 \xi _1 ) \right] \nonumber
\end{eqnarray}
associated with the corner node at ${\vec{\xi}}=$ (1,1,1)
and the function
\begin{equation}
N_9( \xi _a ) = \frac{1}{2} \, (1 + \xi_1) \, (1-\xi _2^2) 
\, (1-\xi _3^2) 
\label{fccFace}
\end{equation}
associated with the face node at ${\vec{\xi}}=$ (1,0,0).
Basis functions, $N_j(\xi _a )$ associated with the other
nodes are generated from (\ref{fccCorner}) and (\ref{fccFace}) 
by the action of the point group.
These functions satisfy the strong requirements for an interpolation
basis.  Each function vanishes outside of the cells containing
the corresponding node, and it goes to zero at the opposite
faces of those cells: it is local and continuous.  Also,
taken together they form a partition of unity, as is easily checked.  
These 14 functions comprise only part of the set of 26 polynomials 
of order at most (2,2,1); i.e. the set of polynomials with terms no
higher than $x^2y^2z$ (or $x^2yz^2$, etc.). 
But they are specified uniquely by the three
basis requirements and the fact that they respect the point group;
i.e. a point group operation which leaves a particular node invariant
also leaves the corresponding function invariant.  These fcc shape
functions are most useful for testing purposes such as the computation
of phonon spectra and scattering properties where it is desirable
to maintain as many symmetries as possible.  

Other (novel) 
fcc symmetric bases are possible if one abandons the notion
of self-contained elements.  Typically, equations of motion in
finite elements are assembled element by element.  In CGMD,
the interaction between two nodes decreases exponentially
with their separation.  Since the interactions are not contained
within an element (in fact there is no absolute cutoff to
their range in principle), the equations of motion are not constructed
element by element, and the role of the elements is simply
to guide in the construction of an interpolation basis.
So we can consider an fcc lattice of nodes as four interlaced
simple cubic lattices.  The interpolation function for the
corner node at ${\vec{\xi}}=$ (1,1,1) is given by
\begin{equation}
N_1( \xi _a )  =  {\tst \frac{1}{16} (1+\xi _1) \, (1+\xi _2)
\, (1+\xi _3)  \, ( \xi _1 +\xi _2 +\xi _3 -1 ) }
\label{fccCornerFour} 
\end{equation}
and the functions for the other corners follow from symmetry.
This completely determines a basis set which satisfies the
criteria of locality and continuity. It does not satisfy
the partition of unity requirement in the strictest sense, 
since the uniform displacement mode is over-represented: 
it is represented once for each sublattice.  A constraint
must be introduced that the uniform displacement on each
sublattice is equal to the mean displacement:
\begin{equation}
\bar{{\mathbf{u}}} = \sum _j {\mathbf{u}}^{(p)}_j / N_{node}
   ~~~~ {\mathrm{for~all~sublattices}}~p.
\end{equation}
Note that all the nodes are equivalent.  This is possible
when the nodes are in a Bravais lattice such as fcc, but
it is not true of the fcc basis in (\ref{fccCorner}) and (\ref{fccFace}).

Interpolation functions for other crystal lattices are also
available, either because they exist already in the finite element
literature or because they are easily generated.  We consider a
few cases here.  

Interpolation functions for the bcc lattice may be
constructed in a similar fashion.  They are of
the order (2,2,2).  In particular, the shape
function for the center node (0,0,0) is 
\begin{equation}
N_9( \xi _a )  =  \left(1-\xi _1^2\right)
                  \left(1-\xi _2^2\right)
                  \left(1-\xi _3^2\right)
\label{bccShape9} 
\end{equation}
which is unity at the associated node and
vanishes on each face of the cell.  Then 
the shape functions associated with the
corners of the cell are of the form
\begin{equation}
N_1( \xi _a )  =  {\tst \frac{1}{8} \left[ (1+\xi _1) \, (1+\xi _2)
\, (1+\xi _3) - N_9( \xi _a ) \right] }
\label{bccShape} 
\end{equation}
where this particular function is associated with the node 
at (1,1,1).  The shape functions associated with the other
corners are generated by the appropriate rotations of
this function.  Again, these shape functions satisfy
the criteria of locality, partition of unity and
continuity.

Finally, we consider a two-dimensional case that is relevant
for many of the crystal lattices: the square lattice.  In 
particular, suppose that the CG region will be treated as
a two-dimensional projection of the 3D lattice along
the [001] direction.  The
square lattice has been used extensively in the literature
and we include the minimal interpolation functions here for
reference:
\begin{equation}
N_1( \xi _a )  =  {\tst \frac{1}{4} (1+\xi _1) \, (1+\xi _2) }
\label{squareCorner} 
\end{equation}
These interpolation functions are not only useful for two-dimensional 
projections of the lattices treated above (simple cubic, 
fcc and bcc with [001] projection), but also other
more complicated lattices such as the diamond cubic
lattice with [001] projection.

\subsection{Shape Functions in Reciprocal Space}

In order to make use of the reciprocal space
representation of the dynamical matrix, it is necessary
to have the Fourier transform of the shape functions.
The Fourier transform can be computed numerically,
of course, using fast Fourier transform (FFT) 
techniques.  In some cases it is also possible to
compute the Fourier transform analytically.  In
this section, we derive the atomic-index
Fourier transform of the linear interpolation
function in one dimension.  The result is immediately
applicable to the 4-node square and the 8-node
brick in two and three dimensions, respectively.
The result is given in Eq.\ (\ref{linearFT}) below,
and readers who are not interested in the derivation
are free to skip to Section \ref{subsec-zeromodes}.

Consider the symmetric linear
interpolation function on a regular mesh in one dimension: 
\begin{equation}
N_j(x) = \left\{ 
\begin{array}{ll}
1 - \left| \frac{x-x_j}{x_{j+1}-x_j} \right| ~~~ & {\mathrm{for}}~|x-x_j|<x_{j+1}-x_j \\
0 ~~~ & {\mathrm{otherwise}}.
\end{array}
\right.
\label{triangleN}
\end{equation}
Let $a$ be the lattice constant, and $N_{per} = (x_{j+1}-x_j) / a$.
We first note the useful identities:
\begin{eqnarray}
\sum _{\mu = 0}^{N_{per}} e^{i k a \mu} & = & e^{i k a N_{per}/2} \,
   \frac{ \sin \left[ k a (N_{per} + 1)/2 \right]}{ \sin \left( k a /2 \right)} 
\label{id1} \\
\sum _{\mu = -N_{per}}^{N_{per}} e^{i k a \mu} & = &
   \frac{ \sin \left[ k a (2N_{per} + 1)/2 \right]}{ \sin \left( k a /2 \right)} 
\label{id2}
\end{eqnarray}
that follow from the well known formula to sum geometric series
[$1+z+z^2+\ldots + z^N = (1-z^{N+1})/(1-z)$] 
together with de Moivre's formula.

We first transform the atomic index $\mu$ of the shape function
to the Fourier conjugate variable $k$:
\begin{equation}
N_j(k) = \sum _{\mu} N_{j\mu} \, e^{i k a \mu}  .
\end{equation}
The shape function $N_{j\mu}$ is expressed as the sum of two
terms in Eq.\ (\ref{triangleN}): the transformation of 
the first term, just equal to
unity, is given by Eq.\ (\ref{id2}), but the transformation
of the second is more involved.  It is calculated as follows:
\begin{eqnarray}
\sum _{\mu = -N_{per}}^{N_{per}} |\mu | \, e^{i k a \mu} 
   & = & (2/a) \partial _k \Im \left\{ \sum _{\mu = 0}^{N_{per}} 
               e^{i k a \mu} \right\} \\
   & = & (2/a) \partial _k \Im \left\{ e^{i k a N_{per}/2} \,
      \frac{ \sin \left[ k a (N_{per} + 1)/2 \right]}{ \sin \left( k a /2 \right)}
       \right\} \\
   & = & (2/a) \partial _k \left[ \sin ^2 ( k a N_{per}/2) \, \cot ( k a / 2 ) +
      \frac{1}{2} \sin ( k a N_{per} ) \right] \\
   & = & N_{per} \, \frac{ \sin \left[ k a (2N_{per} + 1)/2 \right]}{ \sin 
          \left( k a /2 \right)} -
          \frac{ \sin ^2 ( k a N_{per} / 2 )}{ \sin ^2 ( k a / 2 ) } .
\end{eqnarray}
Combining the two contributions we find
\begin{equation}
N_j(k) = \frac{1}{N_{per}} \, \frac{ \sin ^2 ( k a N_{per} / 2)}{\sin ^2 ( k a / 2 )} e^{i k x_j}
\label{linearFT}
\end{equation}
where $N_{per}$ is the number of lattice sites per CG cell.
In higher dimensions, $N_{per}$ would be replaced by
$N_{per}^x$, $N_{per}^y$ and $N_{per}^z$.
This result applies to the regular CG lattice.  The corresponding
formula for a general one-dimensional CG lattice is much more
complicated, and using a numerical FFT to calculate it is
generally recommended.

\subsection{The Center-of-Mass Mode}
\label{subsec-zeromodes}

The stiffness matrix definition involves two matrix inverses.  
This is somewhat ill-defined because $D _{\mu \nu}$
is singular, due to the zero modes.  The zero modes are the
zero energy phonons at the $\Gamma$ point in reciprocal space
that are associated with translation invariance of the center of mass.
There are $d$ zero energy phonons in any $d$-dimensional system
corresponding to uniform translation in each of the $d$ directions. 
These zero modes make the matrix singular.  
Since we have imposed the criterion that the center-of-mass mode
should be represented within the set of interpolation functions,
the singularity is superficial.  There are two
inverses in Eq.\ (\ref{Kdef}), the matrix $K_{ij}$ is finite
after a suitable regularization.
Indeed, the alternate
derivation of the stiffness matrix given in Appendix \ref{app-CGderiv}
is free from any zero mode problems, so it must be possible
to devise a suitable regularization scheme.
An obvious example is
\begin{equation}
K_{jk} = \lim _{\epsilon \ra 0} \left(
\sum_{\mu \nu} f_{j\mu } \, 
\left( D _{\mu \nu} + \epsilon I _{\mu \nu}\right) ^{-1}
\, f_{k\nu } \right)^{-1}
\label{regKdef}
\end{equation}
where $I _{\mu \nu}$ is the identity matrix.

The regularization (\ref{regKdef}) is conceptually simple, but in
practice a small but finite $\epsilon$ must be used, and error is
introduced into the contribution of the long wavelength modes.
The error can be controlled through the choice of an $\epsilon$
which is small enough that frequencies of interest are not
affected appreciably, but large enough that the matrix is
numerically well conditioned.  This regularization is cheap
and adequate for many purposes.

We have developed an alternative resolution of the zero mode problem, 
which gives an exact formula for the stiffness matrix with a well defined
double inverse (\ref{Kdef}).  Let 
$(v_a)_{\mu}$ be the $\mu$th component of the
$a$th zero mode of $D_{\mu \nu}$; i.e.\ $\sum _{\nu} D_{\mu \nu}
(v_a)_{\nu} = 0$ for $a=1,2,3$.
Define the zero mode matrices as
\begin{eqnarray}
\epsilon _{\mu \nu} & = & \sum _a (v_a)_{\mu} (v_a)_{\nu} \\
\epsilon ^{\prime} _{ij} & = & \sum _{a,\mu,\nu}  
\frac{N_{i\mu}(v_a)_{\mu} \, N_{j\nu}(v_a)_{\nu} }{| N v_a |^2}
\label{zeroMatrices}
\end{eqnarray}
Using these matrices we construct the projected shape matrix
\begin{eqnarray}
\tilde{N}_{j\mu} & = & \sum _k {\mathcal P}_{jk}\, N_{k\mu} \\
{\mathcal P}_{jk} & = & \delta _{jk} - \epsilon ^{\prime} _{jk}
\end{eqnarray}
Then the stiffness matrix (\ref{Kdef}) is given by the matrix equation
\begin{equation}
K  =  (\tilde{N}\,\tilde{N}^T) \, 
   \left[ \tilde{N} \left( D + c \, \epsilon \right)^{-1}  \tilde{N}^T +
c^{\prime} \epsilon ^{\prime} \right] ^{-1} \, (\tilde{N}\,\tilde{N}^T)
\end{equation}
The nonzero numbers $c$ and $c^{\prime}$ are arbitrary, 
but should be comparable
to the eigenvalues of $D_{\mu \nu}$ to make the matrices well-conditioned.  
This formula works by shifting the zero eigenvalues in the atomic
space and those in the nodal space by $c$ and $c^{\prime}$, respectively,
without affecting the other eigenvalues.  The projection matrices 
undo this shift.  They are needed within the brackets to stifle the
crossterms between the zero modes and the nonzero modes for
incommensurate meshes.  For commensurate meshes, this formula
simplifies to
\begin{eqnarray}
K_{{\mathrm commensurate}} & = & \left( N \, N^T \right) 
   \left[ X - X \cdot \epsilon ^{\prime} \right] 
\left( N \, N^T \right) \\
X & = & 
   \left[ N \left( D + c \, \epsilon \right)^{-1}  N^T \right] ^{-1} 
\end{eqnarray}
where $X \cdot \epsilon ^{\prime}$ is a symmetric matrix since
$\sum _{\mu} N_{i\mu}(v_a)_{\mu}$ are eigenvectors of $X_{ij}$.


The zero modes are not integrated out, so a 
short ranged $D _{\mu \nu}$ 
results in a short ranged $K_{ij}$.  On the other hand, a nearest neighbor
$D _{\mu \nu}$ does not generally produce a nearest neighbor $K_{ij}$,
except where the mesh is atomic sized.
The stiffness matrix elements typically decrease exponentially with
separation, so the effective interaction is short ranged but not
nearest neighbor.  This is an important point, since it is this
quality that improves the CGMD phonon spectrum.

\section{(Quasi-)Harmonic Crystals}
\label{sect-harmxtal}

Various properties of harmonic crystals have been computed
within CGMD as a validation of the methodology.

\subsection{Phonon Spectra}
\label{subsec-argon}

The CGMD phonon spectrum offers a good first test of the model.
Consider a regular, but not necessarily commensurate CG mesh.
The equations of motion for the Hamiltonian (\ref{cgHamHarm}) are 
\begin{equation}
M_{ij} \, \ddot{u}^a_j = - K_{il}^{ab} \, u_l^b ,
\end{equation}
where $\ddot{u}^a_j$ is the nodal acceleration of the $j^{th}$
node in the $a^{th}$ direction.  Substitution of a plane-wave
normal mode $u_j^a(t) = u_0^a 
e^{i {\mathbf{k}} \cdot {\mathbf{x}}_j - i \omega t}$
produces the secular equation
\begin{equation}
M({\mathbf{k}}) \, \omega ^2 \, \delta _{ab} = K^{ab}({\mathbf{k}}) 
\end{equation}
where $M({\mathbf{k}})$ and $K^{ab}({\mathbf{k}})$ are 
the Fourier transform of the mass and stiffness matrices,
respectively.  
The form of the mass matrix for a monatomic
lattice allows further simplification:
\begin{equation}
m^2 \omega ^2 \, \delta _{ab} = 
  \left[ M \, (N D^{-1} N^T)^{-1}_{ab} \right] ({\mathbf{k}}) 
\label{monosec}
\end{equation}
where we have used Eqs.\ ({\ref{Mdefmono}) and (\ref{Kdef2}).
For incommensurate meshes,
we have calculated the right-hand side of Eq.\ (\ref{monosec}) in real space,
then found its Fourier transform and solved the secular
equation for the phonon frequencies.  One such spectrum
is plotted in Fig.\ 1 of Ref.\ \onlinecite{CGMD}.
In principle this could be done in any case, but the
computational cost limits the size of systems that can
be treated in this manner.  We have done calculations with
billions of atoms per CG cell, but in order to do this it
is necessary to eliminate the real-space representation of
the dynamical matrix.

For commensurate meshes with uniform mesh size, we may go to
reciprocal space and the formulas simplify considerably.  
It is even possible to derive analytic formulas for the
spectra in some cases.  Suppose the CG mesh contains $N_{node}^a$
nodes in the $a^{th}$ dimension for a total length of $L_a$.
The CG shape functions in reciprocal space may be expressed
in terms of a Bravais lattice character $\chi$ for the CG mesh 
and a CG element structure factor $S$:
\begin{eqnarray}
N({\mathbf k},{\mathbf k}\p) & = &  
   \sum _{j,\mu} e^{i {\mathbf k} \cdot {\mathbf x}_j -
   i {\mathbf k}\p \cdot {\mathbf x}_\mu} \, N_{j\mu} \\
   & = &
   \chi ({\mathbf k}-{\mathbf k}\p) \,
   S({\mathbf k}\p) \\
\left| \chi (\delta {\mathbf k}) \right| & = & \prod _{a=1}^3 \left(
   \frac{ \sin (\delta k_a \, L_a /2 )}{\sin (\delta k_a \, L_a 
  /(2 N_{{\mathrm{node}}}^a))}
    \right) \\
S({\mathbf k}\p) & = & 
   \sum _{\mu \in \Omega _j} e^{ - i {\mathbf k}\p \cdot 
    ({\mathbf x}_\mu - {\mathbf x}_j)} \, N_{j\mu}
\end{eqnarray}
where $\Omega _j$ is any one element from the CG mesh.  
Note that in the atomic limit $S$ is just a delta function in the
first Brillouin zone.
The CG phonon spectrum for a monatomic lattice is given by
\begin{eqnarray}
\omega ^2 ({\mathbf k}) & = & \left[ \frac{1}{m} \sum _{k\p} 
   \left| \chi ( {\mathbf k}-{\mathbf k}\p ) \right|^2 \,
   \left| S ( {\mathbf k}\p ) \right|^2 \right]  \times 
\label{CGspectrum} \\
   & & ~~
   \left[ \sum _{k\p} \rule{0mm}{4mm}^{\prime}
   \left| \chi ( {\mathbf k}-{\mathbf k}\p ) \right|^2 \,
   \left| S ( {\mathbf k}\p ) \right|^2 
   \left[ D({\mathbf k}\p) \right] ^{-1} \right] ^{-1}
\nonumber
\end{eqnarray}
where $D({\mathbf k}\p)$ is a $3n\times3n$ matrix where $n$ is the
number of atoms in the unit cell, and the two
inverses are matrix inverses.  The frequencies are the eigenvalues
of the resulting matrix.  As in Eq.\ (\ref{monosec}),
the first term represents the mass
matrix in reciprocal space divided by $m^2$; 
the second term is the middle factor
of the stiffness matrix.
In the atomic limit, the formula reduces to the
usual expression, $D(k)/m$.

\subsection{Analytic Formula for CG Spectrum}
\label{subsec-analyt}

The CGMD spectrum (\ref{CGspectrum}) may be computed in closed form 
for a monatomic solid with a commensurate CG mesh in one dimension. 
We presented the analytic expression 
for the spectrum with nearest neighbor interactions and 
linear interpolation in 
Ref.\ \onlinecite{CGMD}, Eq.\ (12):
\begin{equation}
\omega (k) =  2 \sqrt{\frac{K}{m}} \left(
\frac{
\sum _{p} \sin ^{-4} (\txthalf ka + 
\frac{\pi p}{N_{per}})}
{\sum _{p} \sin ^{-6} (\txthalf ka + 
\frac{\pi p}{N_{per}})}
\right) ^{1/2}
\label{exactFreq}
\end{equation}
where the sums over $p$ run from 0 to $N_{per}-1$,
$N_{per} = \natom/\nnode$ is the number
of atoms per cell and $K$ is the nearest-neighbor
spring constant. 
This formula shows the contribution of many modes of the underlying
crystal to each CGMD mode, resulting from the choice of interpolation
functions which have many normal mode components.  Near the center
of the CG Brillouin zone, a single mode ($p=0$) dominates the sums
(\ref{exactFreq}).  This dominance reflects the fact that long
wavelength modes are well represented on the CG mesh. 
Near the boundary of the CG zone 
[$k \approx N_{{\mathrm node}}\pi /(Na)$], 
many modes contribute.  The many modes are needed
because periodicity forces the slope of the spectrum to zero,  
and the modes act in concert
to keep the CGMD spectrum close to the true spectrum
which is not smooth at the boundary.
For comparison, the formula for the lumped mass FEM
spectrum is 
\begin{equation}
\omega ^{{\mathrm{lump}}} (k) =  2 \sqrt{\frac{K}{m}} 
\frac{1}{N_{per}} \sin \left| \txthalf k N_{per} a  \right| ,
\label{exactlumpFreq}
\end{equation}
the formula for the FEM spectrum with the distributed (consistent) 
mass matrix is 
\begin{equation}
\omega ^{{\mathrm{dist}}} (k) =  2 \sqrt{\frac{K}{m}} 
\frac{1}{N_{per}} \frac{\sin \left| \txthalf k N_{per} a  \right|}{\sqrt{1-\frac{2}{3} \sin ^2 \left( \txthalf k N_{per} a  \right)}} 
\label{exactdistFreq}
\end{equation}
and the formula for the exact MD spectrum is 
\begin{equation}
\omega ^{{\mathrm{MD}}} (k) =  2 \sqrt{\frac{K}{m}} 
\sin \left| \txthalf k a  \right|  .
\label{exactMDFreq}
\end{equation}
The coarse-grained mass and stiffness matrices
conspire to produce a Pad\'e approximant of the true spectrum,
and thereby achieve the $\OO(k^4)$ improved relative error,
compared to the $\OO(k^2)$ relative error of the two
FEM spectra.

The remainder of this subsection is devoted to the derivation
of the analytic formula for the CGMD spectrum (\ref{exactFreq}).
We make use of the formula for the Fourier transform
of the symmetric linear interpolation function on a 
regular mesh in one dimension, given above in Eq.\ (\ref{linearFT}):
\begin{equation}
N_j(k') = \frac{1}{N_{per}} \, \frac{ \sin ^2 ( k' a N_{per} / 2)}{\sin ^2 ( k' a / 2 )} e^{i k' x_j}
\nonumber
\end{equation}
where in this formula only the atomic index has been transformed.
The Fourier transform of the index $j$ is straightforward, 
and the spectrum could then be derived using Eq.\ (\ref{CGspectrum}).
We take a different approach.  The spectrum is given by
\begin{eqnarray}
\omega (k) & = & \sqrt{\frac{K(k)}{M(k)}} \\
 & = & \sqrt{\frac{1}{m}} \, 
       \sqrt{\frac{\left(N N^T\right)_k}{\left(N D^{-1} N^T\right)_k}}
\label{freqEqn}
\end{eqnarray}
where we have made use of the formulas for the stiffness and mass
matrices, Eqs.\ (\ref{Kdef2}) and (\ref{Mdefmat}), respectively.
The subscript $k$ denotes the Fourier transform of wavenumber $k$,
and we note that the Fourier transforms of both the nodal indices
and the atomic indices take on values of the form $2\pi n/L$,
but for the atomic indices 
$-\frac{1}{2}L/a < n \le \frac{1}{2}L/a$
whereas the Fourier transform of the nodal index lives in a
reduced Brillouin zone, $-\frac{1}{2}\nnode < n \le \frac{1}{2}\nnode$,
where $\nnode$ is the number of nodes (in one dimension).

Evidently, we need to calculate quantities of the form
$\left( N X N^T \right) _k$, where the matrix $X$ is
either the identity matrix or the inverse of the dynamical matrix.
Such quantities are calculated in the following way:
\begin{eqnarray}
\left( N X N^T \right) _k & = & N_{{\mathrm{nodes}}}^{-1} 
  \sum _{i,j} e^{-i k (x_i-x_j)} \left( N X N^T \right) _{ij} \label{step1} \\
  & = & N_{per}^{-1} \sum _{\Delta j=0}^{\nnode-1} \sum _{k'} e^{-i (k-k') \Delta j N_{per} a } \,
    \times \nonumber \\
  & & ~~~
    X(k') \left[ 
    \frac{ \sin ^2 ( k' a N_{per} / 2)}{\sin ^2 ( k' a / 2 )} 
    \right] ^2 \label{step2} \\
  & = & N_{per}^{-1} \, \sin ^4 ( \frac{1}{2} k' a N_{per}) \, \times \label{step3} \\
   & & ~~~~~  \sum _{p=0}^{N_{per}-1} 
     \frac{X\left(k+ \frac{2 \pi p}{N_{per}a}\right)}{\sin ^{4} 
     \left( \frac{1}{2} k a + \frac{\pi p}{N_{per}}\right)}  \nonumber
\end{eqnarray}
where Eq.\ (\ref{step2}) follows from Eq.\ (\ref{linearFT}), and
we have used $x_i-x_j=(\Delta j)N_{per}a$ where $\Delta j=i-j$.
To get from (\ref{step2}) to (\ref{step3}), 
we have used the fact that the sum over $\Delta j$
gives a delta function in the reduced Brillouin zone; i.e., a sum of delta functions
periodically repeated through the full Brillouin zone.

The CGMD spectrum is then calculated using Eq.\ (\ref{freqEqn}) together
with Eq.\ (\ref{step3}) with $X$ equal to the identity in the numerator
and $D^{-1}$ in the denominator.  The result is
\begin{equation}
\omega (k) =  \sqrt{\frac{1}{m}} \left(
\frac{
\sum _{p} \sin ^{-4} (\txthalf ka + 
\frac{\pi p}{N_{per}})}
{\sum _{p} \sin ^{-4} (\txthalf ka + 
\frac{\pi p}{N_{per}}) D^{-1}(k +
\frac{2\pi p}{N_{per}a})}
\right) ^{1/2}
\label{exactFreqGen}
\end{equation}
where $D(k)$ is the dynamical matrix in $k$-space
and the sums over $p$ run from 0 to $N_{per}$.
For a nearest-neighbor harmonic model
$D(k) = 4 K \sin ^2 (\frac{1}{2} k a )$.
Substitution of this into Eq.\ (\ref{exactFreqGen})
results in the analytic formula for the CGMD spectrum
that appears above (\ref{exactFreq}).
This calculation may be generalized to 3D, where
$D(k)$ is a $3\times3$ matrix and the shape functions
are products of linear interpolations functions
in each of the three dimensions. 

The first test is the phonon spectrum for atoms with harmonic 
interactions coarse grained to a regular, but not necessarily 
commensurate mesh.  The normal
modes are plane waves both on the underlying ring of atoms and
on the CG mesh.  The wave vector ${\mathbf{k}}$ is a good quantum
number for both.  The nonzero terms of the dynamical matrix are 
of the form: $D_{\mu \mu} = 2 K, D_{\mu ,\mu \pm 1} = -K$.
Figure 1 of Ref.\ \onlinecite{CGMD} shows the resulting phonon spectra in four
cases:  exact, CGMD, distributed mass FEM and lumped mass FEM.\cite{lumpDef}
The latter two use the long wavelength elastic constants.
The spectra are for a periodic chain of 1024 atoms with lattice
constant $a$ coarse grained to 30 nodes.

Figure 1 of Ref.\ \onlinecite{CGMD} shows that CGMD gives a 
better approximation to the true 
phonon spectrum than the two kinds of FEM do.  
All three do a good job at the longest wavelengths, as expected, but
CGMD offers a higher order of accuracy.  
The relative error for CGMD is $\OO (k^4)$
while that of the two versions of FEM is only $\OO (k^2)$.
At shorter wavelengths, there are significant deviations from the 
exact spectrum.
The worst relative error of CGMD is about
6\%, better by more than a factor of three than that for 
FEM.  This improvement is made possible by the longer-ranged
interactions of CGMD as compared to FEM.  
The continuity condition satisfied by linear interpolation is
enough to ensure that the hydrodynamic modes ($k\sim 0$) are
well modeled, but the lack of continuity of the derivatives shows
up as error in the spectrum of the modes away from the zone center.
This error vanishes for the smooth, nonlocal basis consisting
of the longest wavelength normal modes.
It turns out that the CGMD error at the CG zone boundary is relatively
small (less than 1\%) for technical reasons.  Also note that even
though the number of atoms varies from cell to cell in the
incommensurate mesh, the CGMD spectrum is free of anomalies.
Other computations have shown that CGMD with linear interpolation is 
well behaved on irregular meshes, as well.  

\begin{figure}[tbp]
\includegraphics[width=0.45\textwidth]{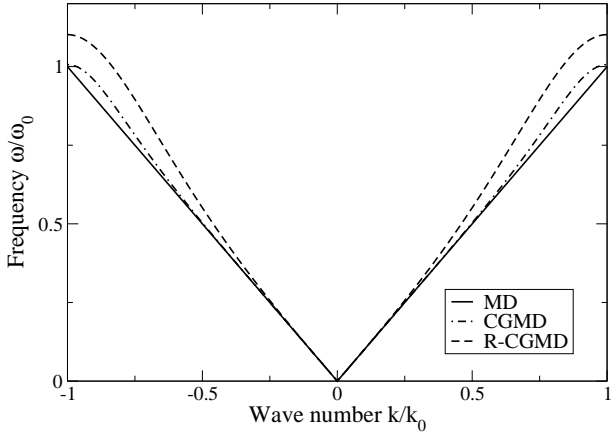}
\caption{A comparison of the acoustic wave spectra for
CGMD, R-CGMD and MD with frequency vs.\ wavenumber plotted.
The units are $k_0 = \pi / (N_{per} a)$ and $\omega _0 = k_0 a \sqrt{K/m}$.
The MD spectrum, corresponding to a simple 1D ball and spring
model, is the ideal case.  The CGMD and R-CGMD spectra are
computed on a regular mesh with $N_{per}=32$ atoms per
cell.  Both full CGMD and its rigid
approximation, R-CGMD, are in good agreement with the MD
spectrum for $k\sim0$; i.e., at long wavelengths.  At short
wavelengths near the zone boundary, the CGMD spectrum is
more accurate than the R-CGMD spectrum, a property that 
we attribute to the relaxation effects accounted for by
CGMD but eliminated in the rigid approximation.
\label{fig-rcgmd}}
\end{figure}

We are now in a position to investigate the effect of
the CGMD relaxation terms eliminated in R-CGMD.  They make
the CGMD stiffness matrix non-local and therefore add to the
cost of CGMD.  What is the benefit of this additional
computational complexity?  Using the same procedure outlined
above, we have computed the R-CGMD spectrum:
\begin{eqnarray}
\omega (k) 
 & = & \sqrt{\frac{1}{m}} \, 
       \sqrt{\frac{\left(N D N^T\right)_k}{\left(N N^T\right)_k}} \\
 & = &  \sqrt{\frac{1}{m}} \left(
\frac{
\sum _{p} \sin ^{-4} (\txthalf ka + 
\frac{\pi p}{N_{per}}) \,
D(k + \frac{2\pi p}{N_{per}a})
}
{\sum _{p} \sin ^{-4} (\txthalf ka + 
\frac{\pi p}{N_{per}}) 
} \right) ^{1/2}
\label{exactFreqGenRCGMD}
\end{eqnarray}
where the two lines are to be compared with the CGMD
results (\ref{freqEqn}) and (\ref{exactFreqGen}), respectively.
The CGMD and R-CGMD spectra are plotted together with the MD
spectrum in Fig.\ \ref{fig-rcgmd}.  It is clear that
both approaches work well for long wavelengths ($k\sim 0$).
In fact, it is reasonable that the relaxation should be
unimportant in this regime since the displacement field
is varying slowly on the scale of the mesh, so the 
lowest energy configurations of the MD best fits to
the interpolated displacement field should be close
to having the atoms at their linearly interpolated 
positions.  At short wavelengths (near the CG zone
boundary), the story is different, however.  The
displacement field is varying at the scale of the 
mesh, and the atoms can reduce the energy through
relaxation.  This is evident in the improved value
of the zone boundary frequency for CGMD (0.67\% error)
vs.\ R-CGMD (10.2\% error). As shown in Fig.\ \ref{fig-rcgmdErr},
these values are typical, and close to the asymptotic value 
for coarse meshes.  This difference in the
performance of CGMD and R-CGMD is evident in many
properties sensitive to the coarse-grained lattice
dynamics at the zone boundary, such as the scattering
properties we consider below.  It is interesting to
note that while the magnitude of the error is
appreciably different, it is
small in both cases.  This suggests that there may
be ways to formulate an approximation to CGMD that
is intermediate between CGMD and R-CGMD, both in
terms of the detail of the physics that is described
and the computational cost.  This is a topic we will
address in the future.

\begin{figure}[tbp]
\includegraphics[width=0.45\textwidth]{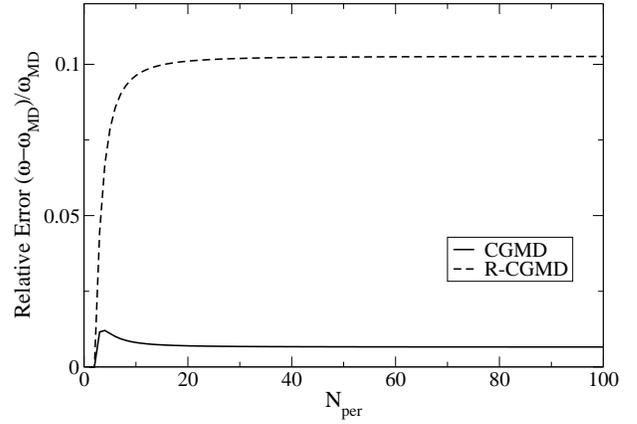}
\caption{A comparison of the error in the acoustic wave frequency
at the zone boundary [$k=\pi/(N_{per}a)$] for
CGMD and R-CGMD as a function of the level of coarse graining,
as expressed by $N_{per}$.  At $N_{per}=1$, there is no error
in either frequency.  At all other values of $N_{per}$, the 
error in the CGMD frequency is less than that of R-CGMD,
asymptotically going to 0.66\% and 10.3\%, respectively.
\label{fig-rcgmdErr}}
\end{figure}

\subsection{Numerical Calculation of Argon CG Spectrum}

It was shown in Ref.\ \onlinecite{CGMD} that the CGMD phonon
spectrum is closer to the true spectrum than that of FEM 
for a one dimensional chain of atoms with nearest-neighbor interactions.
We now compare phonon spectra for a three dimensional real material:
solid argon.  Some of these results were reviewed
briefly in Ref.\ \onlinecite{PSS}. We also treat tantalum below.  
We find that CGMD again offers an improved spectrum.

Solid argon crystallizes in the fcc structure where it is 
well-described by the Lennard-Jones potential
\begin{equation}
V = 4\epsilon \left( \frac{1}{(r/\sigma)^{12}} - \frac{1}{(r/\sigma)^{6}} \right) 
\end{equation}
where $\epsilon = 1.63\times 10^{-21}$~J and $\sigma = 3.44$ \AA. \cite{Grindlay}  
The elastic constants for this potential are given by 
$C_{11} = 105.3\epsilon/\sigma^3=4.21$~GPa and $C_{12} = C_{44} = 60.18 \epsilon/\sigma^3=2.41$~GPa.

We use the four-fold symmetrized 8-node brick interpolation functions  
defined on the fcc Bravais lattice for both CGMD and FEM.  
The fcc lattice is generated by the unit cell vectors 
${\mathbf{a}}_1=(0,\frac{a}{2},\frac{a}{2})$,
${\mathbf{a}}_2=(\frac{a}{2},0,\frac{a}{2})$, and
${\mathbf{a}}_3=(\frac{a}{2},\frac{a}{2},0)$.
As discussed above, the conventional rhombohedral
interpolation functions of the 8-node brick consist
of products of one-dimensional linear interpolation functions:
\begin{eqnarray}
N_j({\mathbf{x}}) & = & \prod _{a=1}^3 \tilde{N} 
\left[2 {\mathbf{b}}_a \cdot 
  ({\mathbf{x}}- {\mathbf{x}}_{j}) - 1 \right] 
\label{asym8node}
\end{eqnarray}
as in Eq.\ (\ref{BravaisShape})
These interpolation functions break the cubic symmetry.
Even though the effect of the symmetry breaking is
small, it complicates the analysis of the spectrum,
causing deviations from the true spectrum that have
nothing to do with the intrinsic accuracy of the
methods.

In order to eliminate the small symmetry-breaking
effects completely, we introduce a symmetrized
version of the interpolation functions that is
useful for spectrum calculations.  We construct 
a multiplet of interpolated fields, where each
component of the multiplet is an image of the 
mesh associated with Eq.\ (\ref{asym8node}) under
the action of the point group.  For the fcc
lattice with rhombohedral cells, there are four
inequivalent transformations of the mesh, so
our field becomes a four component vector.  The
frequency values are then averaged over these
four components.  This is equivalent to the
standard group theoretic operation of averaging
over the orbit in order to restore symmetry.
The four inequivalent group operations are
the $C_2$ elements
\begin{equation}
g =
\left(
\begin{array}{ccc}
\pm1 & 0 & 0 \\
0 & \pm1 & 0 \\
0 & 0 & \pm1 
\end{array}
\right)
\label{symmg}
\end{equation}
with $-1$ appearing an even number of times such that $g$ is 
a proper rotation with $\det (g)=1$.  Another way to view
this symmetrization procedure is that a symmetrized
stiffness matrix is used
\begin{equation}
K_{ab}^{sym}({\mathbf{k}}) =
\frac{1}{4} \sum _g g_{aa'} K_{a'b'}(g{\mathbf{k}}) g_{bb'}
\label{Ksym}
\end{equation}
where the sum is over the four diagonal matrices $g$ (\ref{symmg}).
If the meshing did not break the point group symmetry,
there would be no need for symmetrization and, indeed,
the sum in Eq.\ (\ref{Ksym}) would reduce to a single
term.  The symmetrization procedure is simply a
way to restore the symmetry broken by the mesh in order
to facilitate analysis and comparison of the spectra.

We now undertake the actual calculation of the stiffness
and mass matrices.  The FEM matrices are computed as follows.
The mass matrix takes one of two forms.  
The mass matrix computed strictly from the interpolation
functions is known as the distributed or consistent mass matrix.  
It is given by
\begin{eqnarray}
M_{ij}^{{\mathrm{dist}}} & = & 
  \int \! d^3x \, \rho \, N_i({\mathbf{x}}) \, N_j({\mathbf{x}}) \\
  & = & \frac{1}{8} \, m \, N_{per}^3
  \int _{-1}^1 d^3\xi \, N_i(\xi _a) \, N_j(\xi _b) \\
   & = & m \, N_{per} \left( \frac{2}{3} \right)^{3-l}  
                                 \left( \frac{1}{6} \right)^{l}
\label{FEdistM}
\end{eqnarray}
where $l$ ($0\le l \le 3$) is the number of edges 
in the shortest path along a single hexahedral cell
connecting nodes $i$ and $j$.  The number $N_{per}$
is a generalization of the one-dimensional case,
where now it is the number of lattice sites along
one dimension of the cell, and for cells with
different dimensions in the three directions
$N_{per}^3$ should be replaced by
$N_{per}^xN_{per}^yN_{per}^z$.
The Fourier transform of $M_{ij}^{{\mathrm{dist}}}$
may then be calculated as a sum over the 27
neighboring nodes (indexed by $n_1,n_2,n_3$ running
from -1 to 1)
\begin{eqnarray}
M^{{\mathrm{dist}}}({\mathbf{k}})
   & = & \frac{8\, m \, N_{per}^3}{27} \sum _{n_a=-1}^{1} \!
       \left( \frac{1}{4} \right)^{\sum |n_a|}  
       \prod _{a=1}^3 \cos \left( N_{per} n_a {\mathbf{a}}_{a} \cdot {\mathbf{k}} \right)
  \nonumber \\
\label{FEdistMk}
\end{eqnarray}
where ${\mathbf{a}}_{b}$ is the $b^{th}$ real space
basis vector.

For many applications, it
is sufficient (and in some cases even more accurate)
to use a diagonal approximation to the
mass matrix known as the lumped mass matrix.  It is given by
\begin{eqnarray}
M_{ij}^{{\mathrm{lump}}} 
 & = & \delta _{ij} \, m \, N_{per}^3 
\label{FElumpM}
\end{eqnarray}
so each element of the diagonal is just equal to the mass
contained in the Voronoi cell about the corresponding node.
The Fourier transform of $M_{ij}^{{\mathrm{lump}}}$ is
\begin{eqnarray}
M^{{\mathrm{lump}}}({\mathbf{k}}) & = & m \, N_{per}^3 
\label{FElumpMk}
\end{eqnarray}
Note that $M^{{\mathrm{dist}}}({\mathbf{k}}=0)=
           M^{{\mathrm{lump}}}({\mathbf{k}}=0)$,
and they are equal to the Voronoi mass as they
should be.

The FEM stiffness matrix is given by 
\begin{eqnarray}
K_{ij;bd}^{FE} & = & 
  C_{abcd} \, b_{aa'} b_{cc'} \int \! d^3x \, \partial _{a'} 
         N_i({\mathbf{x}}) \,  \partial _{c'} N_j({\mathbf{x}}) \\
  & = & \frac{a}{8} \, N_{per}^3 \, C_{abcd} \, b_{aa'} b_{cc'}\! 
  \int _{-1}^1 d^3\xi \, \partial _{a'} N_i({\mathbf{\xi}}) \, 
                         \partial _{c'} N_j({\mathbf{\xi}}) 
  \nonumber \\
\label{FEstiff}
\end{eqnarray}
where the prefactor $a$ is the lattice constant of the
rhombohedral unit cell and the elastic tensor $C_{abcd}$
has been contracted with the reciprocal space metric as
appropriate for the non-orthogonal coordinates.  
This equation has too many
components to present the complete expression here
[cf.\ Ref.\ \onlinecite{Huebner}].  
Nevertheless,
the calculation is elementary algebra, and the results
were used to calculate the Fourier transform.
The result is a $3\times3$ matrix for each value
of $k$: $K_{bd}^{FEM}({\mathbf{k}})$.

The formulas for the CGMD mass and stiffness matrices
in real space for rhombohedral elements are computed
similarly.  The mass matrix is given by
\begin{eqnarray}
M_{ij}^{{\mathrm{CGMD}}}
   & = & m \, N_{per}^3 \, 
  \left( \frac{1-N_{per}^{-2}}{6} \right)^{l} \times
 \nonumber \\
  & & ~~~ 
  \left( \frac{2+N_{per}^{-2}}{3} \right)^{3-l}  
\label{CGMDregM}
\end{eqnarray}
where as in Eq.\ (\ref{FEdistM}),
$l$ ($0\le l \le 3$) is the number of edges 
in the shortest path along a single rhombohedral cell
connecting nodes $i$ and $j$.  
The correspondence of the leading terms
to the terms in the FEM distributed mass matrix (\ref{FEdistM})
is evident, so that CGMD reproduces the FEM distributed mass
matrix in the large-$N_{per}$ limit.
This expression assumes that the mesh consists of trigonal
cells in which the linear dimensions are equal in all
three dimensions, but it could be generalized immediately
to unequal dimensions. The Fourier transform is given
by
\begin{eqnarray}
M^{{\mathrm{CGMD}}}({\mathbf{k}}) & = & m \, N_{per}^3 \,
   \sum _{n_a=-1}^{1} \!
     \left( \frac{1-N_{per}^{-2}}{6}  \right)^{\sum |n_a|}  
         \times 
         \nonumber \\
       & & 
     \left( \frac{2+N_{per}^{-2}}{3} \right)^{3-\sum |n_a|}
       \prod _{a=1}^3 \cos \left( N_{per} n_a {\mathbf{a}}_{a} \cdot {\mathbf{k}} \right)
 \nonumber \\
\label{CGMDregMk}
\end{eqnarray}
where as in Eq.\ (\ref{FEdistMk}), 
the 27 neighboring nodes are indexed by $n_1,n_2,n_3$ running
from -1 to 1.  Note that the CGMD mass matrix also satisfies
the mass sum rule: $M^{{\mathrm{CGMD}}}({\mathbf{k}}=0) = m N_{per}^{3}$.

The CGMD stiffness matrix is calculated according to
Eq.\ (\ref{Kdef2}) using the reciprocal space representation
of the dynamical matrix.  In particular, we calculate
the spectrum using Eq.\ (\ref{freqEqn}), suitably generalized to
a monatomic lattice in three dimensions:  
\begin{eqnarray}
\omega _{ab}^2 ({\mathbf{k}}) & = & 
       \frac{1}{m^2} \, 
       M({\mathbf{k}})\,
       \left(N D^{-1} N^T\right)^{-1}({\mathbf{k}})
\label{freqEqn3a}
\end{eqnarray}
where the actual frequency on each phonon branch
is given by the square root of one of the three
eigenvalues of the $3\times3$ matrix $\omega _{ab}^2 ({\mathbf{k}})$.
Here we have made use of the expression for the mass
matrix of a monatomic lattice (\ref{Mdefmono}).
The denominator is
part of the stiffness matrix,
\begin{eqnarray}
\tilde{K}_{ab} & = & 
  \left(N D_{ab}^{-1} N^T\right)^{-1}({\mathbf{k}}) \\
  & = & \left(N({\mathbf{k}},{\mathbf{k}}')  
        \, D_{ab}^{-1}({\mathbf{k}}') \,
        N^*({\mathbf{k}},{\mathbf{k}}') \right)^{-1}
\end{eqnarray}
where $D_{ab}^{-1}({\mathbf{k}}')$ is the
matrix inverse of the $3\times3$ matrix $D_{ab}({\mathbf{k}}')$.
The outer inverse is a $3\times3$ matrix inverse, as well.
The Fourier transform of the shape functions is found using
Eq.\ (\ref{linearFT}) to be
\begin{eqnarray}
\left| N({\mathbf{k}},{\mathbf{k}}') \right|^2
   & = & 
 N_{per}^3 \, \prod _{b=1}^3 \sum _{p_b=1}^{N_{per}}
  \frac{ \sin ^4 ( {\mathbf{a}}_b \cdot {\mathbf{k}} N_{per} / 2)}{N_{per}^4 \, \sin ^4 ( {\mathbf{a}}_b \cdot {\mathbf{k}} / 2 )} 
   \times \nonumber \\
 & & ~~~
  \, \delta \left( {\mathbf{k}}-{\mathbf{k}}' + \frac{2\pi p_b}{L_b} \right)
\label{linearFT2}
\end{eqnarray}
where $L_b$ is the length of the CG cell in the $b^{th}$ direction.
Upon substitution back into Eq.\ (\ref{freqEqn3a}), we find that 
\begin{eqnarray}
\omega _{ab}^2 ({\mathbf{k}}) & = & 
 \left\{
 \sum _{p_b=1}^{N_{per}}
 \prod _{b=1}^3 
  \left[ \frac{ \sin ^4 ( {\mathbf{a}}_b \cdot {\mathbf{k}} N_{per} / 2)}{N_{per}^4 \, \sin ^4 ( {\mathbf{a}}_b \cdot {\mathbf{k}}_p / 2 )}   \right]
  D_{ab}^{-1}({\mathbf{k}}_p) \right\} ^{-1}
   \times \nonumber \\
 & & ~~~
  M({\mathbf{k}}) / ( m^2 N_{per}^3 )
\label{freqEqn3}
\end{eqnarray}
where ${\mathbf{k}}_p= {\mathbf{k}} + 2 \pi p_b / L_b$.
The mass matrix $M({\mathbf{k}})$ is given by Eq.\ (\ref{CGMDregMk}).
It is in the numerator, what might seem to be the wrong place,
because of the form of the monatomic secular equation (\ref{monosec}).
It is clear from Eq.\ (\ref{freqEqn3}) that CGMD reproduces
the MD spectrum in the long wavelength limit
\begin{eqnarray}
\omega _{ab}^2 ({\mathbf{k}}) & \sim & D_{ab}({\mathbf{k}}) / m
 ~~~ {\mathrm{for}}~ {\mathbf{k}} \sim 0
\label{freqEqnlimit}
\end{eqnarray}
which follows from expanding $\sin(x)=x+\ldots$ for small
arguments.  In the short wavelength limit the many terms
in the sum over $p_b$ contribute to Eqs.\ (\ref{linearFT2}) 
and (\ref{freqEqn3}), ensuring periodicity in the CG reciprocal space.

The spectrum is then computed in each case from the
resulting secular equation at each value of ${\mathbf{k}}$. 
For the true spectrum, the secular equation is
\begin{equation}
\det \left[ \omega ^2({\mathbf{k}}) \, \delta _{ab} - D_{ab}({\mathbf{k}})/m \right] = 0,
\end{equation}
for the CGMD spectrum the secular equation is
\begin{equation}
\det \left[ \omega ^2({\mathbf{k}}) \, \delta _{ab} - \omega _{ab}^2 ({\mathbf{k}}) \right] = 0
\end{equation}
with $\omega _{ab}^2$ given by Eq.\ (\ref{freqEqn3}),
and for the FEM spectra the secular equation is
\begin{equation}
\det \left[ \omega ^2({\mathbf{k}}) \, \delta _{ab} - K_{ab}({\mathbf{k}})/
M({\mathbf{k}}) \right] = 0
\end{equation}
with $M({\mathbf{k}})$ given by Eqs.\ (\ref{FEdistMk}) and (\ref{FElumpMk})
for distributed and lumped mass, respectively.
The determinant of the $3\times3$ matrix gives a cubic
secular equation with three eigenvalues.  The eigenvalues are
real for all of the cases considered here.  The Lennard-Jones
potential for argon was cut off in real space after the
twelfth nearest neighbor shell for both the MD and CGMD
spectra.  
For many applications it would not be necessary to extend
the range this far; however, in this case we wanted to test
a potential extending well beyond nearest neighbors.
The semi-analytic
formulas that we have presented here, such as Eq.\ (\ref{freqEqn3}),
have been used in order to calculate the spectra of extremely
large systems with minimal computational expense, and all of
the plots presented below were calculated using these equations.
It should be emphasized, however, that the spectra could have
been calculated using the basic real-space matrix formulas
or using numerical Fourier transforms.  This has been done
for the smaller systems as a validation.

The spectra have been computed for three levels of coarse-graining:
the atomic limit (1$\times$1$\times$1 or no coarse-graining), a slight
coarse-graining (2$\times$2$\times$2) and a case approaching the continuum 
limit (32$\times$32$\times$32).  
The numbers $n_x\times n_y\times n_z$ indicate the number of
atoms within an unsymmetrized CG cell in each direction,
i.e., $N_{per}=1,~2,~32$, respectively.
These values correspond to cells containing 1, 8 and 32678 atoms,
respectively.
The results are shown in Figs.\ \ref{fig-cg1}, \ref{fig-cg2} and
\ref{fig-cg32}.  

\begin{figure}[t]
\includegraphics[height=6cm]{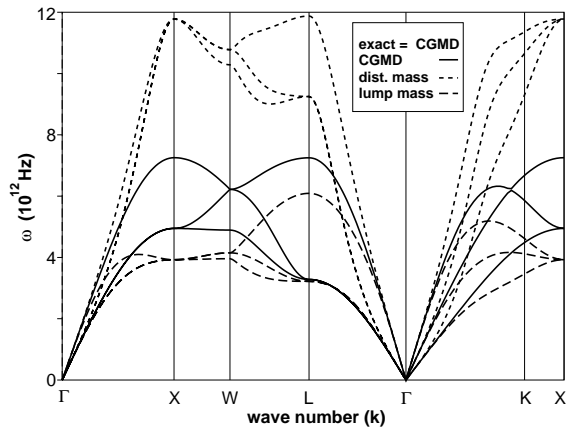}
\caption{The phonon spectra for solid argon in the atomic limit 
are shown as plots of wave frequencies vs.\ wave number
along various high symmetry lines through the Brillouin
zone.\cite{PSS} The high symmetry points are labeled according
to the standard convention for an fcc lattice \cite{AM,groups};
for example, $\Gamma$ is at the center of the zone, ${\mathbf{k}}=0$
and $X=(1,0,0)\pi/a$ where $a$ is the fcc lattice constant.
The 9 curves represent the 3 branches of the spectra for CGMD and two versions
of FEM, one using the conventional distributed mass matrix and
one using a diagonal lumped mass matrix.  In the atomic limit
CGMD reproduces the Lennard-Jones MD spectra exactly, whereas the FEM
spectra show significant error, especially with the distributed
mass matrix.  
\label{fig-cg1}}
\end{figure}

Elastic wave spectra are of interest because they provide a means
of quantifying the small-amplitude dynamics of the system.  They
represent the energetics of every normal mode of vibration
of a system of atoms.  In coarse-grained dynamics, the wave 
spectra provide an excellent way to quantify the fidelity of 
the coarse-grained model.  Since
the normal modes of a crystal are plane waves, they are uniquely
identified by a wave number ${\mathbf{k}}$ and a branch index,
for example indicating a transverse optical mode or a longitudinal
acoustic mode.  The normal modes correspond to a lattice of points in
reciprocal space (${\mathbf{k}}$-space) inside a bounded region
known as the Brillouin zone.  It is not possible to plot
$\omega ({\mathbf{k}})$ for ${\mathbf{k}}$ throughout the three
dimensional Brillouin zone, so typically $\omega ({\mathbf{k}})$
is plotted along lines, in particular high symmetry lines,
through the 3D Brillouin zone.\cite{AM,groups}  
This convention has been used
in Figs.\ \ref{fig-cg1}, \ref{fig-cg2} and \ref{fig-cg32}.  

Consider the spectra in the atomic limit shown in Fig.\ \ref{fig-cg1}.
The CGMD spectrum agrees precisely with the true spectrum.
Of the two FEM spectra, the lumped mass spectrum is closer to
the true spectrum.  This is sensible because the mass is localized
to the nodes in the atomic limit, since each node represents one
atom.  Overall, the lumped mass frequencies are lower than the 
true frequencies, whereas the distributed mass frequencies are higher.
This ordering remains true regardless of the level of coarse-graining.

\begin{figure}[t]
\includegraphics[height=6cm]{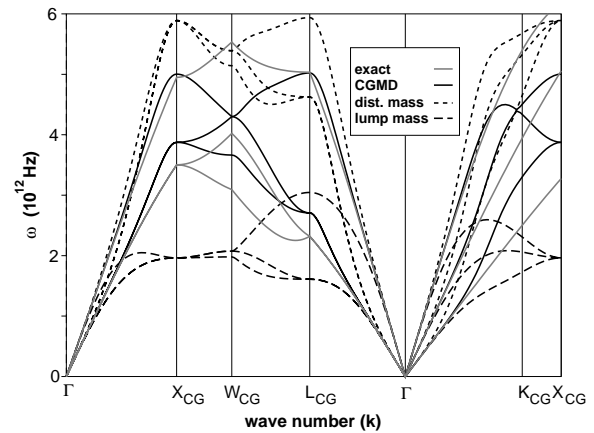}
\caption{The phonon spectra for solid argon on a mesh with 
slight coarse graining
are shown as plots of wave frequencies vs.\ wave number
along various high symmetry lines through the Brillouin
zone (cf.\ the caption to Fig.\ \ref{fig-cg1}). 
The 12 curves represent the 3 branches of the spectra for MD, CGMD and two versions
of FEM, one using the conventional distributed mass matrix and
one using a diagonal lumped mass matrix.  
Each cell of the CG mesh contains 8 atoms.
For this first level of coarse graining
the CGMD spectra is in better agreement with the MD spectra than either
of the FEM spectra.  Of the two FEM spectra, the lumped mass
spectra is somewhat better.
\label{fig-cg2}}
\end{figure}

In the continuum limit shown in Fig.\ \ref{fig-cg32}, the CGMD spectrum
no longer agrees exactly with the true spectrum, but it is still
a better approximation than either FEM spectrum.  It is clear that
in the continuum limit, the distributed mass produces the better
spectrum of the two finite element cases.  This again makes sense,
because the mass is becoming more evenly spread throughout the
CG cell.  Still, the CGMD spectrum is significantly better than
the distributed mass FEM spectrum.

\begin{figure}[t]
\includegraphics[height=6cm]{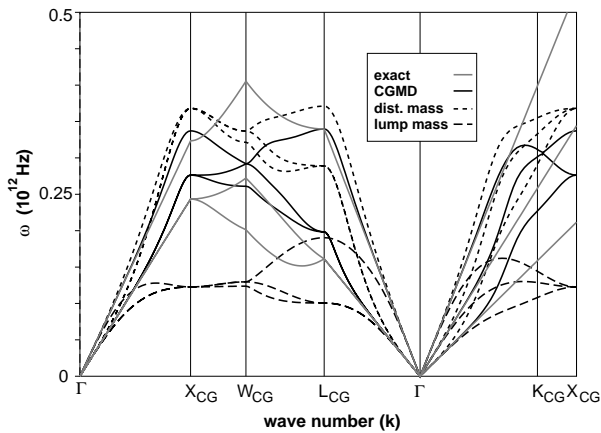}
\caption{The phonon spectra for solid argon on a mesh 
approaching the continuum limit (32768 atoms per cell)
are shown as plots of wave frequencies vs.\ wave number
along various high symmetry lines through the Brillouin
zone\cite{PSS} (cf.\ the caption to Fig.\ \ref{fig-cg1}). 
The 12 curves represent the 3 branches of the spectra for MD, CGMD and two versions
of FEM, one using the conventional distributed mass matrix and
one using a diagonal lumped mass matrix.  
With this significant level of coarse graining
the CGMD spectra is again in better agreement with the MD spectra than either
of the FEM spectra.  Of the two FEM spectra, now the distributed mass
spectra is better.
\label{fig-cg32}}
\end{figure}

The intermediate case is shown in Fig.\ \ref{fig-cg2}. 
Already the distributed mass FEM is in better agreement
with the MD spectra than the lumped mass FEM is.
It is remarkable that cell containing as few as 8
atoms are beginning to exhibit continuum behavior.
This one-two-many qualitative dependence is typical
of many large-$N$ expansions, where the large-$N$ 
limit quickly becomes a good approximation to the
real system behavior, and even for $N$ as low as
two or three it is a good approximation.  The
CGMD spectra are again in better agreement with
the MD spectra than either of the FEM spectra are.

It should be emphasized that the shape of the FEM
spectra is the same in the three plots.  Continuum
elasticity is scale invariant, and the changes in
in FEM spectra are a simple rescaling of frequency
and wave number.  This scaling is clearly evident in Eq.\ (\ref{FEdistMk})
for the FEM distributed mass.  The scale, $N_{per}$,
enters through the prefactor $N_{per}^3$ scaling the
frequency and the factor of $N_{per}$ in the argument 
of the cosine scaling the wave number.
The same scaling of the wave number is present
in the stiffness matrix, but its prefactor goes
like $N_{per}$ rather than $N_{per}^3$.  As a 
result, the frequency ($\sim \sqrt{K/M}$) and
the wave number scale like $1/N_{per}$.  In the
linear portion of the spectra near $k=0$, the
two effects cancel, but the spectra are modified
significantly near the zone boundary.  Thus, when
we discuss the comparison of the MD spectra with the
FEM spectra and find better agreement
with the lumped mass FEM for small cells and 
better agreement with distributed mass FEM for
large cells, it is not that the FEM spectra are
changing.  The MD spectra are changing from
discrete atomic behavior to continuum behavior.
It is the natural scale dependence of true lattice
dynamics.  CGMD has this scale dependence arise
naturally, as well, and so it is able to a large 
extent to track the changes in the MD spectra with cell size.

If we look at the differences between the spectra in more
detail, we can start to understand how the underlying
physics gives rise to these differences.  On all three
plots, the transverse phonons are degenerate (have the
same frequency) along the lines connecting the
zone center and the middle of the zone
faces, $\Gamma-X$ in the [100] direction and 
$\Gamma-L$ in the [111] direction.\cite{groups}  
This degeneracy follows from the $C_4$ and $C_3$ symmetry along those 
lines, respectively.  The lines along the zone boundary
and the $\Gamma-K$ line have reduced symmetry (at most $C_2$), 
and the transverse phonons are not degenerate in those cases.  
There are some isolated points of increased symmetry.  The
$W$ point is an example.  One transverse phonon and the
longitudinal phonon are degenerate at $W = (2,1,0)\frac{\pi}{2a}$, 
as can be seen in the MD curve in Fig.\ \ref{fig-cg1}.  
This happens because $W$ is the mid-point on a line 
between $X$ on two adjacent Brillion zones, and the
longitudinal mode at one $X$ becomes the transverse
mode at the other $X$, and vice-verse.  The two branches
must cross at the mid-point, and hence they are degenerate
at $W$.  This is no longer true for the MD phonons 
in the coarse grained systems,
since W$_{{\mathrm{CG}}}$ no longer has this mid-point
property.  As a result, the frequency of the longitudinal
MD mode increases going from $L$ to $W$ ($W$ is farther
from $\Gamma$), whereas it decreases for CGMD and the
two versions of FEM since the longitudinal frequency 
is dropping to meet the transverse frequency.  This
is the most pronounced discrepancy in the spectra.
It is particularly bad for the lumped mass FEM
as the cell becomes large.  CGMD is in better
agreement with the two transverse MD modes at $W$,
while the distributed mass FEM, which tends to be
high overall, is in better agreement with the
high longitudinal MD frequency at $W$.  In general,
the CGMD spectra are in substantially better agreement
with the MD spectra than either of the FEM spectra are.

It is not obvious, but even in the long wavelength limit (near the
$\Gamma$ point), the CGMD spectrum is better than either of the
two FEM spectra.  The relative error for the CGMD spectrum is of
order $\OO (k^4)$, whereas it is $\OO (k^2)$ for the two FEM
cases.  This improved error was demonstrated above with the
formulas for the 1D frequencies, and it continues to hold
in 3D.  An order $\OO (k^2)$ relative error is the naive 
expectation, since the phonon dispersion relation is linear,
with the leading corrections of order $\OO (k^3)$ due to 
symmetry.  The higher order error for CGMD is due to a subtle
cancellation between the mass and stiffness matrices.  
This cancellation can be seen from the formula for the general 1D CGMD
spectra (\ref{exactFreqGen}) and its 3D generalization
\begin{eqnarray}
\omega _{ab}^2 ({\mathbf{k}}) & = & 
 \frac{1}{m}
 \prod _{b=1}^3 \frac{
   \sum _{p_b=1}^{N_{per}} \sin ^{-4} ( \frac{1}{2} {\mathbf{a}}_b 
    \cdot {\mathbf{k}}_{p_b})
     }{
   \sum _{p_b=1}^{N_{per}} \sin ^{-4} ( \frac{1}{2} {\mathbf{a}}_b 
     \cdot {\mathbf{k}}_{p_b}) \, D_{ab}^{-1}({\mathbf{k}}_{p_b}) 
     }
\label{freqEqn3b}
\end{eqnarray}
where ${\mathbf{k}}_{p_b} = {\mathbf{k}} + \frac{2 \pi p_b}{N_{per}^b a} 
{\mathbf{b}}_b$ and ${\mathbf{b}}_b$ is the reciprocal lattice basis
element.
Note that this formula is equivalent to Eq.\ (\ref{freqEqn3}),
which is the equation we actually used to compute the
CGMD spectra.  The two equations differ because they make
use of different expressions for the mass matrix.

It is interesting again to compare the spectra for CGMD and R-CGMD,
now for the 3D phonons.  In the atomic limit ($N_{per}^b=1$), 
the two agree with each other and with the exact MD spectrum.  
For coarsened lattices ($N_{per}^b>1$),
near the $\Gamma$ point (${\mathbf{k}}=0$), CGMD and R-CGMD are
in good agreement (R-CGMD not plotted here), as was observed in the 
1D case and shown in Fig.\ \ref{fig-rcgmd}.  
We may compare the formulas for the spectra, Eq.\ (\ref{freqEqn3b}) 
and its R-CGMD counterpart,
\begin{eqnarray}
\omega _{ab}^2 ({\mathbf{k}}) & = & 
 \frac{1}{m}
 \prod _{b=1}^3 \frac{
   \sum _{p_b=1}^{N_{per}} \sin ^{-4} ( \frac{1}{2} {\mathbf{a}}_b 
    \cdot {\mathbf{k}}_{p_b})\, D_{ab}({\mathbf{k}}_{p_b}) 
     }{
   \sum _{p_b=1}^{N_{per}} \sin ^{-4} ( \frac{1}{2} {\mathbf{a}}_b 
     \cdot {\mathbf{k}}_{p_b}) 
     }
\label{freqEqn3brcgmd}
\end{eqnarray}
The two equations for the spectra
may be expanded about the $\Gamma$ point, and both exhibit the
improved relative error, $\OO(k^4)$.  Near the zone boundary,
R-CGMD behaves more like the distributed mass FEM case.
It is here that the relaxation physics built into CGMD
has its most pronounced effect, especially at the high
symmetry point $L$ on the acoustic branch, where 
the full CGMD error is very small.

\subsection{The Finite Temperature Tantalum CG Spectrum}

We have calculated the elastic wave spectrum for a variety of materials.  
As a second example we present results for the phonon spectra of tantalum at
room temperature.  Tantalum was chosen to demonstrate CGMD in a more
open crystal structure (bcc) and in a system using many-body interatomic
potentials.  We use the Finnis-Sinclair many-body potential for
tantalum \cite{FS} with the improved Ackland-Thetford core 
repulsion.\cite{Ackland}  The elastic constants for this potential
are $C_{11} = 266.0$~GPa, $C_{12} = 161.2$~GPa, and $C_{44} = 82.4$~GPa.
We calculate the finite temperature dynamical matrix in a conventional
MD simulation consisting of 2000 atoms in a lattice of 10$\times$10$\times$10
bcc unit cells with periodic boundary conditions.  The system is equilibrated
to $T=300\pm0.1$K and $P= 0 \pm 10^{-3}$~GPa through scaling of
the box size and velocities every 100 time steps until the target temperature
and pressure were attained and then an additional 5000 steps without
rescaling to ensure equilibration.  The equilibrium lattice constant at
this temperature was found to be 3.3129~\AA, expanded by 0.2\% from
the $T=0$K value of 3.3058~\AA.

Subsequently, the dynamical matrix was calculated every 1000 time steps
averaged over every atom in the simulation.  With the Finnis-Sinclair
potential it is possible to use an analytic expression for the dynamical
matrix, since it is possible to take two derivatives of the energy analytically
with respect to atomic displacements.  The expression is given in
Ref.~\onlinecite{FS}.
In principle we are computing an ensemble average, which we have
implemented by averaging over the equivalent lattice sites of the system
and over multiple time steps (relying on ergodicity for the equivalence of
ensemble and temporal averages in equilibrium).  In all, we have averaged
over a total of 10 snapshots of the system and imposed the cubic (O$_h$)
point group symmetry by averaging over the point group operations.  
The range of the dynamical matrix includes out to 
the sixth nearest neighbor shell in tantalum at T=300K (the range of the 
pairwise functions entering the potential includes the first and second 
nearest neighbor shells).

\begin{figure}[t]
\includegraphics[height=6cm]{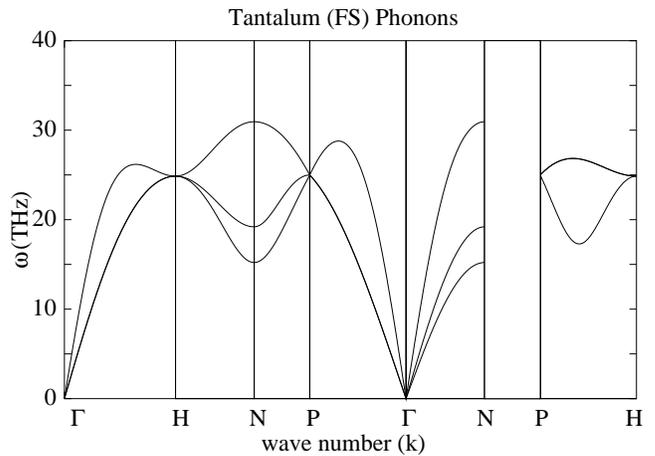}
\caption{The room temperature phonon spectra for 
solid tantalum on a mesh with no coarse graining
are shown as plots of wave frequencies vs.\ wave number
along various high symmetry lines through the Brillouin
zone (for comparison to the solid argon case shown in Fig.\ \ref{fig-cg1}). 
As in the case of argon, the CGMD phonon spectrum agrees
exactly with the MD spectrum when the mesh is 
refined to the atomic limit.  It is common practice to 
leave the gap between the second
N and P points since that part of the 
spectrum is already represented to the left.
\label{fig-cgTa1}}
\end{figure}

The results for tantalum are very similar to those for solid
argon in terms of quality.  The elastic wave spectra are
plotted in Figs.\ \ref{fig-cgTa1} and \ref{fig-cgTa2}.  
The first figure shows the spectrum of elastic waves
on a fully refined mesh.  The CGMD and MD spectra agree
exactly and are overlapping.  The second figure shows
the spectrum on a mesh consisting of 8 atoms per
rhombohedral cell, as described in the argon case.  Here
the spectra do not agree exactly, but the results
are comparable in quality to those from the argon simulations.
In comparing to the corresponding argon plots 
(Figs.\ \ref{fig-cg1} and \ref{fig-cg2}) it should be noted 
that the high symmetry k points are somewhat different
due to the differences in the bcc and fcc crystal structures.
For example, the $\Gamma$-H line is in the [100] direction 
and the $\Gamma$-P line is in the [111] direction for bcc Tantalum.  
Again CGMD performs better than
either of the two FEM models (data not shown) in reproducing the MD spectra.
The many-body potential and the more open crystal structure
do not have a significant impact on the quality of the results.

\begin{figure}[t]
\includegraphics[height=6cm]{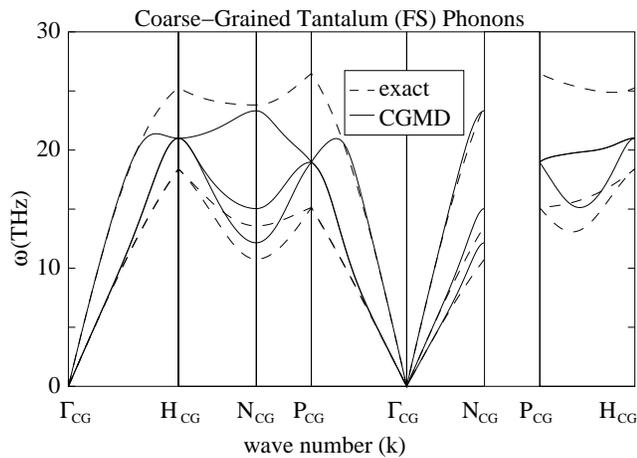}
\caption{The room temperature phonon spectra for 
solid tantalum on a mesh with 8 atoms per mesh cell
are shown as plots of wave frequencies vs.\ wave number
along various high symmetry lines through the Brillouin
zone (cf.\ Fig.\ \ref{fig-cg2}). 
As in the case of argon, the CGMD phonon spectrum agrees
very well with the MD spectrum in the limit of long wavelengths
(near the $\Gamma$ point), and it agrees reasonably well
throughout the coarse-grained Brillouin zone.
\label{fig-cgTa2}}
\end{figure}

\subsection{Dynamics and Scattering}

Most of the applications that we envision for CGMD are dynamical
and have varying mesh size.
For example, in studies of crack propagation it may be advantageous
to introduce a coarse-grained model of the far-field 
regions away from the crack. \cite{CLS}  
For these applications it is important that elastic waves generated 
at the crack tip do not reflect from the coarse-grained region and
perturb the crack propagation.\cite{crackRefl,crackRefl2,AdelmanDoll,Cai,WeinanE}  
CGMD offers the chance to allow the
longest wavelength modes to propagate much farther into the periphery
within incurring a commensurately greater computational expense.  
Other coarse-grained models of the periphery, such as
hybrid FEM/MD schemes, may offer the same advantage.  In energy-conserving
CGMD, the short wavelength modes are reflected from the CG region,
but this process is sufficiently dispersive that shock waves are
smoothed out and the potential wave-reflection pathologies are
averted.
The unphysical wave reflection may also produce a nonzero Kapitza 
resistance at the interface, which can lead to an unphysical temperature
gradient across the interface.  Of course, a stationary system started in
thermal equilibrium remains at a constant, uniform temperature given
a reasonable measure of temperature in the CG region, but a system
driven out of equilibrium may exhibit unphysical gradients on time
scales shorter than the relaxation time.

Given the potential problems associated with wave reflection, we have
developed a methodology to quantify the problem.
The natural measure of the ability of waves to propagate from an atomistic
region into a CG region is the $S$-matrix of scattering theory,
or in one dimension, the transmission and reflection coefficients, 
$\TT$ and $\RR$, respectively.  
The basic approach to scattering problems is to look for solutions
of the equations of motion of the form of an incoming plane wave
and an outgoing spherical wave,
\begin{equation}
{\mathbf{u}}(\mathbf{r},t) \sim 
  \frac{1}{(2\pi)^{3/2}} 
  \left[ 
  e^{i\mathbf{k}\cdot\mathbf{r} - i w \omega t}
  + f_{\mathbf{k}}(\hat{\mathbf{k}}) 
   \frac{e^{i k r - i w \omega t}}{r} \right]
\label{scattAsymp}
\end{equation}
where this asymptotic form of the displacement field holds well
outside the scattering region.  The $S$-matrix and the scattering
cross section may be determined from $f_{\mathbf{k}}(\hat{\mathbf{k}})$.
For CGMD, the scattering region
is the region where the stiffness matrix differs from the 
MD dynamical matrix.  A tremendous amount of theoretical
analysis has been developed for scattering problems.\cite{scatterRef}
In lattice dynamics, scattering is complicated by the anisotropy
of the lattice.  The asymptotic form given above (\ref{scattAsymp})
is only applicable to isotropic scattering, but the formalism
is readily generalized.  To the best of our knowledge, 3D
scattering cross sections have not been computed for any of the
proposed solutions to the wave reflection problem.

We restrict our discussion to the one-dimensional case,
for which the analysis is more straight-forward.
We have calculated these scattering properties for CGMD and
FEM, for comparison.  The reflection coefficients are computed
in the same way for both.  The asymptotic region is described
by harmonic MD on a regular lattice, and the normal modes are the 
well-known plane wave solutions.  As in Eq.\ (\ref{scattAsymp}),
the asymptotic form of the displacements is known for each
frequency:  
\begin{equation}
u_j(t) =
\left\{
\begin{array}{ll}
A \left( 
  e^{i k x_j - i \omega t}
  + r \, e^{-i k x_j - i \omega t}  \right) & {\mathrm{for~j\le 1}} \\
A \, t \, e^{i k x_j - i \omega t} & {\mathrm{for~j\ge N}} 
\end{array}
\right.
\label{scattAsymp1D}
\end{equation}
where the reflection and transmission coefficients are given
by $R=|r|^2$ and $T=|t|^2$, respectively.  We have assumed
that the scattering region is contained between $x_1$ and
$x_N$, in the sense that these points bound the CG region
of the mesh and are separated by more than the range of the 
MD potential from any coarse-grained cell.  This requirement
guarantees that 
the form of $u_j(t)$ in the relation (\ref{scattAsymp1D}) 
is a strict equality and not just an asymptotic relation like
(\ref{scattAsymp}).
The wave number $k$
is determined by the frequency, $m \omega^2 = D(k)$, where
$D(k)$ is the MD dynamical matrix.  The leading
coefficient $A$ just determines the amplitude and is irrelevant
for our purposes, so we set $A=1$.  
In principle, the displacement field could
have components with many different frequencies, but since the
problem is linear, we may restrict to a single frequency without
loss of generality.  Note that while we are considering the
harmonic problem with a short-range MD potential, this analysis 
could be generalized to 
non-linear wave propagation and non-linear or long-range scattering
using the LSZ scattering formalism. \cite{Peskin} The
incoming and outgoing waves forming the asymptotic
boundary conditions (\ref{scattAsymp1D}) would need to
be suitably dressed.  Then the scattering cross section
could be expressed in terms of the one-point irreducible
Feynman graphs.  This extension could be interesting, but
the linear case will suffice for our purposes.

The equations of motion are given by
\begin{equation}
M_{ij} \ddot{u}_j(t) = - K_{il} u_{l}(t)
\end{equation}
and we look for solutions with angular frequency $\omega$,
\begin{equation}
u_{j}(t) = e^{i k x_j - i \omega t} + v_j e^{- i \omega t}
\end{equation}
such that the asymptotic boundary conditions (\ref{scattAsymp1D})
are satisfied: 
\begin{equation}
v_j =
\left\{
\begin{array}{ll}
      r \, e^{-i k x_j } & {\mathrm{for~j\le 1}} \\
  (t-1) \, e^{ i k x_j } & {\mathrm{for~j\ge N}} 
\end{array}
\right.
\label{scattAsymp1Dv}
\end{equation}
where again these boundary conditions are a strict equality.
The equations of motion for $v_j$ are
\begin{equation}
\left( K_{ij} -\omega ^2 M_{ij} \right) v_j = 
 - \left( K_{ij} -\omega ^2 M_{ij} \right) e^{i k x_j}
\label{veom}
\end{equation}
In principle, there are many ways to solve the equations of
motion (\ref{veom}) with the boundary conditions (\ref{scattAsymp1Dv});
in practice, we found this problem to be rather subtle.  A similar
scattering problem must have been solved before, but we have not 
been able to find a solution in the literature.  The approach
we take here is to note that we can relate the solution in the
outer regions to the solution at the boundary points
\begin{eqnarray}
v_{1-n} & = & e^{i n k a} v_1 \\
v_{N+n} & = & e^{i n k a} v_N 
\end{eqnarray}
where $n\ge0$.  Here $a$ is the lattice constant. 
Using this trick, the problem is reduced to the calculation
of $v_1$,\ldots,$v_N$ using the following $N$ equations:
\begin{eqnarray}
\left( K_{ij} -\omega ^2 M_{ij} \right) v_j & = & 
 - \left( K_{ij} -\omega ^2 M_{ij} \right) e^{i k x_j} 
\label{veomBC}
 \\ & &  ~~~~
 + \sum _{n=1}^\infty K_{i(1-n)} e^{i n k a} v_1 
 \nonumber \\ & &  ~~~~
 + \sum _{n=1}^\infty K_{i(N+n)} e^{i n k a} v_N
 \nonumber
\end{eqnarray}
where in practice the sums just run out to the range of the 
potential.  The implicit sums over $j$ run from $1$ to $N$.
Then the scattering coefficients are determined by
$\RR=|r|^2$ and $\TT=|t|^2$ with
\begin{eqnarray}
r & = & e^{i k x_1} v_1 \\
t & = & e^{-i k x_N} v_N + 1
\end{eqnarray}
which follow from Eq.\ (\ref{scattAsymp1Dv}).

In Fig.\ \ref{fig-refl} we plot the reflection coefficient 
$\RR (k)$ for scattering from a CG region of 72 nodes representing 
652 atoms in the middle of an infinite harmonic chain of atoms.  
The reflection coefficients for CGMD, lumped mass FEM and distributed
mass FEM are plotted.
The lattice is shown in Fig.\ \ref{fig-scattLatt}.  The cell size 
increases smoothly in the CG region, as it should, 
to a maximum of $N_{max}=20$ atoms per cell.  
In all three cases shown $\RR$ becomes exponentially small in the 
long wavelength limit, and it goes to
unity as the wavelength becomes smaller than the 
mesh spacing---a coarse mesh cannot support short wavelength modes.
The threshold occurs approximately at $k=\pi /(N_{max} a)$,
the natural cutoff corresponding to a wavelength of 
$\lambda = 2 N_{max} a$.  The threshold for CGMD takes place
almost exactly at this point because the CGMD phonon
frequencies are more accurate than those of the two
versions of FEM.  According to three dimensional
scattering theory in the limit of wavelengths much
larger than the size of the scattering region, the
scattering cross section should vary like $\sigma \sim k^4$.  
This favorable transmission of long wavelengths 
is the well known Rayleigh scattering that gives us
a blue sky.\cite{Rayleigh}  In these one dimensional scattering calculations,
the trend is for $\RR\sim k^\beta$ where the exponent $\beta$
is roughly $\beta \approx 4\pm1$ for FEM and $\beta \approx 8\pm2$ for CGMD.
We hypothesize that the difference is due to the improved 
accuracy of CGMD at long wavelengths.  Using the Born 
approximation, the scattering
strength should go roughly like $r\sim k^{\gamma}$, where $\gamma$ is
the order of the error, or $\gamma=2$ in FEM and 
$\gamma=4$ in CGMD.  Then the reflection coefficient would
go like $\RR = |r|^2 \sim k^{2\gamma}$, giving $\beta \approx 4$ for 
FEM and $\beta \approx 8$ for CGMD in rough agreement with the
numerical solution; however, we should stress that this hypothesis
has not been proved mathematically and the resonance structure
of the scattering curve leads to large uncertainties in the $\beta$ fit.
If we fit to the tops of the peaks at long wavelengths in scattering
from an abruptly changing mesh, the uncertainty is reduced to
$\beta \approx 4\pm0.2$ and $8\pm0.2$ for FEM and CGMD, respectively.

\begin{figure}[tbp]
\includegraphics[width=0.45\textwidth]{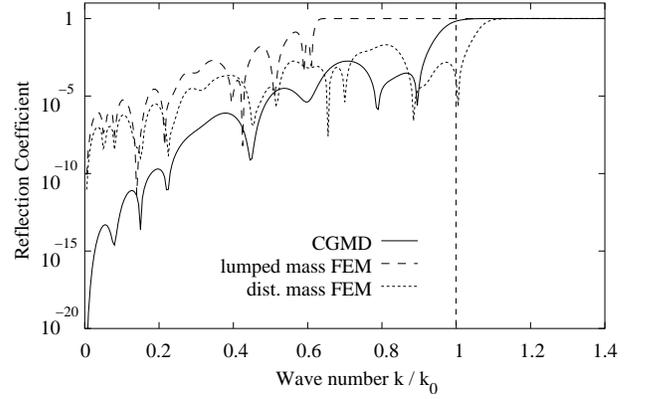}
\caption{A comparison of the reflection of elastic waves from
a CG region in three cases: CGMD and two varieties of FEM.
Note that the reflection coefficient is plotted on a log scale.
A similar graph plotted on a linear scale is shown in Ref. \onlinecite{CGMD}.
The dashed line marks the natural cutoff [$k_0= \pi / (N_{max} a)$],
where $N_{max}$ is the number of atoms in the largest cells.  
The bumps in
the curves are scattering resonances.  Note that at long wavelengths
CGMD offers significantly suppressed scattering.
\label{fig-refl}}
\end{figure}

A series of bumps is visible in each of the curves in the 
transmissible region.  Most of these bumps were not visible
in plot of the reflection coefficients in Fig.\ 2 of 
Ref.\ \onlinecite{CGMD}
with a linear scale.  The log scale used in Fig.\ \ref{fig-refl} 
brings out these features in regions with extremely low 
scattering.  These bumps are scattering resonances, wavelengths
at which the scattering cross section is increased because the
frequency of the incoming wave is in resonance with an internal
mode of the scattering region.  Of course, they are much more
peaked on a linear scale, where their width is an indication of
the lifetime of the state.  The curvature of the peaks in the
log-linear plot is low, indicating short-lived resonances.  The
height of the peaks is an indication of the scattering strength.
If a peak were high and narrow, it would indicate a strongly
scattering localized mode, which would be pathological behavior
in a concurrent multiscale simulation.  In general, 
weak scattering with broad resonances (if any) is desirable.
The wave reflections in CGMD a weaker and the resonances
much less strong than in FEM, although the distributed mass
FEM actually has a 10\% higher threshold than CGMD because its
frequencies are higher.

\begin{figure}[tbp]
\includegraphics[width=0.45\textwidth]{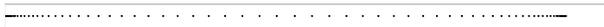}
\caption{A plot of the mesh used for the scattering calculations.
The atomic scale mesh, partially visible at the ends, extends
infinitely to the right and left.  The mesh is commensurate with
the underlying atomic lattice, and the largest cell size is
$N_{max}=20$ atoms.
\label{fig-scattLatt}}
\end{figure}

It is also interesting to consider the transmission coefficient,
plotted in Fig.\ \ref{fig-trans} for CGMD and the two versions
of FEM.  There is a simple relationship between the transmission
and reflection coefficients, $\TT=1-\RR$, so in principle the
calculation of one is equivalent to the other.  However,
because the log-linear plot brings out features near zero
while suppressing features near unity, the two plots show
different information.  
The transmission
coefficient drops exponentially above the threshold, similar to
quantum mechanical tunneling through a rectangular potential barrier
or the transmission of evanescent waves in optics.  As in those
cases the transmission coefficient also decreases exponentially 
as the size of the scattering region is increased.  
One interesting feature of the transmission
coefficient curves is the absence of resonances.  
The peaks are absent because the CG region lacks the degrees of 
freedom that would cause resonances at these high frequencies.

\begin{figure}[tbp]
\includegraphics[width=0.45\textwidth]{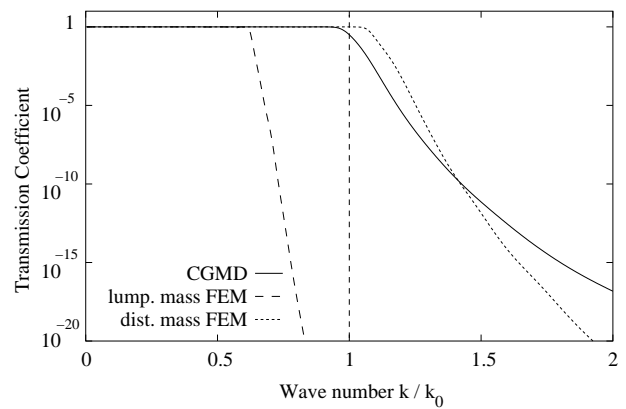}
\caption{A comparison of the transmission of elastic waves through 
a CG region in three cases: CGMD and two varieties of FEM.
The dashed line marks the natural cutoff [$k_0= \pi / (N_{max} a)$].  
Note that the bumps evident in the plot of the reflection
coefficient are absent from the transmission coefficient.
\label{fig-trans}}
\end{figure}

We have calculated the scattering on many different coarse-grained
regions.  The general features of the reflection coefficient curves
remain as the mesh is varied, but the details change.  One of the
most pronounced changes happens if the mesh varies too abruptly.
In this case, strong scattering resonances may occur near the
threshold, even for CGMD, as shown in Fig.\ \ref{fig-scattAbrupt}. 
Note the linear scale in this plot. The mesh used for the 
comparison between FEM and CGMD, shown in 
Fig.\ \ref{fig-scattLatt}, was generated using a $\tanh$ function
for the cell size to ensure smoothness.  For comparison, we have
plotted in Fig.\ \ref{fig-scattAbrupt} the reflection coefficient 
for a CG region of the same size but consisting almost exclusively 
of cells of size $N_{max}=20$, and it is clear that the abrupt
change in mesh size leads to much stronger resonances.  The
increased reflection of waves with $k$ at resonance could lead
to an unphysical size scale, and smooth meshes should be used
to prevent this undesirable behavior.
Apart from ensuring
smoothness, we have not optimized the mesh, and it may be possible
to reduce the scattering further still through a more optimized
mesh.

\begin{figure}[tbp]
\includegraphics[width=0.45\textwidth]{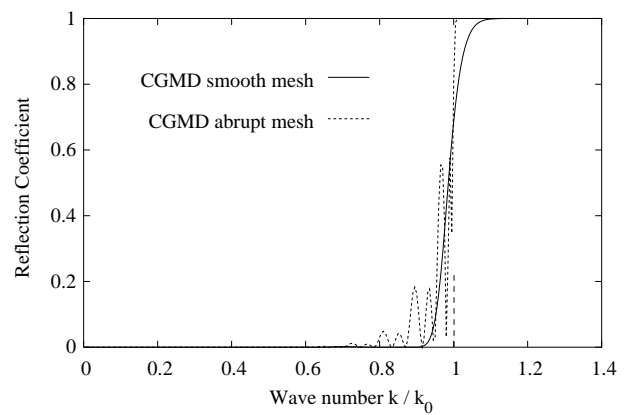}
\caption{A comparison of the reflection of elastic waves from 
a CG region for a mesh with smoothly varying cell size and 
one with an abrupt change in cell size, both computed in
CGMD.
The dashed line marks the natural cutoff [$k_0= \pi / (N_{max} a)$].  
Note that  on the linear scale the resonances are not visible
for the smooth mesh, but are quite pronounced for the abruptly
changing mesh.
\label{fig-scattAbrupt}}
\end{figure}

It should also be noted that the plots of wave reflection on a
log-linear scale provide a sensitive test of the numerical
formulation.  The CGMD data plotted in Figs.\ \ref{fig-refl} 
are not the same data plotted in Fig.\ 2 of
Ref.\ \onlinecite{CGMD}.  The data were noisy at the level
of $10^{-10}$, and we ultimately recalculated the stiffness
matrix using Eq.\ (\ref{stiffA1}) in order to remove the noise.

The reflection coefficient provides a measure to quantify
the 1D scattering properties of CGMD.  What information does
it give about the reflection problem.  The actual amount of
reflection in any application is the product of the elastic
wave power spectrum and the reflection coefficient.  Suppose
the application of interest (e.g.\ crack propagation) can
only tolerate scattering at 1\% of the natural level.  If
99\% of the power is at wavelengths greater than $\lambda _c$,
then there will be an acceptable level of scattering if the
mesh size is less than or equal to $\lambda _c$.

In applications like the crack propagation problem, it may
be important to consider non-linear effects as well.  In
the anharmonic MD crystal, waves of sufficiently large
amplitude will steepen into shock waves.  The wave velocity
increases with the pressure, so that a wavefront with a slow
rise to a higher pressure will steepen into a step-like 
shock wave in which the abruptness of the rise is ultimately
limited by dissipative processes at the front.  As a result, 
compressive waves that are generated at the crack tip evolve 
into shock waves that have a strong impact on the crack if
they return to it due to the boundary conditions.  The
reflection coefficient is a property of the system in the
small amplitude, harmonic limit, and as such does not give
any information about the behavior of shock waves.  Shock
waves, of course, have an abrupt rise and hence have power
at short wavelengths localized at the wavefront.  When a
shock wave is incident on the CG mesh, the short wavelengths
are reflected.  Since the mesh spacing increases gradually,
this reflection disperses the power at the front; i.e., the
shortest waves are reflected first, then the next shortest and
so on.  The shock wave is dispersed and much of the power
flows out to the CG mesh, so the reflected wave is a low
amplitude wave that does not steepen into a shock wave.  Thus,
while some short wavelength components are reflected, they
are not shock waves and do not appear to have an appreciable
impact on the processes in the MD region.  The majority of
the power is carried out into the CG region, effectively
delaying its return to the MD region by the transit time
across the CG region.

This dispersion and delay in wave reflection due to the CG region
is the way CGMD and FEM/MD hybrid motheods solve the reflection
problem.  Several other solutions have been proposed that make use 
of absorbing boundaries.\cite{AdelmanDoll,Cai,WeinanE,Liu,MRS}  
Those techniques have much lower scattering of short wavelengths 
and hence a lower Kapitza resistance at zero temperature.  They
also involve considerable computer memory usage and 
considerable coding overhead.  At this point it is clear that
several approaches exist that solve the wave reflection 
problem in principle, but it is not yet clear which will
offer the ease of use and scalability that will be demanded
for widespread use in large-scale simulation.

%
%
%

\section{Conclusion}
\label{sect-conc}

Coarse-grained molecular dynamics provides a means of concurrently coupling
MD with a coarsened description of the mechanics similar to FEM.  The
practical formulation relies heavily on the properties of a crystal
lattice, and it is therefore suited to solids.  The formulation
discussed here is based on a Hamiltonian, a conserved energy for
the system, and is free from ghost forces.  
We have applied CGMD to three dimensional systems with interatomic
potentials that are many-body in nature and
extend well beyond nearest neighbors.
In this article, we have elaborated on how CGMD is implemented to 
include anharmonic (non-linear) and finite temperature effects.  
It has been shown
that these effects are described by analytic matrix formulas that
may be precomputed prior to the simulation.  The formalism is
useful from the point of view of both providing the means to 
take the calculations beyond the harmonic description and to
be able to estimate the errors that are made in resorting to
a harmonic model.  In the process we have shown how CGMD is
related to other concurrent multiscale methodologies such as
the Quasicontinuum technique.

We have also reported on several calculations done with CGMD in
order to understand its properties.  The elastic wave spectra for
solid argon and tantalum have been computed in MD, CGMD and FEM for
comparison.  In some cases it has been possible to derive
largely analytic formulas for these spectra and to analyze
their differences.  In this investigation the MD spectrum
has been taken as the ideal, and the CGMD and FEM models
have each been formulated to give within its own framework
the optimal agreement with MD.  Several interesting results
were found, including some reported briefly in previous
publications.  First, both CGMD and FEM agree well with MD
in the long wavelength limit, as expected.  It was shown
that CGMD provides a better description, in that the error
is $\OO(k^4)$, than FEM with errors of the order $\OO(k^2)$.
Second, throughout the Brillioun zone, the CGMD errors are
smaller than those of FEM, especially at certain points on
the zone boundary.  Of course, the CGMD spectrum is exact
in the limit of one atom per mesh cell.

We have also used elastic wave spectra to examine the effect
of the internal relaxation terms present in CGMD.  This 
relaxation is the difference between forcing the atomic
displacements to agree exactly with the interpolated
displacement field and allowing fluctuations about the
interpolated displacement field provided the interpolated
field remains the best fit to the atomic displacements.
We have shown how the relaxation effects can be eliminated 
to produce a rigid approximation to CGMD (R-CGMD), 
a formalism similar to the Quasicontinuum technique.  
We have compared the CGMD and
R-CGMD elastic wave spectra and found that the two agree
well in the long wavelength limit, but that CGMD provides
a better description of elastic waves of short and moderate
wavelengths.  The internal relaxation is able to give a
better description of the energetics of waves changing
rapidly on the mesh.

Much remains to be done with CGMD, and we are actively
developing the model.  Many questions of numerical 
efficiency still need to be addressed.  A controlled
means of bandwidth reduction for the CGMD stiffness
matrix is needed.  Also an efficient and consistent
treatment of wave absorption is an open challenge.  We
have not discussed the implementation of CGMD on
parallel platforms; the decomposition of the MD region
into parallel domains is straightforward, but the
decomposition of the CG region is less obvious and is
linked to the question of the stiffness bandwidth.  
Finally, the formulation of CGMD presented here provides
the foundation for use in full-scale applications, a 
subject to which we plan to return soon.

\acknowledgments

We would like to acknowledge that much of this work was done 
at the Naval Research Laboratory and supported by ONR and DARPA.  
Support for this work was also provided by DOE/BES
and the LLNL Laboratory Directed Research and Development 
program.
This work was performed in part under the auspices of the 
U.S. Department of Energy 
by University of California, Lawrence Livermore National Laboratory 
under Contract W-7405-Eng-48.
R.~E.~R.\ would like to thank J. Feldman for insightful discussions,
and to thank A. El-Azab, B. Kraczek, B. Sadigh and many others for 
encouragement to finish this article.

\appendix

\section{Alternate Derivation of CG Hamiltonian}
\label{app-CGderiv}

In this Appendix we compute the CG Hamiltonian using a more
rigorous and straightforward, albeit laborious, approach.
The choice to put this computation in an appendix rather
than the main text was made because it repeats a calculation
done in the main text. It gives different formulas for the
mass and stiffness matrices, which are quite useful, and
the positioning in an appendix is not intended to indicate
that these derivations and formulas are less important than
those in the main text.
For this computation, as in Section \ref{subsec-harmLatt}, we will 
concentrate on calculating the contribution of the potential 
to the partition function (\ref{CGpot1}):
\begin{eqnarray}
Z_{{\mathrm{pot}}}(u_k; \beta) & = & 
 \int \!  du \, \Delta \,
 e^{-\frac{1}{2} \beta u_{\mu} D_{\mu \nu} u_{\nu} } 
\label{CGpotA1}
\end{eqnarray}
where once again we combine the atomic and
spatial indices to form the 3$\natom$-dimensional 
configuration space for the atoms and the
3$\nnode$-dimensional space for the CG displacements.  
Notice that we have left $\Delta$ as a product of
$\delta$ functions, rather than using the Fourier
representation as we did above:
\begin{equation}
\Delta = \prod _{j=1}^{3\nnode}
\delta \left( u_j - f_{j\mu} u_{\mu} \right) .
\label{realConstraint}
\end{equation}
Suppose that the constraints were of a particularly
simple form:
\begin{equation}
\Delta ^{{\mathrm{simple}}} = \prod _{j=1}^{3\nnode}
\delta \left( u_j - \delta_{j\mu} u_{\mu} \right)
\label{toyConstraint}
\end{equation}
Then the evaluation of the integral would be easy.
The first $3\nnode$ atomic displacements would
not be integrated, but rather just set to the corresponding
$u_j$ value.  The remaining $3(\natom-\nnode)$ degrees of
freedom would be integrated by completing the square and
evaluating the Gaussian integrals.  The only problem is
that the constraints are not of the simple form 
(\ref{toyConstraint}).

We must introduce some mathematical formalism to 
transform the constraints into a form analogous to
the simple kind (\ref{toyConstraint}).
The approach we take in this derivation is based
on an explicit factorization of the 3$\natom$-dimensional space
into the direct product of the subspace spanned by the 
constraints and the orthogonal subspace.  In order
to accomplish this factorization we introduce two
projection matrices in 3$\natom$-dimensional space
\begin{eqnarray}
P^{{\mathrm{CG}}}_{\mu\nu} & \equiv &
   f_{j\mu} \left( f_{j\lambda}f_{k\lambda} \right) ^{-1}  f_{k\nu} 
   \label{PCG1} \\
   & = & N_{j\mu} \left( N_{j\lambda}N_{k\lambda} \right) ^{-1}  N_{k\nu} 
   \label{PCG2} \\
   & = & N_{j\mu} f_{j\nu} 
   \label{PCG3} \\
P^{\perp}_{\mu\nu} & \equiv & \delta _{\mu\nu} - P^{{\mathrm{CG}}}_{\mu\nu}
   \label{pperp}
\end{eqnarray}
where repeated indices are summed, as usual, and
the inverses are $3\nnode\times3\nnode$ matrix
inverses.
These are projection matrices in the sense that 
$P^{{\mathrm{CG}}}_{\mu\nu} f_{j\nu} = f_{j\mu}$
and $P^{\perp}_{\mu\nu}  f_{j\nu} = 0$.
Since $N_{j\mu}$ is a linear combination of $f_{k\nu}$,
it likewise holds that 
$P^{{\mathrm{CG}}}_{\mu\nu} N_{j\nu} = N_{j\mu}$
and $P^{\perp}_{\mu\nu}  N_{j\nu} = 0$.
The matrices are useful because $P^{{\mathrm{CG}}}$ 
projects onto the constrained subspace:
\begin{eqnarray}
P^{{\mathrm{CG}}}_{\mu\nu} u_{\nu} & = & 
N_{j\mu} f_{j\nu} u_{\nu} \\
  & = & N_{j\mu} u_j
\end{eqnarray}
where we have used Eq.\ (\ref{PCG3}).
Thus, whenever $P^{{\mathrm{CG}}}$ acts on
the configuration space its result depends
only on the nodal displacements; it is completely
independent of the unconstrained degrees of 
freedom.  This is just what we need.

Using the fact that the two projection matrices sum to the identity, 
$\delta _{\mu\nu} = P^{\perp}_{\mu\nu} + P^{{\mathrm{CG}}}_{\mu\nu}$,
we can rewrite the potential part of the partition function
\begin{eqnarray}
Z_{{\mathrm{pot}}} & = & 
 \int \!  du \, \Delta \,
 e^{-\frac{1}{2} \beta u_{\mu} 
\left( P^{\perp}_{\mu\rho} + P^{{\mathrm{CG}}}_{\mu\rho} \right) 
  D_{\rho \sigma} 
\left( P^{\perp}_{\sigma\nu} + P^{{\mathrm{CG}}}_{\sigma\nu} \right) 
u_{\nu} } 
\nonumber \\
 & = & 
 \int \!  du \, 
 e^{-\frac{1}{2} \beta \left( u_{\mu} D^{\perp}_{\mu\nu} u_{\nu} +
2 u_{j} D^{\times}_{j\mu} u_{\mu}  \right) } \times
\nonumber \\
& & ~~~~~~~~~~ \Delta \, e^{-\frac{1}{2} \beta 
u_{j} N_{j\mu} D_{\mu\nu} N_{k\nu} u_{k}
} 
\label{CGpotA2}
\end{eqnarray}
where we have introduced the notation of the
orthogonal and cross components of the dynamical
matrix:
\begin{eqnarray}
D^{\perp}_{\mu\nu} & = & 
P^{\perp}_{\mu\rho} D_{\rho \sigma} P^{\perp}_{\sigma\nu} \\
D^{\times}_{j\mu} & = & P^{\perp}_{\mu\rho} D_{\rho \nu} N_{j\nu} 
\end{eqnarray}
Note that there is no $D^{CG}$ defined or used in Eq.\ (\ref{CGpotA2}).

We can now compute the integral (\ref{CGpotA2})
with a slight trick.  We can integrate over the
constrained degrees of freedom
\begin{eqnarray}
Z_{{\mathrm{pot}}} & = & 
 C \int \!  d^{\perp}u \, 
 e^{-\frac{1}{2} \beta \left( u_{\mu} D^{\perp}_{\mu\nu} u_{\nu} +
2 u_{j} D^{\times}_{j\mu} u_{\mu}  \right) }
\nonumber \\
& & ~~~~~~~~~~ \times \, e^{-\frac{1}{2} \beta 
u_{j} N_{j\mu} D_{\mu\nu} N_{k\nu} u_{k}
} 
\label{CGpotA3}
\end{eqnarray}
where the argument of the exponential is independent
of the constrained subspace, so $C$ is just a constant
(independent of $\beta$) and hence irrelevant.
Now we can restore the constrained subspace
in order to make the integral easier
by inserting an integral that equals unity:
\begin{eqnarray}
1 & = & \int \!  d^{CG}u \, 
  \left( \frac{2\pi}{\varepsilon \beta} \right) ^{-3\nnode/2}
 e^{-\frac{1}{2} \beta u_{\mu} \varepsilon P^{{\mathrm{CG}}}_{\mu\nu} u_{\nu} }
\end{eqnarray}
where $\varepsilon$ is an arbitrary constant that we are free to
determine below.
This integral is unity because $P^{{\mathrm{CG}}}$ is just
the identity matrix on the constrained subspace.
Inserting this expression into Eq.\ (\ref{CGpotA3}), we have
\begin{eqnarray}
Z_{{\mathrm{pot}}} & = & 
 C \int \!  du \, 
 e^{-\frac{1}{2} \beta \left[ u_{\mu} \left( D^{\perp}_{\mu\nu} +
\varepsilon P^{{\mathrm{CG}}}_{\mu\nu} \right) u_{\nu} + 
2 u_{j} D^{\times}_{j\mu} u_{\mu}  \right] } \times
\nonumber \\
& & ~~~ \left( \frac{2\pi}{\varepsilon \beta} \right) ^{-3\nnode/2}
  \, e^{-\frac{1}{2} \beta u_{j} N_{j\mu} D_{\mu\nu} N_{k\nu} u_{k} } 
\label{CGpotA4}
\end{eqnarray}
where now we are left with an Gaussian integral over all space.
Most importantly, the matrix in the Gaussian
\begin{eqnarray}
\tilde{D}_{\mu\nu} & = & D^{\perp}_{\mu\nu} +
\varepsilon P^{{\mathrm{CG}}}_{\mu\nu} 
\label{Dtilde}
\end{eqnarray}
is non-singular provided $\varepsilon\ne0$, as it must be
since $Z_{{\mathrm{pot}}}$ has been well defined at each
step of the calculation.

Now we complete the square to transform Eq.\ (\ref{CGpotA4})
into a purely quadratic Gaussian integral using the
shift
\begin{eqnarray}
\tilde{u}_{\mu} & = & u_{\mu} + \tilde{D}_{\mu\nu}^{-1} D^{\times}_{j\nu} u_j
\end{eqnarray}
we find
\begin{eqnarray}
Z_{{\mathrm{pot}}} & = & 
C \left( \frac{2\pi}{\beta} \right) ^{3\deltaN /2}
  \frac{\left( \varepsilon \right)^{3\nnode/2}}{\left(\det \tilde{D}\right)^{1/2}}
  \, e^{-\frac{1}{2} \beta u_{j} K_{jk} u_{k} } .
\label{CGpotA5}
\end{eqnarray}
where $\deltaN=\natom-\nnode$. 
We find the important result that the CG stiffness matrix is given by
\begin{equation}
K_{jk} = N_{j\mu} D_{\mu\nu} N_{k\nu} 
   - D^{\times}_{j\mu} \tilde{D}_{\mu\nu}^{-1} D^{\times}_{k\nu}
\label{stiffA}
\end{equation}
and each matrix in this expression is well defined.
In principle, this expression holds for any non-zero $\varepsilon$;
in practice, it is advantageous to choose $\varepsilon$
to be at the upper end of the eigenspectrum of $D_{\mu\nu}$
so that $\tilde{D}_{\mu\nu}$ is well conditioned.

The same analysis can be carried out for the mass matrix.
The result is
\begin{eqnarray}
M_{jk} & = & N_{j\mu} m_{\mu} N_{k\mu} 
   - M^{\times}_{j\mu} \tilde{M}_{\mu\nu}^{-1} M^{\times}_{k\nu}
\label{massA} \\
       & = & m N_{j\mu} N_{k\mu} ~~~{\mathrm{for~monatomic~systems}}
\end{eqnarray}
In this case, $\varepsilon$ is chosen to be the average mass.

\section{Effective Potential}
\label{app-EffPot}

The CGMD formalism we have developed is an effective theory
in the sense that the short wavelength modes are integrated
out to determine the effective interaction of the long
wavelength modes.  In field theory, an effective potential
is computed that is somewhat different in character.  The
typical approach would be to define the coarse-grained
fields as an expectation value of the corresponding 
combination of microscopic degrees of freedom.  Note that
this differs from the approach we have taken in that we
constrain the coarse-grained fields to equal that 
combination of microscopic degrees of freedom: they are
identical and not an ensemble average.  It is only the
degrees of freedom that are integrated out that are
treated as an ensemble average.  We have taken this
approach because at least in principle there can be
bifurcation points in the trajectories of the coarse-grained
degrees of freedom that would be eliminated by defining
them as ensemble averages.  Nevertheless, the conventional
effective potential approach has a certain theoretical
elegance, and it could be useful in some contexts.
We will therefore give a brief discussion of the CGMD 
effective potential and investigate its usefulness.
Hopefully, in the process we will eliminate any confusion
that might arise between the two approaches.

Again we start with the Helmholtz free energy in the presence
of a source $F(\mathbf{J}_k)$:
\begin{equation}
Z(\mathbf{J}_k) = e^{-\beta F(J_k)} ~=~ \int du~dp \ 
  \exp \left( -\beta H^{MD} - \mathbf{J}_k \cdot N_{k\mu} 
\mathbf{u}_{\mu} \right)
\end{equation}
Just as in spin systems the magnetization is the derivative of the
Helmholtz free energy with respect to the external field, the
expectation value of the coarse-grained displacement is the
derivative of this CGMD Helmholtz free energy with respect to
the conjugate source:
\begin{equation}
-\left. \frac{\partial F}{\partial \mathbf{J}_k} \right| _{\beta} =
\langle N_{k\mu} \mathbf{u}_{\mu} \rangle = \mathbf{u}_k
\end{equation}
One could then go further and take the Legendre transform
to find the Gibbs free energy $G(\mathbf{u}_k)$
\begin{equation}
G = F + \mathbf{u}_k \cdot \mathbf{J}_k
\end{equation}
for which 
\begin{equation}
\left. \frac{\partial G}{\partial \mathbf{u}_k} \right| _{\beta} =
\mathbf{J}_k
\end{equation}
and so is a minimum with respect to $\mathbf{u}_k$ at equilibrium
when $\mathbf{J}_k=0$.

The challenge is to derive an expression for $G(\mathbf{u}_k)$
given that it is expressed in terms of $F(\mathbf{J}_k)$. A
calculation of $\mathbf{J}_k(\mathbf{u}_j)$ is needed.
Fortunately, the formalism is well developed in field theory
\cite{Peskin}, and the result is that $G(\mathbf{u}_k)$ is
the generating function for one-point irreducible (1PI)
Feynman diagrams.  The 1PI graphs are those that cannot
be divided into two disconnected diagrams by breaking a
single internal line.  This is a subset of the connected
diagrams we have considered for CGMD (cf.\ Fig.\ \ref{fig-feynman}),
so the effective potential approach does lead to a simplified
formalism.  We have not explored this model in depth.  It
might be interesting to do so, but we believe that the
approach presented in the body of the text is more meaningful
for coarse-grained solid mechanics problems.

%
%


\vspace{-3mm}


\begin{thebibliography}{99}

\bibitem{Moriarty} J.~A. Moriarty, J.~F. Belak, R.~E. Rudd, P. Soderlind, 
F.~H. Streitz and L.~H. Yang,  
J. Phys.: Condens. Matter {\bf 14}, 2825 (2002).

\bibitem{Hao} S. Hao, B. Moran, W. K. Liu and G. B. Olson, 
J. Comput.-Aided Mater. Design {\bf 10}, 99 (2003).

\bibitem{PSS} R. E. Rudd and J. Q. Broughton, 
Phys. Stat. Sol. (b) {\bf 217}, 251 (2000).

\bibitem{turb} W. D. McComb, 
{\em The Physics of Fluid Turbulence},
(Clarendon Press, Oxford, 1990);
A. A. Townsend,
{\em The Structure of Turbulent Shear Flow},
(Cambridge Univ. Press, Cambridge, 1976), 2nd ed.

\bibitem{Clementi}  E. Clementi, Philos. Trans. R. Soc. London, Ser. A 
{\bf 326}, 445 (1988).

\bibitem{Bimberg} V. A. Shchukin and D. Bimberg,
Rev. Mod. Phys. \textbf{71}, 1125 (1995).

\bibitem{EQDprl} R. E. Rudd, G.~A.~D. Briggs, A.~P. Sutton, 
G. Medeiros-Ribeiro and R.~S. Williams,  
Phys. Rev. Lett. {\bf 90}, 146101 (2003).

\bibitem{PLiu} P. Liu, Y. W. Zhang and C. Lu,
J. Appl. Phys. {\bf 94}, 6350 (2003).

\bibitem{Roukes} A. N. Cleland and M. L. Roukes, 
Appl. Phys. Lett. {\bf 69}, 2653 (1996). 

\bibitem{JMSM} R. E. Rudd and J. Q. Broughton,
J. Modeling and Simulation of Microsystems \textbf{1}, 29 (1999).

\bibitem{DTM99} R. E. Rudd and J. Q. Broughton, 
in Proc. DTM'99, Paris, France, B Courtois, et al, eds. (SPIE, Bellingham WA, 1999), Vol. 3680, pp.~104-13.

\bibitem{IJMCE} R. E. Rudd, 
Intl. J. on Multiscale Comput. Engin. {\bf 2}, 203 (2004)

\bibitem{MRS01} R. E. Rudd, 
Mat. Res. Soc. Symp. Proc. {\bf 677}, AA1.6.1-12 (2001).

\bibitem{MBKV} J. Q. Broughton, C. A. Meli, P. Vashishta, and R. K. Kalia,
Phys. Rev. B {\bf 56}, 611 (1997).


\bibitem{FaridWhite} F. F. Abraham, R. Walkup, H. Gao, M. Duchaineau, 
T. Diaz de la Rubia, M. Seager,
Proc. Natl. Acad. Sci. USA, {\bf 99}, 5783 (2002).

\bibitem{CoLS} F. F. Abraham, J. Q. Broughton, E. Kaxiras, and N. Bernstein,
Comput. in Phys. {\bf 12}, 538 (1998); Europhys. Lett. (1998).

\bibitem{CGMD} R. E. Rudd and J. Q. Broughton, 
Phys. Rev. B {\bf 58} R5893 (1998).

\bibitem{Alder} B. J. Alder and T. E. J. Wainwright, 
Chem. Phys. {\bf 26}, 1208 (1957).

\bibitem{AT} M. P. Allen and D. J. Tildesley,
{\em Computer Simulation of Liquids}, 
(Clarendon Press, Oxford, 1987).

\bibitem{earlyFEM} J. H. Argyris and S. Kelsey,
Aircraft Engin. {\bf 26} and {\bf 27} (1954-1955);
R.W. Clough,
Proc. Second ASCE Conf. on Electronic Comput.,
Pittsburgh PA, pp. 345-378 (1960).

\bibitem{FE} See for example: O. C. Zienkiewicz and R. L. Taylor, 
{\em The Finite Element Method}, 
4th ed. (McGraw-Hill, New York, 1991), Vols.\ I and II; \\
T. J. R. Hughes, {\em The Finite Element Method: Linear Static
and Dynamic Finite Element Analysis}, (Dover, Mineola, 2000).

\bibitem{KGF} S. Kohlhoff, P. Gumbsch, and H. F. Fischmeister, 
Philos. Mag. A {\bf 64}, 851 (1991).

\bibitem{CLS} J. Q. Broughton,  F. F. Abraham, N. Bernstein and E. Kaxiras, 
Phys. Rev. B {\bf 60}, 2391 (1999).

\bibitem{Phillips} E. B. Tadmor, M. Ortiz, and R. Phillips, 
Philos. Mag. A {\bf 73}, 1529 (1996).

\bibitem{KnapOrtiz} J. Knap and M. Ortiz, 
J. Mech. Phys. Solids {\bf 49}, 1899 (2001).

\bibitem{MillerTadmor} R. E. Miller and E. B. Tadmor,
J. Comput.-Aided Mater. Design {\bf 9}, 203 (2002). 

\bibitem{finiteTQC} L. Dupuy, R. E. Miller, E. B. Tadmor and R. Phillips, private communication.

\bibitem{Curtin} W. A. Curtin and R. E. Miller,
Modelling Simul. Mater. Sci. Eng. {\bf 11}, R33 (2003).

\bibitem{Liu}  H. S. Park, W. K. Liu,
Comput. Meth. Appl. Mech. Engng. {\bf 193} 1733 (2004).

\bibitem{Curtarolo} S. Curtarolo and G. Ceder,
Phys. Rev. Lett. {\bf 88}, 255504 (2002).

\bibitem{UNL} Z.-B. Wu, D. J. Diestler, R. Feng and X. C. Zeng,
J. Chem. Phys. {\bf 119}, 8013 (2003).

\bibitem{NW} G. J. Wagner and  W. K. Liu,
J. Comput. Phys. {\bf 190}, 249 (2003).

\bibitem{Park} H. S. Park, E. G. Karpov, W. K. Liu, 
Comput. Methods Appl. Mech. Engrg. {\bf 193}, 1713 (2004).


\bibitem{CADD} L. E. Shilkrot, R. E. Miller, W. A. Curtin,
Phys. Rev. Lett. {\bf 89}, 025501 (2002).
 
\bibitem{footnote-Drange} To be precise, the CGMD equations of motion agree 
with the atomistic equations of motion for those atoms separated
by more than the range of the potential from the coarse-grained
region.

\bibitem{footnote-adiabat} In
practice, it may be difficult to determine a priori whether
the missing modes would vary adiabatically.  This remains an
area where additional theoretical work would be helpful.

\bibitem{LandauLifschitz} L. D. Landau and E. M. Lifshitz,
  {\em Theory of Elasticity}, 3rd ed. (Pergamon, Oxford, 1986) Ch. 3.

\bibitem{Braginsky} V. B. Braginsky, V. P. Mitrofanov and I. V. Panov, 
  {\em Systems with Small Dissipation},
  (Univ. Chicago Press, Chicago, 1986).

\bibitem{Dupuycalc}  We have learned recently that L. Dupuy and coworkers 
have derived a similar expression independently (private communication).

\bibitem{PIMD} D. Chandler and P. G. J. Wolynes, 
Chem. Phys. {\bf 74}, 4078 (1981);
M. E. Tuckerman, G. J. Martyna, M. L. Klein, B. J. J. Berne, 
Chem. Phys. {\bf 99}, 2796 (1993). 

\bibitem{ICCN} R. E. Rudd,
``The Atomic Limit of Finite Elements in the Simulation of Micro-Resonators,''
in Proc. of the 3rd Intl. Conf. on Modeling and Simulation of Microsystems
(MSM2000), San Diego, Ca., March 27-29, 2000, M. Laudon and B. Romanowicz,
eds. (Compuatitional Publications, Boston, 2000), pp.\ 465-468.
 
\bibitem{WilliamsonZunger} A. J. Williamson, L. W. Wang, and A. Zunger,
Phys. Rev. B {\bf 62}, 12963 (2000).

\bibitem{GaussianInt} See for example, Appendix A of Ref.\ \onlinecite{Ramond}.

\bibitem{footnote-infinity} The somewhat disconcerting fact that this integral 
is infinite is ultimately irrelevant because the multiplicative 
infinity cancels in the calculation of the internal energy.

\bibitem{nonLoc} A. C. Eringen and D. G. B. Edelen,
Int. J. Eng. Sci. {\bf 10}, 233 (1972).

\bibitem{nonLoc2} S. B. Altan and E. C. Aifantis,
Scr. Met. Mater. {\bf 26}, 319 (1992).

\bibitem{Fdiagrams} 
C. Itzykson and J.-M. Drouffe, {\em Statistical Field Theory},
(Cambridge Univ. Press, Cambridge, 1991), Vol. II; \\
R. Kubo, M. Toda, and N. Hashitsume,
  {\em Statistical Physics II}, 
  (Springer-Verlag, Berlin, 1978).

\bibitem{Ramond} P. Ramond, {\em Field Theory: A modern primer},
(Addison-Wesley, Redwood City, 1989), 2nd ed.

\bibitem{SCPA} N. R. Werthamer, in {\em Rare gas solids}, 
  M. L. Klein and J. A. Venables, eds. 
  (Academic Press, New York, 1976) pp.265-300. 
 
\bibitem{Wallace} D. C. Wallace, {\em Thermodynamics of Crystals},
(Dover, New York, 1972).

\bibitem{footnote-regular} The use of $\tilde{D}_{\mu \nu}$ 
makes explicit the 
regularization that is implied in each occurrence of
$D_{\mu \nu}^{-1}$ above.  We return to this point in
Section \ref{subsec-zeromodes}.

\bibitem{Peskin} M. E. Peskin and D. V. Schroeder,
{\em An Introduction to Quantum Field Theory},
(Addison-Wesley, New York, 1995).

\bibitem{MRS} R. E. Rudd, 
Mat. Res. Soc. Symp. Proc. {\bf 695}, T10.2 (2002).

\bibitem{footnote-Nose} The
addition of Nose-Hoover thermostats violates 
energy conservation in the conventional sense, 
but still conserves a generalized Hamiltonian.

\bibitem{Zwanzig} R. Zwanzig, {\em Nonequilibrium Statistical Mechanics},
(Oxford Univ. Press, Oxford, 2001).

\bibitem{vVerlet} W. C. Swope, H. C. Andersen, P. H. Berens, K. R. Wilson,
J. Chem. Phys. {\bf 76}, 637 (1982).

\bibitem{BH} M. Born and K. Huang, 
{\em Dynamical Theory of Crystal Lattices},
(Oxford Univ. Press, Oxford, 1998).

\bibitem{Huebner} K. H. Huebner, E. A. Thornton, and T. G. Byrom,
{\em The Finite Element Method for Engineers}, 3rd ed., 
(Wiley, New York, 1995), p. 144.

\bibitem{serendip} J. G. Ergatoudis, B. M. Irons, and O. C. Zienkiewicz,
Int. J. Solids Struct. {\bf 4}, 31 (1968).

\bibitem{lumpDef} The FEM lumped mass matrix is
a diagonal approximation of the true distributed mass matrix
(cf.\ Ref.\ \onlinecite{FE}).

\bibitem{Grindlay} J.~Grindlay and R.~Howard, ``On the Lattice Dynamics and
  Specific Heat of the Rare-Gas Solids,''  in {\em Lattice Dynamics}, Proc.\
  Int.\ Conf.\, Copenhagen Denmark, R.~F.~Wallis, ed (Pergamon, Oxford, 1965),
  p.~129.

\bibitem{AM} See for example, N. W. Ashcroft and N. D. Mermin, 
{\em Solid State Physics},
(Saunders College Press, Philadelphia, 1976), Fig. 22.13.

\bibitem{groups} T. Inui, Y. Tanabe and Y. Onodera,
{\em Group Theory and Its Applications in Physics},
(Springer-Verlag, Berlin, 1996), Section 11.6.

\bibitem{FS} M. W. Finnis and J. E. Sinclair, 
Philos. Mag. {\bf 50}, 45 (1984).

\bibitem{Ackland} G. J. Ackland and R. Thetford, 
Philos. Mag. {\bf 56}, 15 (1987).

\bibitem{crackRefl} F. F. Abraham, D. Brodbeck, R. A. Rafey, W. E. Rudge,
Phys. Rev. Lett. {\bf 73}, 272 (1994).

\bibitem{crackRefl2} B. L. Holian and R. Ravelo,
Phys. Rev. B {\bf 51}, 11275 (1995).

\bibitem{AdelmanDoll} S. A. Adelman and J. D. Doll,
J. Chem. Phys. {\bf 61}, 4242 (1974).

\bibitem{Cai}  W. Cai, M. de Koning, V. V. Bulatov and S. Yip,
Phys. Rev. Lett. {\bf 85}, 3213 (2000).

\bibitem{WeinanE} W. E and Z. Huang,
Phys. Rev. Lett. {\bf 87}, 135501 (2001).

\bibitem{scatterRef} J. R. Taylor, {\em Scattering Theory},
(Krieger, Malabar, Florida, 1983).

\bibitem{Rayleigh} Lord Rayleigh, Phil. Mag. {\bf XLI}, 107, 274 (1871);
Phil. Mag. {\bf XLVII}, 375 (1899).

\end{thebibliography}
\end{document}